\documentclass[pre,amsmath,amssymb,twocolumn,nofootinbib,floatfix]{revtex4-1}
\usepackage[colorlinks=true,linkcolor=blue,urlcolor=blue,citecolor=blue]{hyperref}
\usepackage{graphicx}\usepackage{bm,color}\usepackage{times}
\definecolor{labelkey}{cmyk}{.4,.2,0,0}

\renewcommand{\log}{\ln}

\renewcommand{\doi}[2]{\href{http://dx.doi.org/#1}{#2}}
\newcommand{\arxiv}[1]{\href{http://arxiv.org/abs/#1}{#1}}

\newcommand{\ca}[1]{{\cal #1}}
\newcommand{\eq}[1]{(\ref{#1})}
\newcommand{\Eq}[1]{Eq.~(\ref{#1})}
\newcommand{\Eqs}[1]{Eqs.~(\ref{#1})}
\newcommand{\nn}{\nonumber}

\newcommand{\rmd}{\mathrm{d}}
\newcommand{\rme}{e}

\renewcommand{\log}{\ln}
\newcommand{\be}{\begin{equation}}
\newcommand{\ee}{\end{equation}}
\newcommand{\beq}{\begin{equation}}
\newcommand{\eeq}{\end{equation}}
\newcommand{\bea}{\begin{eqnarray}}
\newcommand{\eea}{\end{eqnarray}}

\usepackage{graphicx}

\newcommand{\fig}[2]{\includegraphics[width=#1]{#2.pdf}}

\newcommand{\Fig}[1]{\includegraphics[width=\columnwidth]{#1.pdf}}
\newcommand{\Figpng}[1]{\includegraphics[width=\columnwidth]{#1.png}}
\newcommand{\figpng}[2]{\includegraphics[width=#1]{#2.png}}
\newlength{\bilderlength}

\begin{document}

\title{First passage in an interval for fractional Brownian motion}

\author{Kay J\"org Wiese}
 \affiliation{CNRS-Laboratoire de Physique Th\'eorique de l'Ecole Normale
  Sup\'erieure, 24 rue Lhomond, 75005 Paris, France, PSL   University, Sorbonne Universit\'e.}

\begin{abstract}
Be $X_t$ a random process starting at $x \in [0,1]$ with absorbing boundary conditions   at both ends of the interval. Denote $P_1(x)$ the probability to first exit at the upper boundary.  For   Brownian motion, $P_1(x)=x$, equivalent to $P_1'(x)=1$. For  fractional Brownian motion with Hurst exponent $H$,  we establish that $P_1'(x) = {\cal N} [x(1-x)]^{\frac1H -2} \rme^{\epsilon {\cal F}(x)+ {\cal O}(\epsilon^2)}$, where  $\epsilon=H-\frac12$. The function ${\cal F}(x)$ is analytic, and well approximated by its Taylor expansion, ${\cal F}(x)\simeq 16 (C-1) (x-1/2)^2 +{\cal O}(x-1/2)^4$, where $C= 0.915...$ is the Catalan-constant. 
A similar result holds for moments of the exit time starting at $x$.
We then consider the span of $X_t$, i.e.\ the size of the (compact) domain  visited up to time $t$. For Brownian motion, we derive an analytic expression for the probability  that the span reaches 1 for the first time,   then generalized to fBm.
Using large-scale numerical simulations with system sizes   up to $N=2^{24}$ and a broad range of $H$,  we confirm our analytic results.  
There are important finite-discretization corrections which we quantify. They are 
most severe for small $H$, necessitating to go to the large systems mentioned above. 
\end{abstract}
\maketitle

\section{Introduction}

A key problem in  stochastic processes are  the first-passage properties \cite{FellerBook,RednerBook} in a finite domain,  say the unit interval $[0,1]$.
For   Brownian motion,  the probability to exit at the upper boundary $x=1$, starting at $x$ is  
\be\label{1}
P^0_{1}(x) = x\ .
\ee
Another key observable is the exit time, starting at $x$, which 
behaves as   $\left< T_{\rm exit}(x)\right>_0 \sim x(1-x)$. 
Many physical situations, however, cannot be described by Brownian motion. An example is a polymer translocating through a nano-pore. While the motion of the polymer as a whole is a Markov process, the effective process for its position in the pore is  non-Markovian
\cite{Muthukumar1999,LubenskyNelson1999,StormStormChenZandbergenJoannyDekker2005,PanjaBarkema2010,PanjaBarkemaBall2007,ZoiaRossoMajumdar2009}.
\begin{figure}[b]
\vspace*{-0.5cm}
\Figpng{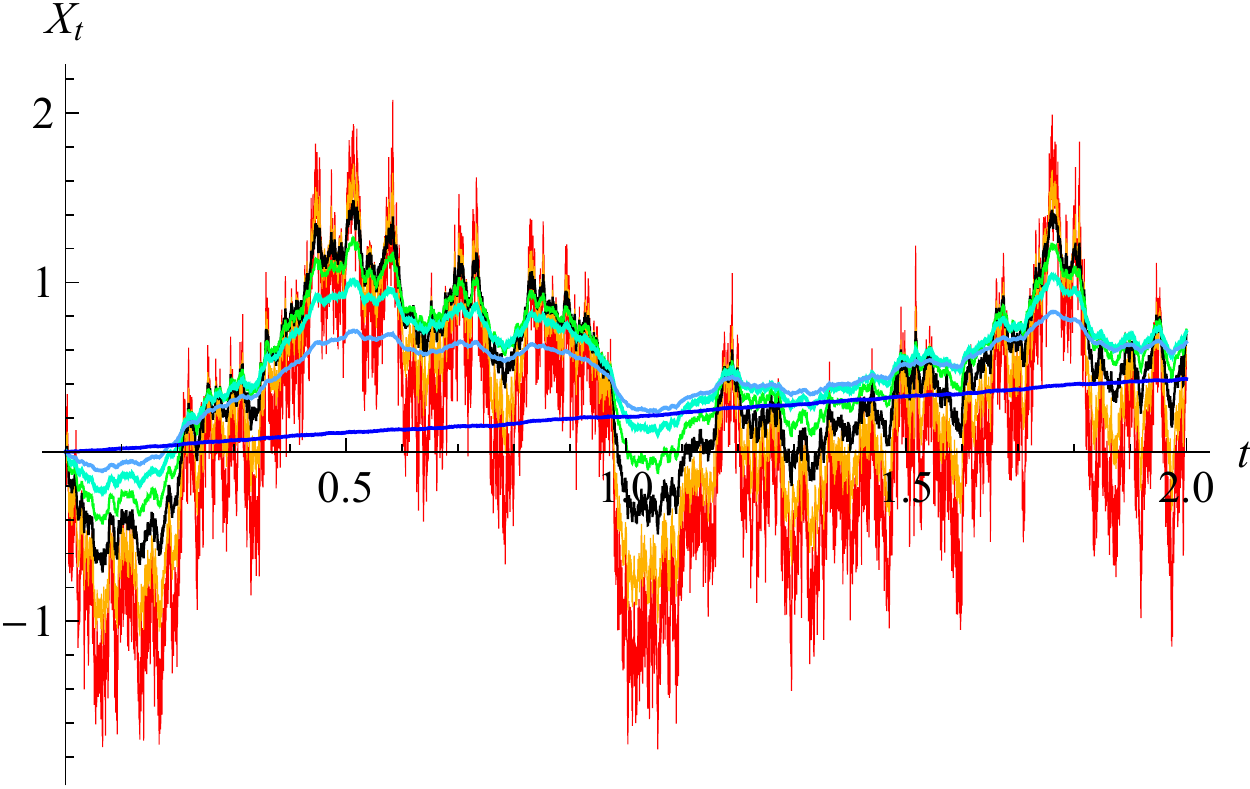}
\caption{Realizations of a fBm for $H=0.25$ (red, roughest curve), $H=0.375$ (orange), $H=1/2$,   Brownian (black), $H=0.625$ (green), $H=0.75$ (cyan), $H=0.875$ (bright   blue) to $H=1$ (dark blue, straight line), using the algorithm of Davies and Harte \cite{DaviesHarte1987,DiekerPhD,DiekerMandjes2003}.}
\label{f:XtdiffH}
\end{figure}
The questions posed above become much  more involved for the latter. The simplest   generalization is fractional Brownian motion (fBm): It is the unique   process that retains from Brownian motion  Gaussianity, scale and translational invariance both in space and time, and that    is drift-free. FBm was introduced in its final form by Mandelbrot and Van Ness~\cite{MandelbrotVanNess1968}. It is indexed by the Hurst exponent $H$, with $0<H\le 1$ (see Fig.~\ref{f:XtdiffH}). As   Gaussian process, it is   specified by its second moment, 
\be\label{eq:covariance}
\left< X(t_1) X(t_2) \right> = t_1^{2H}+t_2^{2H}-|t_1-t_2|^{2H}\ .
\ee
FBm is important as it successfully models  a variety of natural processes \cite{DecreusefondUstunel1998,MetzlerKlafter2000}: a tagged particle in   single-file diffusion ($ H\,{=}\,0.25 $) \citep{KrapivskyMallickSadhu2015,SadhuDerrida2015}, the integrated current in diffusive transport ($ H\,{=}\,0.25 $) \cite{SadhuDerrida2016}, polymer translocation through a narrow pore ($ H\,{\simeq}\,0.4 $) \cite{ZoiaRossoMajumdar2009,DubbeldamRostiashvili2011,PalyulinAlaNissilaMetzler2014}, anomalous diffusion \cite{BouchaudGeorges1990}, values of the log return of a stock  ($H\,{\simeq}\,0.6\; {\rm to }\; 0.8 $) \cite{CutlandKoppWillinger1995,Sottinen2001}, hydrology ($H\,{\simeq}\,0.72\;{\rm to}\;0.87 $) \cite{MandelbrotWallis1968}, a tagged monomer in a polymer   ($ H\,{=}\,0.25$) \cite{GuptaRossoTexier2013}, solar flare activity ($H\,{\simeq}\,0.57\;{\rm to}\;0.86$) \cite{Monte-MorenoHernandez-Pajares2014}, the price of electricity in a liberated market ($ H\,{\simeq}\,0.41 $) \cite{Simonsen2003}, telecommunication networks ($H\,{\simeq}\,0.78\;{\rm to}\;0.86$) \cite{Norros1995}, telomeres inside the nucleus of human cells ($H\,{\simeq}\,0.18\;{\rm to}\;0.35$) \cite{BurneckiKeptenJanczuraBronshteinGariniWeron2012}, or diffusion inside crowded fluids ($ H\,{\simeq}\,0.4 $) \cite{ErnstHellmannKohlerWeiss2012}.

There are   yet no analytical methods to  treat the  questions posed above for $H$ other than $1/2$ (Brownian motion) and $H=1$ (a straight line with a random slope). 
To remedy this, we developed tools \cite{WieseMajumdarRosso2010,DelormeWiese2015,DelormeRossoWiese2017,DelormeWiese2016b,DelormeWiese2016,DelormeThesis}
which allow us to answer this question analytically, in a Taylor expansion around $H=1/2$, i.e.\ in 
\be
\epsilon = H - \frac12\ .
\ee 
These methods have proven feasible and precise up to second order in $\epsilon$ \cite{SadhuDelormeWiese2017}, where they allowed us to distinguish the three classical arcsine laws. 
\begin{figure}[t]
\fig{8cm}{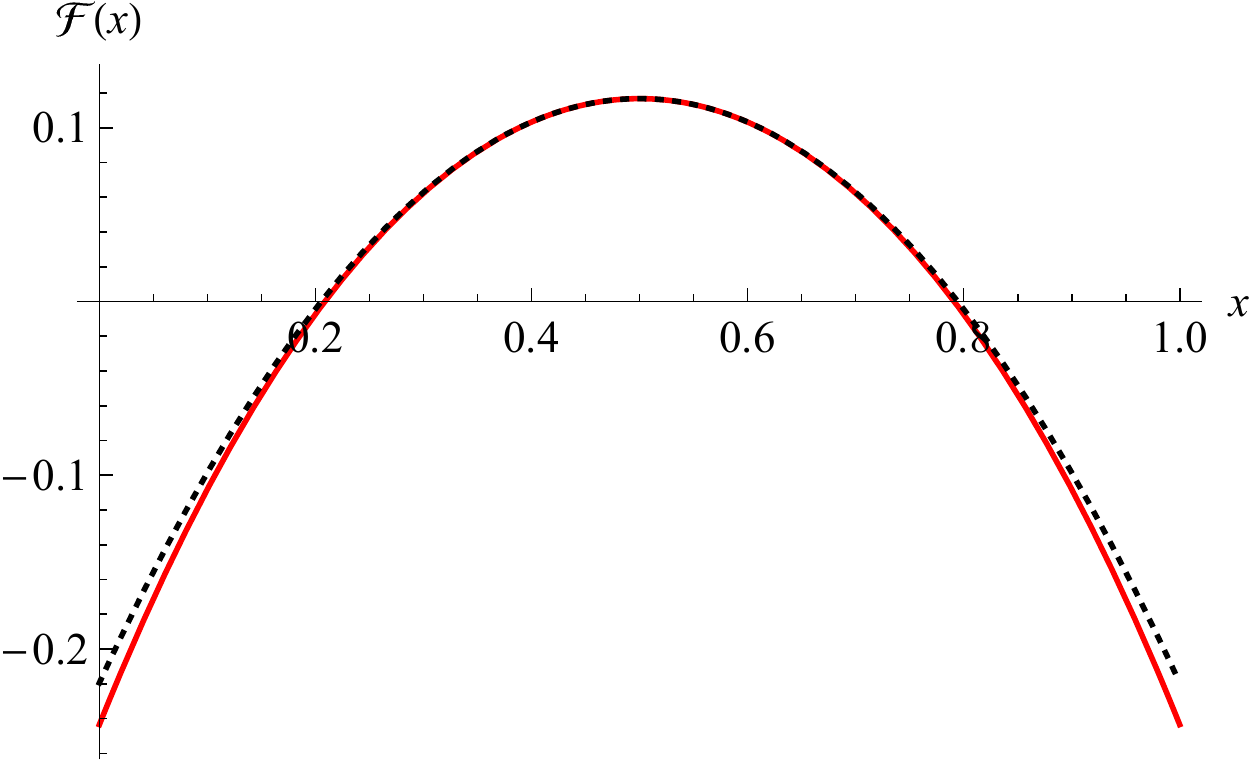}
\caption{The scaling function ${\cal F}(x)$ defined in Eq.~(\ref{Fofx}) (solid, red), normalized s.t.\ that $\int_0^1 {\cal F}(x)\,\rmd x=0$. The dashed line is the quadratic term given in \Eq{43-bis}. }
\label{f:Fofx}
\end{figure}In this article, we generalize the exit probability and distribution of exit times to fractional Brownian motion (Fig.~\ref{f:XtdiffH}). 
It had earlier been argued \cite{MajumdarRossoZoia2010b} that the exit probability at the upper boundary scales  for small  $x$ as
\be
P_1(x) \sim x^\phi\ ,\qquad \phi= \frac {\theta} H\ ,
\ee
 and where   $\theta$ is the persistence exponent. For  fBm  \cite{Majumdar1999,WieseMajumdarRosso2010}
\be
\theta = 1-H\ .
\ee
This led the authors of Ref.~\cite{MajumdarRossoZoia2010b} to    conjecture  that $P_1'(x) \sim  [x(1-x) ]^{\frac{1}{H}-2} $, for all $x$. This however is too simple an approximation \cite{MajumdarRossoZoia2010b}. \begin{figure*}[t]
\fig{8.5cm}{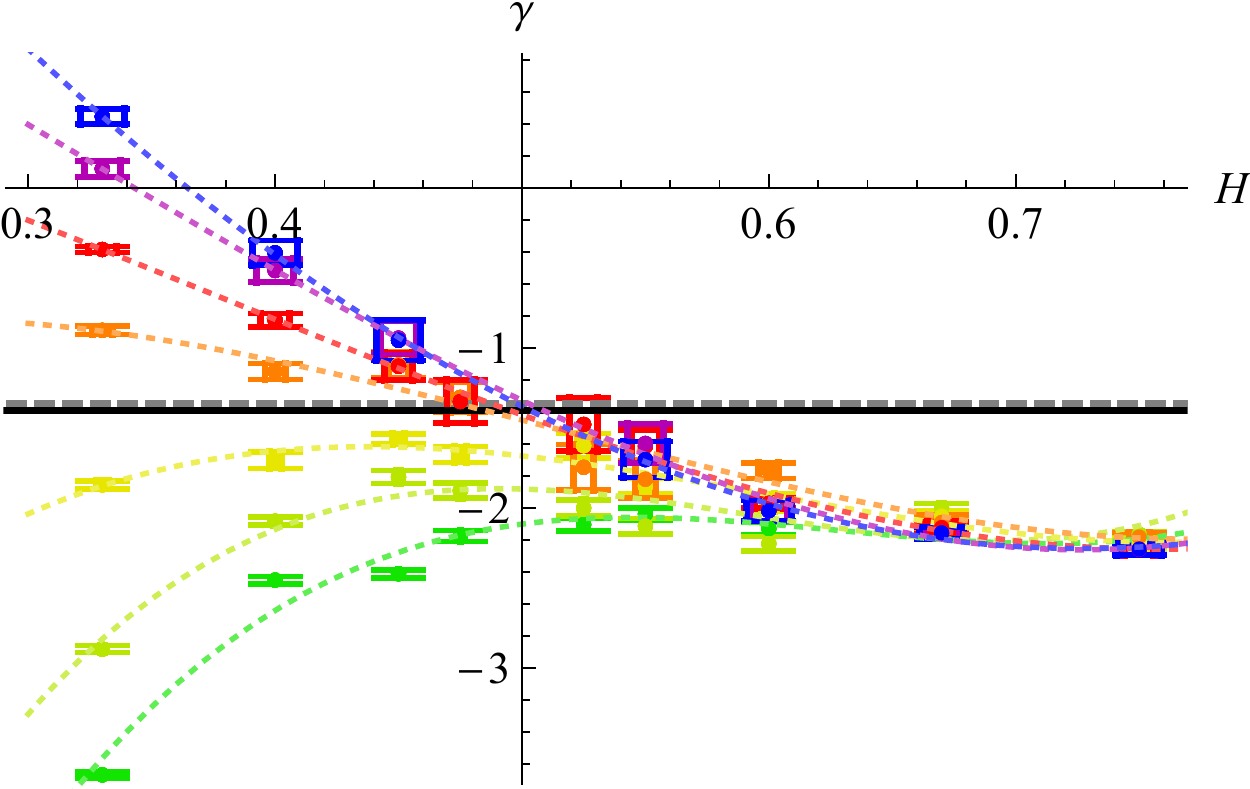}~~~~~~~\fig{8.5cm}{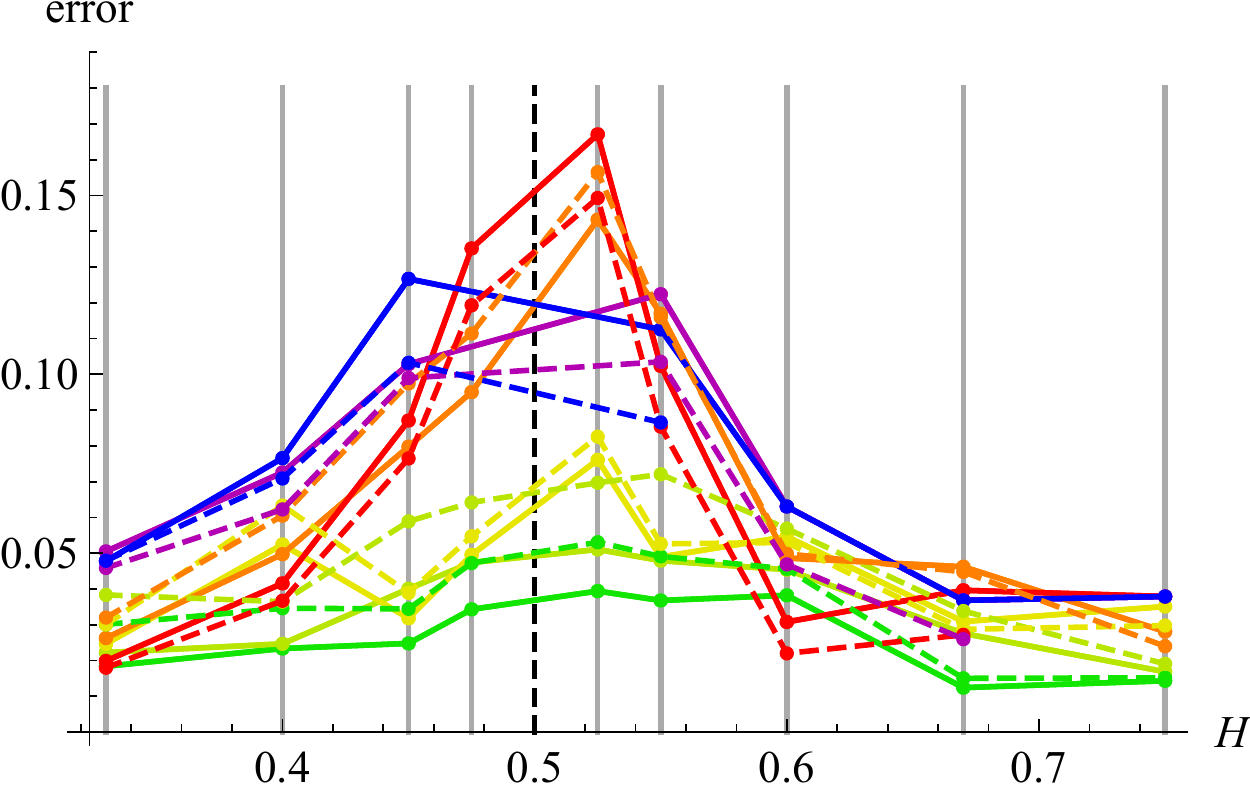}
\caption{The effective measured  curvature $\gamma$, as function of $H$, and system size. The grey dashed line is $\gamma$ as given by Eq.~\eq{gamma-theory}; the black solid line at $-1.39$ is a fit to the curvature of ${\cal F}(x)$ in the range $0.15$ to $0.85$. The color code is (from bottom to top on the left plot) $N=2^{13}$ (dark green), $2^{14}$ (green), $2^{16}$ (olive), $2^{18}$ (orange),  $2^{20}$ (red), $2^{22}$ (dark magenta), $2^{24}$ (blue). The   1-$\sigma$ errors are indicated with bars on the left, and are explicitly shown on the right. There the solid lines are the measured errors obtained as follows: $\gamma$ is estimated  by fitting a parabola to each of the data sets presented on Fig.~\ref{f:all-Fs}, and measuring the variance of the data minus the fit, which allows to estimate the error of the fit, after calibration on white noise. The dashed lines are proportional to the square root of the number of samples, divided by $|\epsilon|$, and calibrated against the estimated data to have a second independent estimate. The grey lines are a guide to the eye, at the location of the $H$-values considered, $H=0.33$, $0.4$, $0.45$, $0.475$, $0.525$, $0.55$, $0.6$, $0.67$, $0.75$.}
\label{f:gamma-estimate}
\end{figure*}
Here we show analytically that  $P_1'(x)$ can be written in the form
\bea\label{8}
P'_1(x) &=& {\cal N} [x(1-x) ]^{\frac{1}{H}-2} 
\rme^{ \epsilon {\cal F}(x) +\ca O (\epsilon^2)}\ ,
\eea
where $\ca F(x)$ is an analytic function,\bea
{\cal F}(x) &=& 4\bigg[12 \zeta '(-1) +\frac{ \log (2)}{3} +\ln\!\big(x(1-x)\big)\nn\\
&& ~~~+\log\! \bigg(\!\Gamma\!
   \Big({ \frac{1}{2}-\frac{x}{2}}\Big)\!\bigg)- \log\! \bigg(\!\Gamma\!
   \Big(1-\frac{x}{2}\Big)\!\bigg) \nn\\
   &&~~~+ \log\! \bigg(\!\Gamma\!
   \Big(\frac{x}{2}\Big)\!\bigg)- \log\! \bigg(\!\Gamma\!
   \Big(\frac{x+1}{2}\Big)\!\bigg)\bigg] \ .
   \label{Fofx-bis}
\eea
Absorbing  the constant into the normalization $\cal N$, its Taylor-expansion reads
\bea
{\cal F}(x) &=& \textstyle 4 -  \frac{20}3 \ln (2) + 8 \big[ \log \! \big( \Gamma 
    (\frac 1 4 ) \big) -  \log \! \big( \Gamma 
    (\frac 3 4 ) \big)  + 6 \zeta'(-1) \big]   \nn\\
&&+ 16 (C-1)
   (x-{\textstyle \frac{1}{2}})^{2}  \nn\\
   && \textstyle +\frac{1}{48} (x-{\textstyle  \frac{1}{2}})^{4}
   \left[\psi ^{(3)} ({\textstyle \frac{1}{4}} )-\psi
   ^{(3)}({\textstyle\frac{3}{4}}) -1536\right]\nn\\
   && +\ca O ({\textstyle  x-\frac{1}{2}})^{6} \nn\\
    &=& 0.116736 - 1.34455 (x-{\textstyle \frac{1}{2}})^{2}  - 0.353774 (x-{\textstyle \frac{1}{2}})^{4}\nn\\
   &&  +\ca O ({\textstyle  x-\frac{1}{2}})^{6} \ .
   \label{43-bis}
\eea
The number $C$ is the Catalan constant
\be
C = \sum_{n=0}^\infty \frac{(-1)^n}{(2n+1)^2}\approx 0.915965594...
\ee
and $\psi^{(3)}(x)= \partial_x^4 \ln(\Gamma(x))$ the polygamma function of order 3. 
As can be seen on Fig.\ \ref{f:Fofx}, the function ${\cal F}(x)$ is well approximated by its second Taylor-coefficient.  Incorporating the forth order term, analytic result and Taylor expansion are indistinguishable on this plot. As a consequence, the most relevant information is   captured by the curvature of ${\cal F}(x)$ at $x=1/2$, 
\be\label{def-gamma}
\gamma := \frac12 {\cal F}''(x)\Big|_{x=1/2} = 16 (C-1)\ .
\ee 
We believe that higher-order terms in $\epsilon$ entering into the exponential of \Eq{8} are also analytic, and well approximated by their low-order (in $x$) Taylor coefficients. We can therefore ask how the effective curvature $\gamma$,   defined by the first equality in \Eq{def-gamma}, changes with $H$. The answer can be read off from Fig.~\ref{f:gamma-estimate}: Consider the top (blue) data on the left plot, obtained for the largest systems.    One sees that $\gamma$ depends on $H$, and that for $H\to \frac12 $ it extrapolates to 
$\gamma = - 1.34 \pm 0.02 $, in agreement with the analytical result \eq{43-bis}.

This article is organized as follows: We first derive   key results for Brownian motion, see section \ref{s:Basic formulas for Brownian Motion with two absorbing boundaries}. Most of them are  known \cite{FellerBook,RednerBook}, except for the span-observables. After a short review of the $\epsilon$-expansion in  section \ref{s:Corrections to the action for fBm}, we derive in section \ref{s:The absorption current at 1-loop order}  the leading-order corrections for fBm for a number of key observables. All our results are checked via extensive numerical simulations in section \ref{s:simul}. We conclude in section \ref{s:conclusion}.

\section{Basic formulas for Brownian Motion with two absorbing boundaries}
\label{s:Basic formulas for Brownian Motion with two absorbing boundaries}
\subsection{Solving the Fokker-Planck equation}

Brownian motion from $x$ to $y$ in time $t$ satisfies the forward Fokker-Planck equation \cite{FellerBook,RednerBook}
\begin{equation}\label{FokkerPlanck}
\partial _t  P_+(x,y,t) = \partial_y^2 P_+(x,y,t)
\end{equation}
The plus refers to surviving paths. 
Its general solution with   absorbing walls at $x=0$ and $x=1$ can be written as
\begin{eqnarray}\label{3}
\lefteqn{P_+(x,y,t)}\nn\\
&=&\frac{1}{\sqrt{4\pi t}} \sum_{n=-\infty}^{\infty}\left(e^{-(x-y+2n)^2/4t} -e^{-(x+y+2n)^2/4t}\right)  \nn\\
&=&\frac{1}{2} \vartheta _3\!\left(\frac{\pi}{2}   (x-y),e^{-\pi ^2 t}\right)-\frac{1}{2}
   \vartheta _3\!\left(\frac{\pi}{2}   (x+y),e^{-\pi ^2 t}\right).~~~~~~~~  
\end{eqnarray}
$\vartheta$ is the elliptic $\vartheta$-function. 
To prove this statement it is enough to remark that the first line satisfies the Fokker-Planck equation (\ref{FokkerPlanck}),  vanishes at $y=0$ and $y=1$, and reduces for $t\to 0$ to a $\delta$-function
\begin{equation}\label{f1}
\lim_{t\to 0}P_{+} (x,y,t) = \delta (x-y)\ .
\end{equation}
Let us introduce the notation
\be\label{mathbbP}
{\mathbb P}(z,t):=\frac{1}{\sqrt{4\pi t}} \sum_{n=-\infty}^{\infty} e^{-(z+2n)^2/4t} = \frac{1}{2} \vartheta _3\!\left(\frac{\pi}{2}   z,e^{-\pi ^2 t}\right).
\ee
In terms of this function, Eq.~(\ref{3}) can be written as
\be\label{17}
{P_+(x,y,t)} = {\mathbb P}(x-y,t)- {\mathbb P}(x+y,t)\ .
\ee
Using the Poisson summation formula, an alternative form for  ${\mathbb P}(z,t)$ is
\be\label{PP-Poisson}
{\mathbb P}(z,t) = \frac12 + \sum_{m=1}^\infty \rme^{-m^2 \pi^2 t} \cos(m \pi z)\ .
\ee
It is useful to   consider its Laplace-transformed version. 
We define the Laplace transform of a function $F (t)$, with
$t\ge 0$, and marked with a tilde  as 
\begin{equation}\label{f56-bis} 
\tilde F (s):= \int_{0}^{\infty}\rmd t\,  e^{-s t} F (t) \ .
\end{equation}
This yields 
\begin{eqnarray}\label{16-bis}
\!\!\!\tilde  P_+(x,y,s) &=& \frac{e^{-\sqrt{s} |x-y|}-e^{-\sqrt{s} (x+y)}}{2
   \sqrt{s}} \nn\\
   && -\frac{\left[\coth  (\sqrt{s})-1\right] \sinh
   (\sqrt{s} x) \sinh (\sqrt{s} y)}{\sqrt{s}}.~~~~~~~
\end{eqnarray}
According to \Eq{17} a form which only depends on $x-y$ and $x+y$ also exists. We use the form \eq{16-bis}, since
the factorization of the second term   facilitates its integration.

\subsection{Boundary currents and conservation of probability}

Conservation of probability reads (the variable $x$ is   the initial condition, here a dummy variable)
\be\label{current-conservation}
\partial_t P_+(x,y,t) +\partial_y J(x,y,t) = 0\ .
\ee
$J$ is the current, which from \Eq{current-conservation} can be identified as 
\be\label{22}
 J(x,y,t) = -\partial_y P_+(x,y,t)\ .
\ee
Due to the Dirichlet conditions at $y=0$ and $y=1$, we have 
\be
\int_0^1 \rmd y \, \partial_t P_+(x,y,t) = J(x,0,t) - J(x,1,t)\ .
\ee
We find 
\bea
J(x,y,t)&=&\frac{\pi }{4}  \vartheta _3^{\prime }\!\left(\frac{\pi}{2}   (x-y),e^{-\pi ^2
   t}\right) \nn\\
   && +\frac{\pi }{4}   \vartheta _3^{\prime }\!\left(\frac{\pi}{2}   (x+y),e^{-\pi
   ^2 t}\right) \ .
\eea
The derivatives of the elliptic $\vartheta$ functions are w.r.t.\ its first argument. 
The probability to exit at time $t$, when starting at $x$ for time $0$ reads
\bea
P_{\rm exit}(x,t)&=&   -J_{\rm tot}(x,t) = J(x,1,t) - J(x,0,t) \nn\\
   &=& \frac{\pi}{2}   \left[\vartheta _3^{\prime }\!\left(\frac{\pi}{2}   (x-1),e^{-\pi ^2
   t}\right)- \vartheta _3^{\prime }\!\left(\frac{\pi  x}{2},e^{-\pi ^2
   t}\right)\right]. \nn\\
\eea
Going to Laplace variables, we find
\be
-1+ s \int_0^1\rmd y\, \tilde P(x,y,s )  = \tilde J(x,0,s) - \tilde J(x,1,s)\ .
\ee
The outgoing currents of the Laplace transform are 
\bea
- \tilde J(x,0,s) &=& \frac{ \sinh \big(\sqrt s  (1-x)\big) }{\sinh(\sqrt s )}\ ,\\
\tilde J(x,1,s) &=& \frac{ \sinh (\sqrt s\, x) }{\sinh(\sqrt s )} \ .
\eea

\begin{figure*}[t]
\fig{8.1cm}{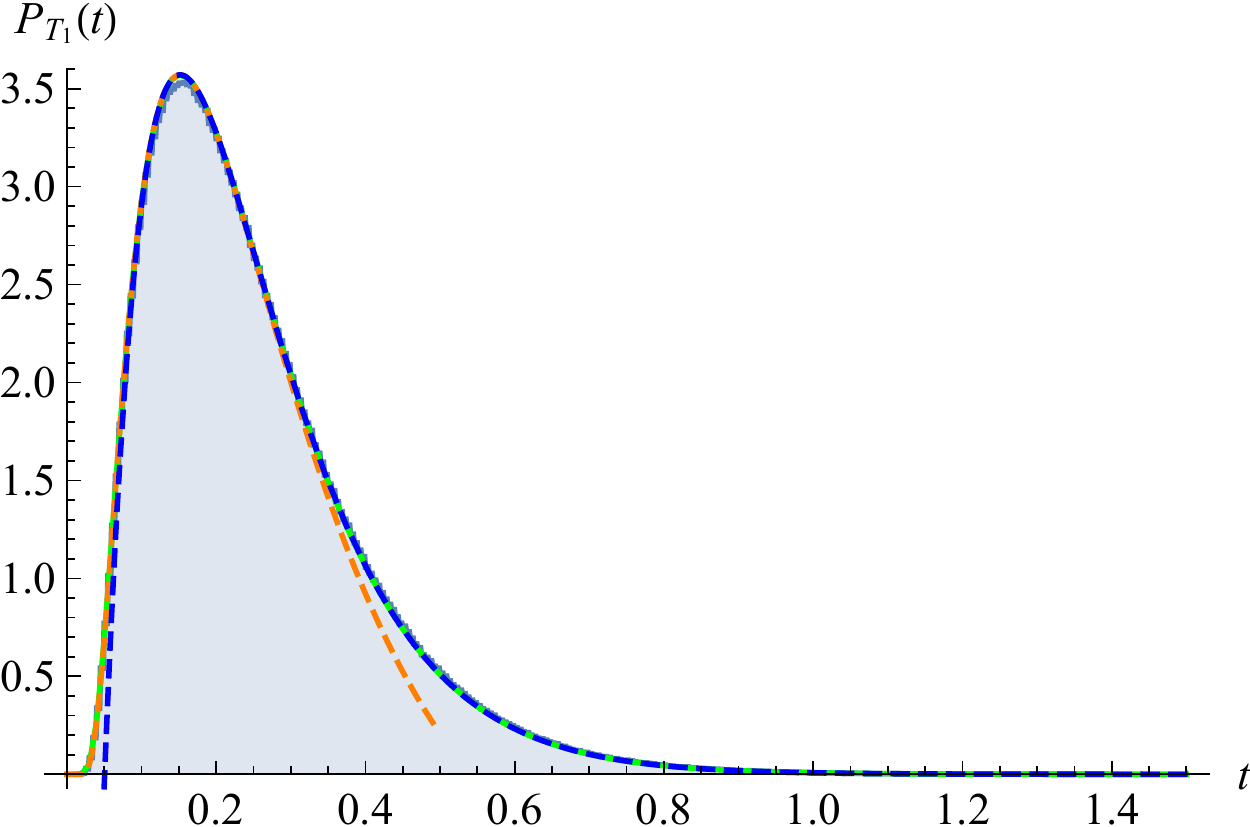}~~~~~~~~~\fig{8.6cm}{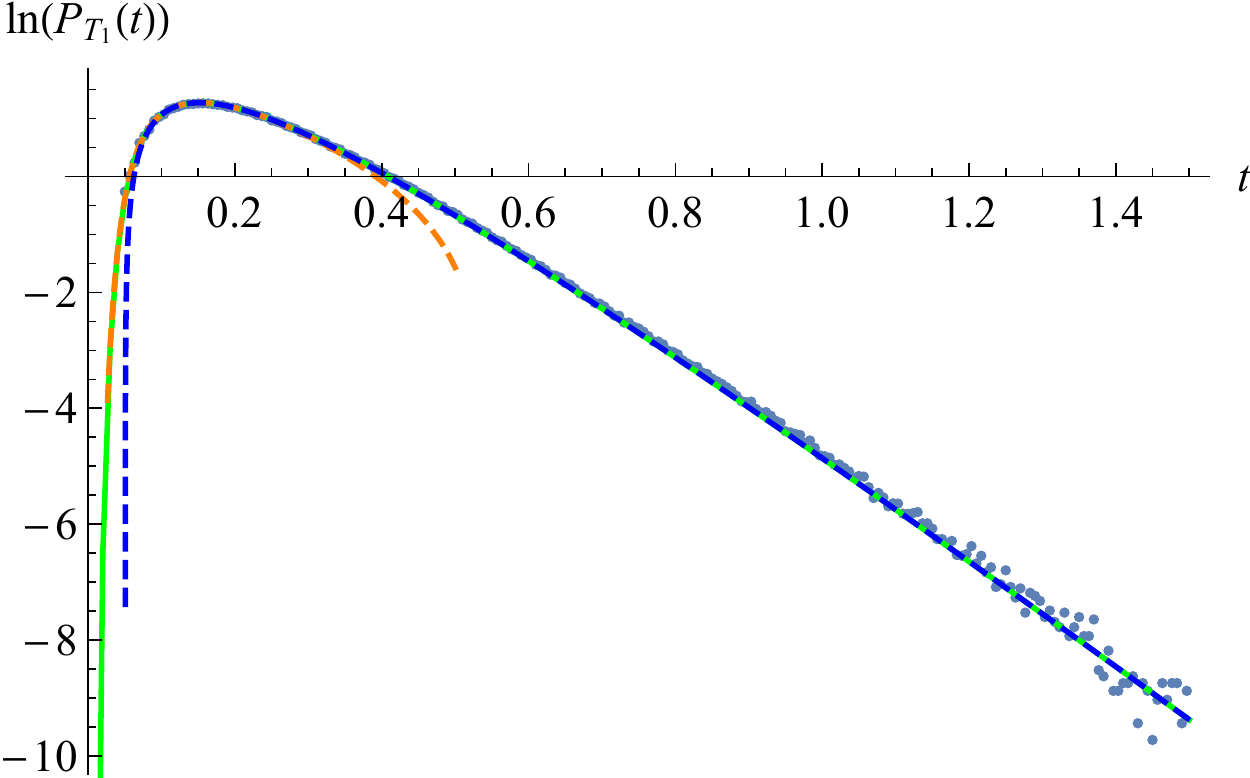}
\caption{The probability that the span reaches 1 for the first time. Grey: RW simulation with $\delta t=10^{-5}$, and $10^6$ samples. Green: the analytic result (\ref{PT1}). 
Orange dashed the small-times asymptote (\ref{PT1-small}); blue dashed the large-time asymptote (\ref{PT1-asymp}). Note a small systematic deviation due to the relatively large time step $\delta t=10^{-5}$. }
\label{f:Pspan}
\end{figure*}

\subsection{Absorption probabilities at $x=0$ and $x=1$}
The absorption probabilities at $x=0$ and $x=1$ are  
\bea\label{14}
P_0(x) &:=& \int _0^\infty \rmd t \, \left[ -  J(x,0,t) \right] \nn\\
&=& \lim_{s\to 0}  \left[ - \tilde J(x,0,s) \right] = 1-x  \ , \nn\\
\label{15}
P_1(x) &:=& \int _0^\infty \rmd t \,    J(x,1,t) = 
\lim_{s\to 0}    \tilde J(x,1,s) = x\ .
\eea

\subsection{Moments of the absorption time, starting at $x$}
Moments of the absorption time are extracted from the Laplace-transformed currents as
\bea\label{Tabs0}
\left< T_{\rm exit}(x)\right>_0 &=&  -\partial_s \left[ \tilde J(x,1,s) - \tilde J(x,0,s)\right] \Big|_{s=0} \nn\\
&= &\frac12 x (1-x)\\
\int_0^1\rmd x \!\!\!\!\!\!&& \left< T_{\rm exit}(x)\right>_0 =  \frac1{12}\ .
\\
\label{Tabs0squared}
\left< T_{\rm exit}(x)^2\right>_0 &=&  \partial_s^2 \left[ \tilde J(x,1,s) - \tilde J(x,0,s)\right] \Big|_{s=0} \nn\\
&=& \frac{1}{12}x (1-x) (1+x-x^2) \\
\int_0^1\rmd x \!\!\!\!\!\!&& \left< T_{\rm exit}(x)^2\right>_0 =  \frac1{60}.
\eea

\subsection{Probabilities for  the span}
The  numerical simulations we will perform later can be stopped when the {\em width} or {\em span} of the process reaches 1. The span is a classical problem treated e.g.\ in \cite{Daniels1941,Feller1951,WeissRubin1976,PalleschiTorquati1989}, but the observable in question seems not to have been considered. Here, we give an analytical result, and validate it numerically. The two series expansions we obtain provide  simple approximate solutions for both small and large times. 

To properly define the problem, we 
note the positive and negative records (a.k.a.\ the running max and min) as 
\bea
M_+(t) &:=& \max_{t'\le t}  X_{t'} \ ,\\
M_-(t) &:=& \min_{t'\le t}  X_{t'} \ .
\eea
The span $s(t)$ is their difference, i.e.\ the size of the (compact) domain visited up to time $t$, 
\be
s(t):= M_+(t) - M_-(t)\ .
\ee
We want to know the probability that $s(t)$ becomes 1 for the first time. We note this time by $T_1$, and its probability distribution by $P_{T_1}(t)$. It can be obtained as follows: 
The  outgoing current at the lower boundary positioned at $m_1$, with the upper boundary  at $m_2$,  and starting at $x$ is
\bea
\lefteqn{\mathbf J(x,m_1,m_2,t)} \nn\\
&&\qquad = \frac{1}{(m_1-m_2)^2}\,J\!\left( \frac{x-m_{1}}{m_{2}-m_{1}},0,\frac{t}{(m_{2}-m_{1})^{2}}\right).~~~~~~~~
\eea
(The scale factor can be understood from the observation that the current is a density in the starting point times a spatial  derivative of a probability.)
The probability that the walk reached $m_2$ before being absorbed at $m_1$ is 
$
\partial_{m_2}   {\mathbf J(x,m_1,m_2,t)} 
$. Finally, 
the probability to have span 1 at time $t$ is this expression, integrated over $x$ between the two boundaries, times a factor of 2. The latter accounts for the term where the two boundaries are exchanged. 
Setting w.l.o.g.\ $m_1=0$ and $m_2=m$, this is written as 
\bea\label{PT1-bare}
P_{T_1}(t) &=& -2 \partial_m \frac1{m^2}\int_0^m \rmd x\, J\!\left( \frac{x}{m},0,\frac{t}{m^2}\right)\bigg|_{m=1}\nn\\
&=& - 2 \partial_m \frac1{m}\int_0^1 \rmd x\, J\!\left( x,0,\frac{t}{m^2}\right)\bigg|_{m=1} \nn\\
&=&  2(1+2 t\partial_t)  \int_0^1 \rmd x\, J\!\left( x,0,t\right)  \ .
\eea
Using Eqs.~(\ref{22}) 
and  (\ref{17}) 
allows us to rewrite the integral as 
\bea
\lefteqn{ \int\limits_0^1 \rmd x\, J\!\left( x,0,t\right) =\int\limits_0^1 \rmd x\, \partial_y \left[ \mathbb P(x-y,t) -\mathbb P(x+y,t) \right] \bigg|_{y=0}} \nn\\
 &=&-2 \int\limits_0^1 \rmd x\, \partial_x   \mathbb P(x,t)   = 2\left[  \mathbb P(1,t) -  \mathbb P(0,t) \right] \ .  \qquad\qquad\qquad
 \eea
\pagebreak[1]
Thus 
\be
P_{T_1}(t) =  4 (1+2 t \partial_t)  \left[  \mathbb P(1,t) -  \mathbb P(0,t) \right]\ .
\ee\begin{figure*}[t]
{\fig{8.1cm}{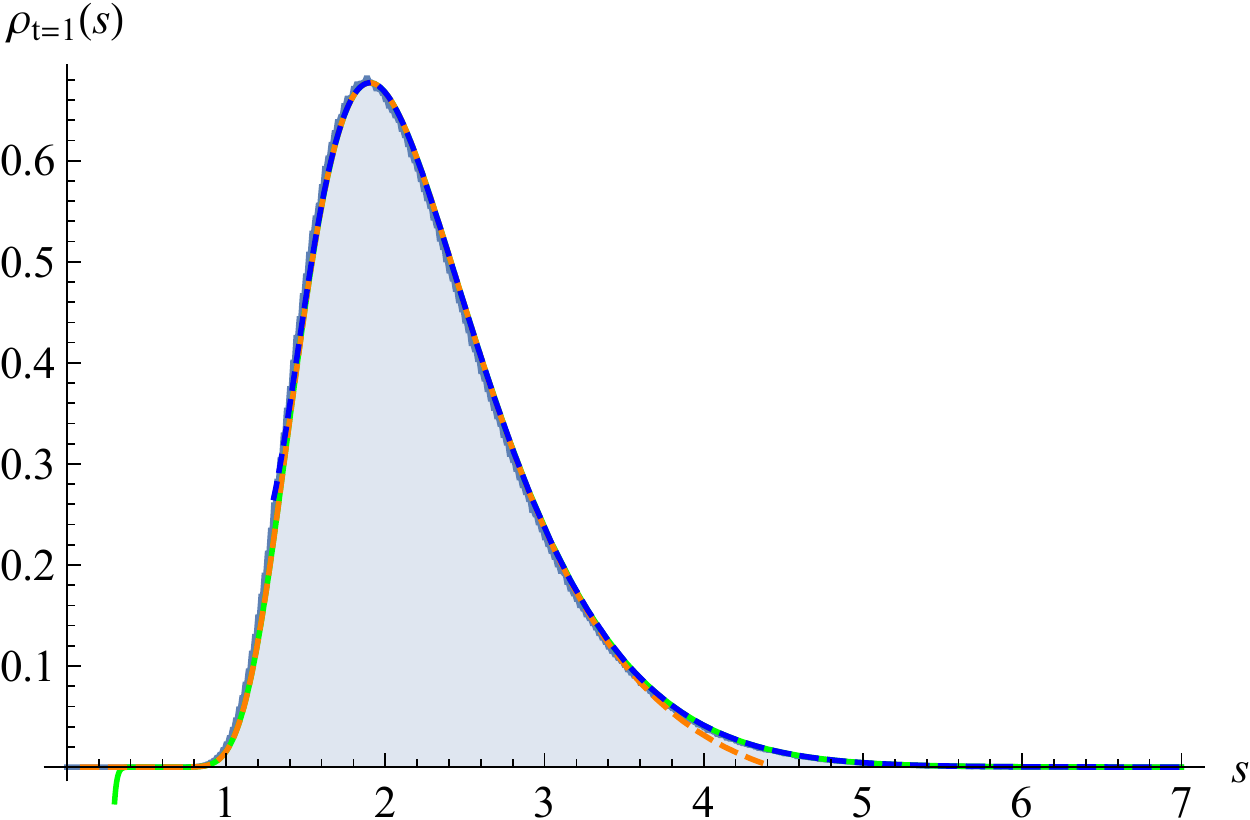}}~~~~~~~~~{\fig{8.6cm}{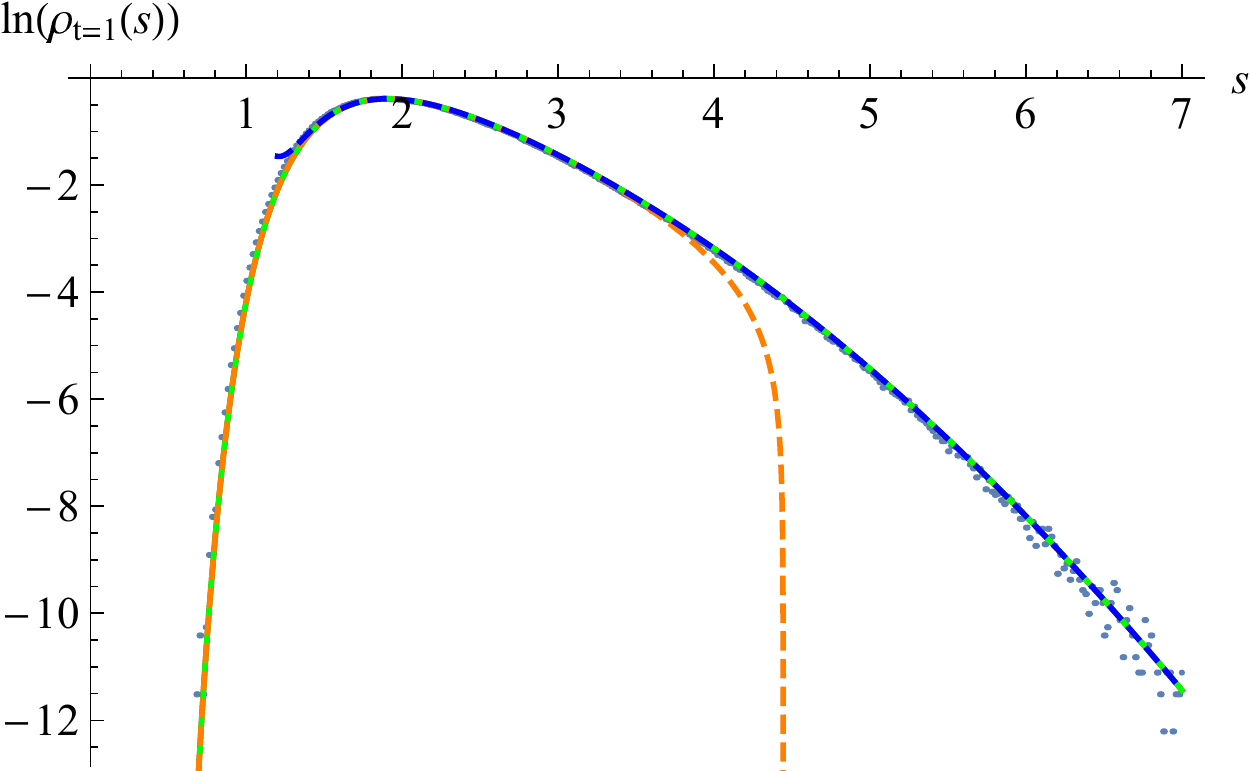}}
\caption{Left: The density of the span at time $t=1$. Grey: RW simulation with $\delta t=10^{-4}$, and $10^6$ samples. Green: the analytic result (\ref{rho-rhospan-ana}). 
Orange dashed the small-$s$ asymptotics (\ref{rho-small}); blue dashed the large-$s$ asymptotics (\ref{rho-large}). Right: {\em ibid.} on a log-scale.}
\label{f:rhospan}
\end{figure*}
Inserting the definition \eq{mathbbP} of $\mathbb P$, we get
\bea\label{PT1}
P_{T_1}(t) &=& 4  \big(  1 +2 t  \partial_{t} \big) \sum_{n=-\infty}^\infty \frac{e^{-\frac{(2 n+1)^2}{4 t}} - e^{-\frac{n^2}{t}}}{\sqrt{4\pi t}}
 \nn\\
 &=& \frac1 {\sqrt{\pi } t^{3/2} }\sum_{n=-\infty}^\infty   (2 n+1)^2 e^{-\frac{(2 n+1)^2}{4 t}}-4 n^2 e^{-\frac{n^2}{t}} \nn \\
   &=& 4 \sqrt{\frac t \pi } \partial_t \left[  \vartheta _2 \!\left(0,e^{-1/
   t}\right)- \vartheta _3 \!\left(0,e^{-1/
   t}\right)\right] \ .
\eea
With the help of the Poisson-formula transformed Eq.~(\ref{PP-Poisson}), this can compactly be written as
\be\label{PT1-Poisson}
P_{T_1}(t) = 8 \sum_{n=0}^\infty \rme^{-\pi^2 (2n+1)^2 t}\left[ 2 \pi^2 (2n+1)^2 t-1 \right] \ .
\ee
This result is compared to a numerical simulation on Fig.~\ref{f:Pspan}. 
Our expansions allow us to give simple formulas for the small and large-$t$ asymptotics, 
\bea\label{PT1-small}
P_{T_1}(t) &\simeq& \frac{2\rme^{-\frac1{4t}}}{\sqrt {\pi }t^{3/2}} + \ca O(\rme^{-\frac1{t}} )\ , \\
\label{PT1-asymp}
P_{T_1}(t) &\simeq&   \rme^{-\pi^2 t} \left[16 \pi^2 t  -8    +\ca O(e^{-8 \pi^2 t}) \right]\ .
\eea 
These expansions work in a rather large, and overlapping domain, as can be seen on Fig.~\ref{f:Pspan}.

Its Laplace transform is \bea\label{tildePT1}
\tilde P_{T_1}(s) &=& 2(1+2 s \partial_s) \sum_{n=-\infty}^{\infty} \int_0^\infty \rmd t\, \frac{   e^{-\frac{n^2}{t}} - \rme^{-\frac{(2 n+1)^2}{4 t}}}{\sqrt{\pi t}}\, \rme^{-st }
\nn\\
&=& \frac1{\cosh(\sqrt s/2)^2}\ .
\eea
Extracting the moments from the Laplace transform   yields
\be
\left< T_1\right> =\frac14\ , \quad \left< T_1^2\right> =\frac1{12} \ , \quad \left< T_1^3\right> = \frac{17}{480}\ ,\quad...
\ee
Finally, let us connect to the classical work on the span \cite{Daniels1941,Feller1951,WeissRubin1976,PalleschiTorquati1989}. We will show how to reproduce formulas (3.7)-(3.8) in \cite{Feller1951}.
The latter give the density $\rho_t(s)$ for the span $s$ at time $t$. 
In our formalism, it can be obtained as 
\be
\rho_t(m_2-m_1) =-\partial_{m_1}\partial_{m_2} \int\limits_{m_1}^{m_2}\rmd x\,  \int\limits_{m_1}^{m_2}\rmd y\, \mathbf P(x,y,m_1,m_2,t)\ , 
\ee
where $\mathbf P(x,y,m_1,m_2,t)$ is the probability to go from $x$ to $y$ in time $t$, without being absorbed by the lower boundary positioned at $m_1$, or the upper boundary positioned at $m_2$.
In terms of the propagator $P_+(x,y,t)$, this can be written as 
\be
\rho_t(s) =\partial_{s}^2 \left[  s\int\limits_{0}^{1}\rmd x\,  \int\limits_{0}^{1}\rmd y\, P_+(x,y,t/s^2)\right] \ .
\ee
Using \Eq{17}, and the series expansions \eq{mathbbP} and \eq{PP-Poisson} yields after integration and simplifications two different representations, 
\bea
\rho_t(s) &=& \frac{4}{\sqrt{\pi t}} \sum_{n=1}^\infty  (-1)^{n+1} n^2 e^{-\frac{n^2 s^2}{4 t} } \nn \\
&=& \frac{16 t}{s^5} \sum_{n=0}^\infty  e^{-\frac{\pi ^2   (2 n+1)^2 t}{s^2}} \Big[2 \pi ^2 (2 n+1)^2 t-s^2\Big]\ .\qquad 
\label{rho-rhospan-ana}
\eea
This is equivalent to Eqs.~(3.7)-(3.8) in \cite{Feller1951}, if one there replaces $t\to 2 t$. (Our covariance \eq{eq:covariance} at $H=1/2$ is $2t$ instead of $t$ as in \cite{Feller1951}.) 
The small and large-$s$ asymptotics are \bea\label{rho-small}
\rho_t(s) &\simeq& \frac{4}{\sqrt{\pi t}}  \left[  e^{-\frac{s^2}{4t} } - 4 e^{-\frac{s^2}{t} }  + \ca O\Big( e^{-\frac{9 s^2}{4t} } \Big)  \right] \ ,\\
\rho_t(s) &\simeq& \frac{16 t}{s^5}   e^{-\frac{\pi ^2  t}{s^2}} \Big[2 \pi ^2 t-s^2\Big] +\ca O\Big(  \rme^{-\frac{9 \pi^2 t}{s^2}}\Big) \ .\qquad 
\label{rho-large}
\eea
Note that in \Eq{rho-small} we have also retained the subleading term for small $s$, which considerably improves the numerical accuracy.  A test is presented on Fig.~\ref{f:rhospan}.

\section{Corrections to the action for fBm}
\label{s:Corrections to the action for fBm}
Here we briefly review the derivation of the effective action for fBm \cite{WieseMajumdarRosso2010,DelormeWiese2015,DelormeWiese2016,SadhuDelormeWiese2017}. 
The exact action for a Gaussian process with correlations ${\cal C}(t_1,t_2)$ is   by definition 
\bea\label{S-eff}
{\cal S}[X] &=& \int_{0<t_1< t_2<T} \dot X_{t_1} {\cal C}^{-1}(t_1,t_2) \dot X_{t_2}\ .
\eea
Here
\bea \label{CC}
\lefteqn{ {\cal C}(t_1,t_2) =\left< \dot X_{t_1} \dot X_{t_2}\right>} \nn\\
&=& 2 \delta (t_1-t_2) 2H|t_2-t_1|^{2 H-1} \nn\\&& + 2 H (2H-1) |t_1-t_2|^{2(H-1)}\nn\\
&=& 2 D_{\epsilon} \left[ \delta (t_1-t_2)  +\frac{ \epsilon}{\left| t_1-t_2\right| } +{\cal O}(\epsilon ^2)  \right] \ .
\eea
The diffusion constant, which depends on the small-time cutoff $\tau$ implicit in the above construction, reads
\be
D_{\epsilon} \equiv 2H \tau^{2H-1} = (1+ 2\epsilon) \tau^{2\epsilon} \ .
\ee
{This scale can be understood as follows: Our procedure yields a random process $X_t$, which is a Brownian process at   times smaller than $\tau$, and an fBm at larger times.}

The functional inverse of \Eq{CC} which enters into the effective action \eq{S-eff} reads
\be\label{159}
{\cal C}^{-1}(t_1,t_2) 
= \frac1{2 D_{\epsilon}} \left[ \delta (t_1-t_2)  -\frac{ \epsilon}{\left| t_1-t_2\right| } +{\cal O}(\epsilon ^2)  \right] .
\ee
This allows us to write the action \eq{S-eff} as the action of Brownian motion,  plus a non-local term
\be
S[X] = {\cal S}_0 + \epsilon {\cal S}_1 + {\cal O}(\epsilon^2)\ ,
\ee
with 
\begin{align}
   {\cal S}_0 &:= \frac1{4D_\epsilon}\int_t \dot X_t^2\\
   \label{162}
  {\cal S}_1 &:=    \int_{t_1<t_2}      \delta {\cal C}^{-1}(t_1,t_2)\dot X_{t_1} \dot X_{t_2}  \ .
\end{align}
Here $ \delta {\cal C}^{-1}(t_1,t_2)$ is the non-local part ($t_1\neq t_2$) of ${\cal C}^{-1}(t_1,t_2) $ defined in \Eq{159}.

We will use the trick to represent the propagator as  $|t|^{-1} = \int_{y>0 }\rme^{-y |t|}$, which allows us to treat a small-time cutoff $\tau$ for a momentum cutoff $\Lambda$. 
The relation between these two cutoffs can be inferred from 
\bea
&&\int_0^T \rmd t\int_0^\Lambda  \rme^{-y t } \rmd y = \ln (T \Lambda) + \gamma_{\rm E} + \ca O(\rme^{- T\Lambda}) \nn\\
&&\qquad \stackrel!= \ln (T/\tau)= \int_\tau^T \frac {\rmd t}t \ .
\eea
This implies that up to exponentially small terms
\be
\Lambda = \frac{\rme^{-\gamma_{\rm E}}}\tau\ . 
\ee
\smallskip

\section{The absorption current at 1-loop order}
\label{s:The absorption current at 1-loop order}
\subsection{General formulas}
We want to calculate  the   current at the upper boundary at time $t$, when starting at $x$ at time 0.
We denote this by calligraphic $\ca J(x,1,t)$, to distinguish it from the Brownian result $ J(x,1,t)$. We follow the procedure outlined in Ref.~\cite{DelormeWiese2016}, which works on the  Laplace-transformed version. The outgoing current at order $\epsilon$ reads \bea\label{calJ}
&&\lefteqn{\tilde {\ca J}(x,1,s) =  \tilde J\Big(x,1,\frac s{ D_\epsilon}\Big) + 2 \epsilon \tilde {\cal A}(x,s) + \ca O(\epsilon^2)}\nn\\
 &&=  \tilde J(x,1,s) +  2\epsilon \left[   \tilde {\cal A}(x,s) - (1+\ln \tau)  s\partial_s\tilde J(x,1,s)\right]\nn\\
 && ~~~+ \ca O(\epsilon^2)\ .
\eea
(The relation for the currents in time has an additional factor of $1/D_\epsilon$.)
The  first-order correction for the current   at 
$y=1$ is
\bea\label{Adef}
\tilde {\cal A}(x,s) =\int_0^{\Lambda}\rmd y &&\int_0^1 \rmd x_1  \int_0^1 \rmd x_2\, \tilde  P_+(x,x_{1},s) \\
&& \times \partial_{x_1} \tilde  P_+(x_1,x_{2},s+y) \,\partial_{x_2}\tilde  J(x_{2},1,s)\ . \nn
\eea
The resulting expression after integration over $x_1$ and $x_2$ is rather lengthy, but  can be simplified to
\begin{widetext}
\vspace*{-.1cm}
\bea\label{Atilde}
\!\!\!\tilde {\cal A}(x,s) &=&\int_0^{\Lambda}\rmd y\, \frac{\sqrt{s} }{2 y^2 \sinh(\sqrt{s+y}) {\text{sinh} (\sqrt{s} )} } \times \nn\\
&& ~~~~\times \bigg[\frac{ \sinh  (\sqrt{s} x )}{\text{sinh} (\sqrt{s} )} \bigg(\sqrt{s+y} \Big(3-4 \cosh (\sqrt{s}) \cosh (\sqrt{s+y})+\cosh
   (2 \sqrt{s})\Big) +  y \cosh \left(\sqrt{s}\right) \sinh \left(\sqrt{s+y}\right)\bigg) \nn\\
   &&  ~~~~~~~~~~~~ -
       {\cosh (\sqrt{s} x)} \bigg( x y
   \sinh \left(\sqrt{s+y}\right)+2 \sqrt{s+y} \Big (\cosh  (\sqrt{s} )-\cosh
    (\sqrt{s+y} )\Big)\bigg) \nn\\
    &&  ~~~~~~~~~~~~ -2 \sqrt{s+y}  { \,\cosh  \Big((1-x) \sqrt{s+y} \Big) }  +2 \sqrt{s+y} \cosh \left(\sqrt{s}\right) \cosh \left(x \sqrt{s+y}\right)\bigg]\ .
\eea
As the integrand vanishes at $x=0$ and $x=1$, 
\be\label{32}
\tilde A(0,s) = \tilde A(1,s)=0\ .
\ee
The integral \eq{Atilde} is  difficult to integrate analytically -- or numerically. We will therefore study moments of $s$, which   allow us to access the exit probability, and the first moments of the exit times. We start with the lowest moment, the exit probability.

\subsection{Absorption probability at the upper boundary}
The limit of $s\to 0$ in the integral   \eq{Atilde}  yields the correction to the probability to exit at the upper boundary, starting at $x$. 
Simplifying  \Eq{Atilde}, we find
\bea\label{Atilde0}
\tilde {\cal A}(x,0) &=& \int_0^{\infty}\rmd y\,  \frac{e^{-x \sqrt{y}} \left(-2 x e^{x \sqrt{y}+\sqrt{y}}+2 x e^{x
   \sqrt{y}}+e^{x \sqrt{y}+\sqrt{y}}-e^{x \sqrt{y}}+e^{2 x
   \sqrt{y}}-e^{\sqrt{y}}\right)}{\left(e^{\sqrt{y}}+1\right) y^{3/2}}\ .
\eea
\end{widetext}
We set the cutoff $\Lambda\to \infty$, as the integral is convergent. 
The current at the lower boundary is by symmetry\be
\tilde A(x,0) = - \tilde A(1-x,0)\ .
\ee
To simplify this expression, we perform two variable transformations. The first sets $y = z^2$. The second $z=-\ln(r)$. This yields
\be
\tilde {\cal A}(x,0) = -2 \int_0^1 \rmd r \,\frac{-r^{1-x}+r^x-2 r x+r+2 x-1}{r (r+1) \log ^2(r)}\ .
\ee
This expression is  still difficult to integrate, due to the logarithms in the denominator. Taking two derivatives simplifies this to\begin{figure}[b]
\figpng{7cm}{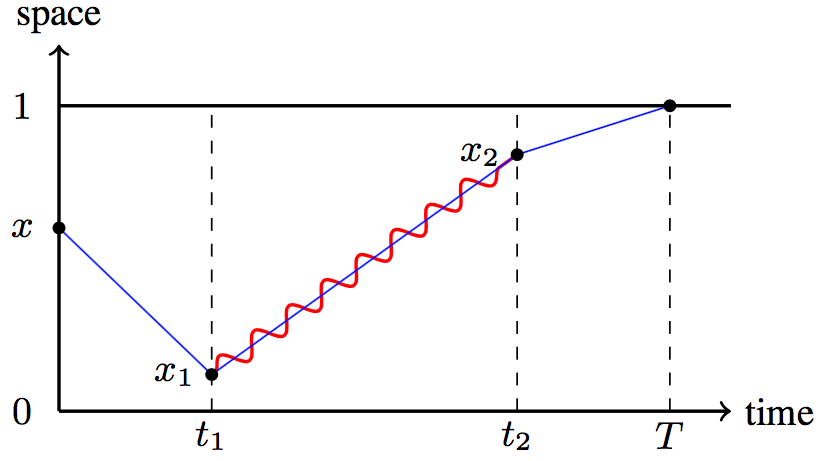}
\caption{Graphical representation of the
path-integral for the order-$\epsilon$ contribution $\tilde {\cal A}(x,s)$ given in Eq.~(\ref{Adef}).}
\label{f:Z3}
\end{figure}\bea
\lefteqn{\partial_{x}^2\,\tilde {\cal A}(x,0) = 2 \int_0^1 \rmd r \, \frac{r^{-x}- r^{x-1}}{r+1}}
\nn\\
& =&   -\frac{2}{x}-\psi \!\left(\frac{1}{2}-\frac{x}{2}\right)+ \psi\!
   \left(1-\frac{x}{2}\right) - \psi\!
   \left(\frac{x}{2}+\frac{1}{2}\right) \nn\\
   && + \psi\!
   \left(\frac{x}{2}+1\right)\ .
\eea
We now have to integrate twice w.r.t.\ $x$, which gives the result plus   terms of the form $a+b x$. The latter can be fixed by \Eq{32}.
The result is \bea
\tilde {\cal A}(x,0) &=& \frac{1}{3} (2 x-1) \Big[\log (2)-3 + 36 \zeta'(-1) \Big] \nn\\
&& -
4 \psi
   ^{(-2)}\!\left(\frac{1}{2}-\frac{x}{2}\right)+4 \psi
   ^{(-2)}\!\left(1-\frac{x}{2}\right)\nn\\
   && +4 \psi
   ^{(-2)} \!\left(\frac{x}{2}\right)-4 \psi
   ^{(-2)}\!\left(\frac{x+1}{2}\right)\ .
\eea
For $P_1'(x)$, we  also need its first derivative \bea
\label{A'}
\partial_x \tilde {\cal A}(x,0) = 
2&\bigg[&12 \zeta '(-1) +\frac{ \log (2)}{3} +\log\! \bigg(\Gamma
   \Big(\frac{1}{2}-\frac{x}{2}\Big)\bigg) \nn\\ &&- \log\! \bigg(\Gamma
   \Big( 1-\frac{x}{2}\Big)\bigg)   + \log\! \bigg(\Gamma
   \Big(\frac{x}{2}\Big)\bigg) \nn\\
   && - \log\! \bigg(\Gamma 
   \Big(\frac{x+1}{2}\Big)\bigg)\bigg]\ .
\eea
The Taylor expansion of $\tilde A(x,0)$  is
\bea\label{expansion69}
\tilde {\cal A}(x,0) &=&x \left[-2 \log (x)+\frac{8}{3} \Big(9 \zeta '(-1)+\log
   (2)\Big)\right] \nn\\
   && +x^2 \log (4)+\frac{x^4 \zeta (3)}{4}+\frac{x^6 \zeta
   (5)}{8}+\frac{9 x^8 \zeta (7)}{128}\nn\\
   && +\frac{17 x^{10} \zeta
   (9)}{384}+{\cal O}  (x^{12} )\ .
\eea
Note the logarithmic term, which can be interpreted as a correction to the power law for $x\to 0 $ in $P'_1(x)$. 
Indeed, scaling suggests  \cite{WieseMajumdarRosso2010,MajumdarRossoZoia2010b,DelormeWiese2016} \be\label{P'scaling}
\ca P_{\rm 1, scaling}'(x) =   [x(1-x) ]^{\frac{1}{H}-2} \frac{\Gamma
   \big(\frac{2}{H}-2\big)}{\Gamma \big(\frac{1}{H}-1\big)^2}\ .
\ee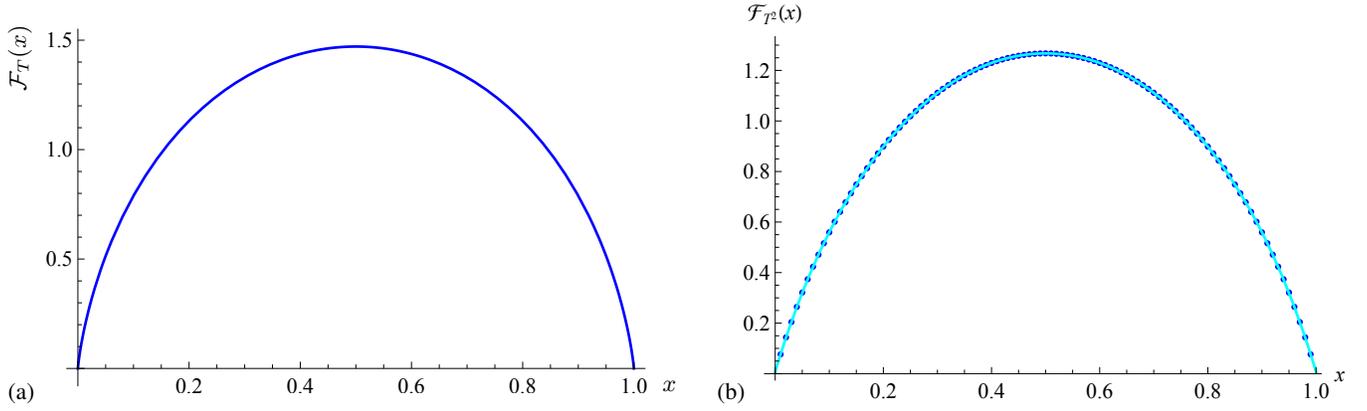
\begin{figure*}\fboxsep0mm
\vspace*{-3mm}
\newcommand{\kbox}[1]{{\!\!{#1}}}
\setlength{\unitlength}{1.0cm}
\kbox{\begin{picture}(8.4,5.3)
\put(0,0){(a)}
\put(0.5,0){\fig{8\unitlength}{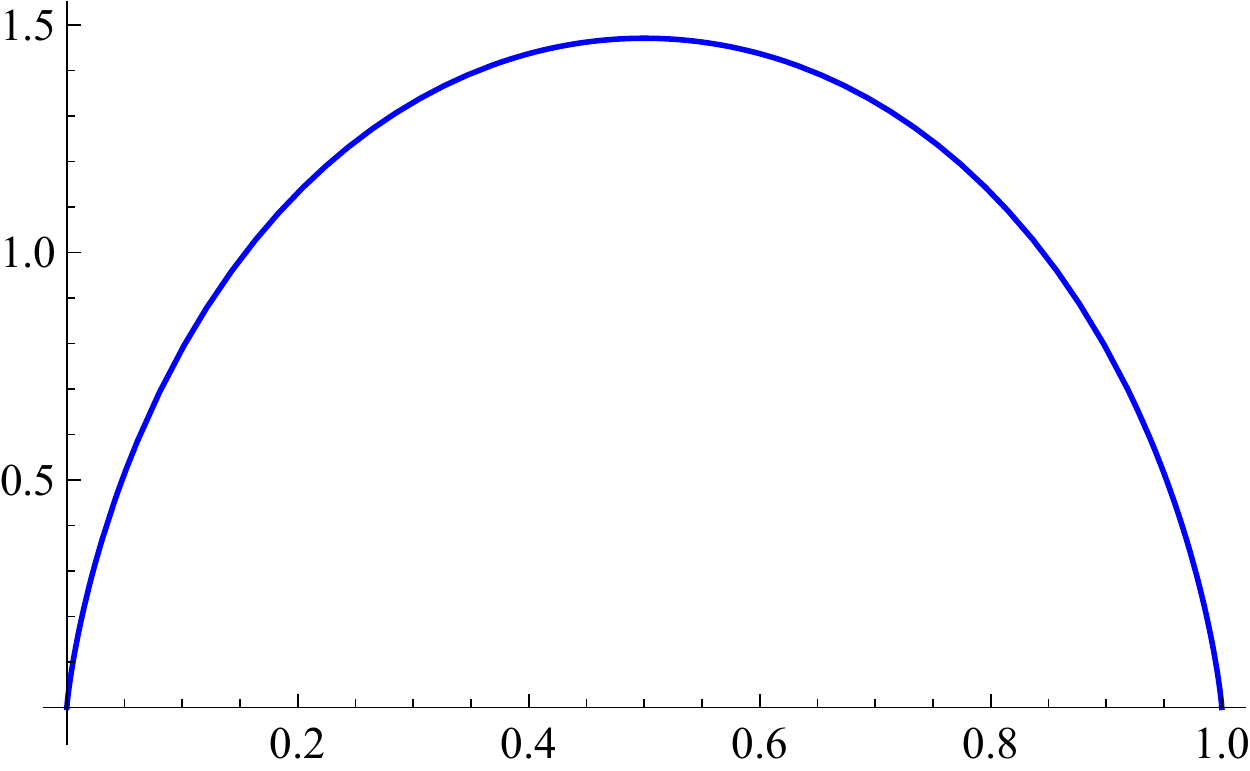}}
\put(0,4.1){\rotatebox{90}{${\cal F}_T(x)$}}
\put(8.7,0.1){$x$}
\end{picture}}~~~~~~~~~~~~
{(b)}\!\!
\fig{8cm}{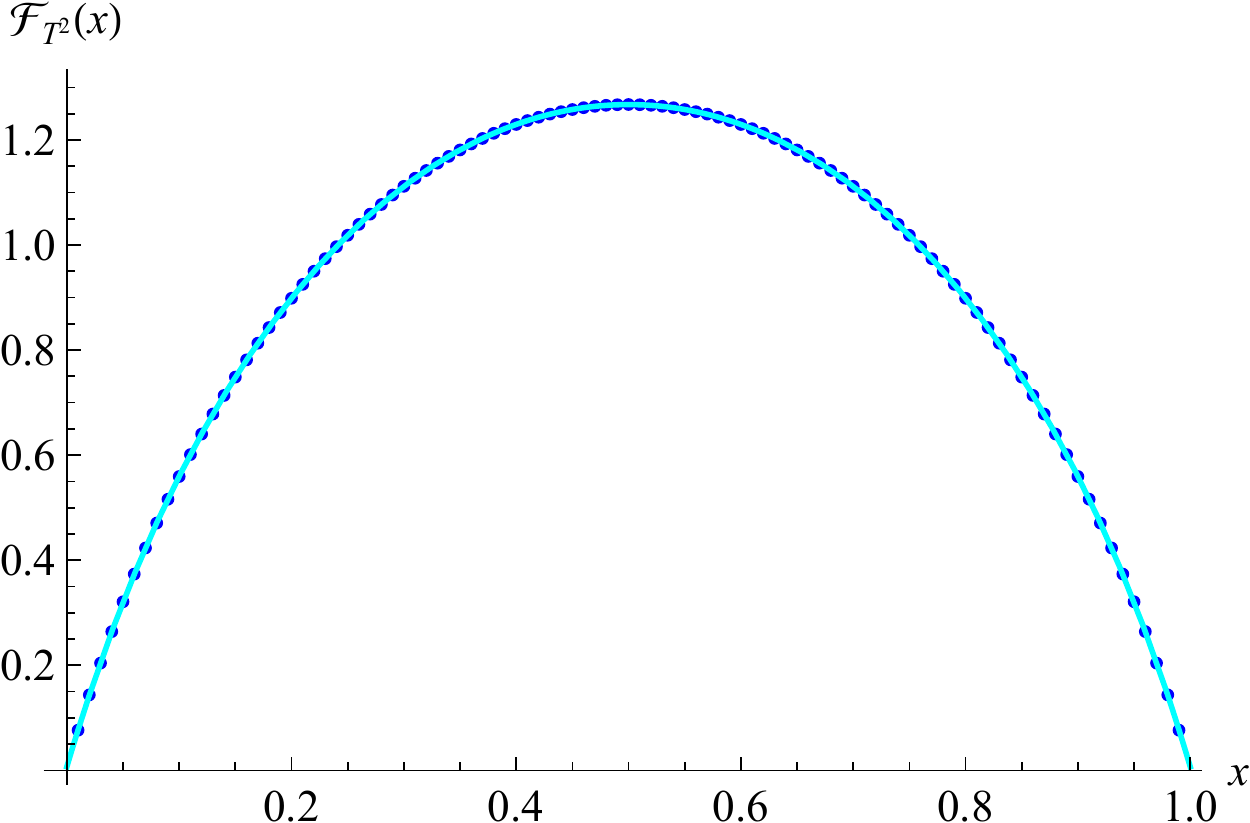}
\caption{(a) The function ${\cal F}_T(x)$, defined in Eq.~(\ref{FTdef}). (b) The function ${\cal F}_{T^2}(x)$, defined in Eq.~(\ref{FT2}). The  dots are the numerically obtained points, the line the fit of Eq.~\eq{54}.}
\label{f:FT}
\end{figure*}The correction to \Eq{P'scaling} at order $\epsilon$ can be written as
\bea\label{40}
\ca P_1'(x) &=& {\cal N} [x(1-x) ]^{\frac{1}{H}-2} 
\rme^{ \epsilon {\cal F}(x) }\ , \\
{\cal F}(x) &=& 2 \partial_x \tilde {\cal A}(x,0)+4 \ln(x)+4 \ln(1-x)+8 \nn\\
&=&  
4\bigg[12 \zeta '(-1) +\frac{ \log (2)}{3} +\ln\big(x(1-x)\big)\nn\\
&& ~~~+\log\! \bigg(\!\Gamma\!
   \Big({ \frac{1}{2}-\frac{x}{2}}\Big)\!\bigg)- \log\! \bigg(\!\Gamma\!
   \Big(1-\frac{x}{2}\Big)\!\bigg) \nn\\
   &&~~~+ \log\! \bigg(\!\Gamma\!
   \Big(\frac{x}{2}\Big)\!\bigg)- \log\! \bigg(\!\Gamma\!
   \Big(\frac{x+1}{2}\Big)\!\bigg)\bigg] \ .
   \label{Fofx}
\eea
We have chosen conventions s.t.\ $\int_0^1\rmd x\, {\cal F}(x)=0$, moving the constant term into the normalization ${\cal N}$. The latter  has to be chosen such that $\int_0^1\rmd x\, \ca P_1'(x)=1$.
The function $\ca F(x)$ is plotted on Fig.~\ref{f:Fofx}.
It has a regular Taylor expansion around $x=0$, 
\bea\label{F-Taylor}
{\cal F}(x) &=&
\frac{4}{3} \left[36 \zeta '(-1)+3+4 \log (2)\right]+x (\log (256)-4)-2
   x^2 \nn\\
   && +\frac{2}{3} x^3 (3 \zeta (3)-2)-x^4+\frac{1}{10} x^5 (15 \zeta
   (5)-8)\nn\\
   && +{\cal O}(x^6)\ .
\eea
For numerical purposes, a Taylor expansion around $x=1/2$ is appropriate (with  error $<0.002$)
\bea
{\cal F}(x) &=&  48 \zeta '(-1)-8 \text{ln$\Gamma $}({\textstyle \frac{3}{4}})+8 \text{ln$\Gamma
   $}({ \textstyle \frac{1}{4}})+4-{\textstyle \frac{20 }{3}}\ln (2) \nn\\
   && +16 (C-1)
   (x-{\textstyle \frac{1}{2}})^{2} \nn\\
   && \textstyle +\frac{1}{48} (x-{\textstyle  \frac{1}{2}})^{4}
   \left[\psi ^{(3)} ({\textstyle \frac{1}{4}} )-\psi
   ^{(3)}({\textstyle\frac{3}{4}}) -1536\right]\nn\\
   && +\ca O ({\textstyle  x-\frac{1}{2}})^{6} \nn\\
   &=& 0.116736 - 1.34455 (x-{\textstyle \frac{1}{2}})^{2}  - 0.353774 (x-{\textstyle \frac{1}{2}})^{4}\nn\\
   &&  +\ca O ({\textstyle  x-\frac{1}{2}})^{6} \ .
   \label{43}
\eea
Validation of 
this function via a numerical simulation  is given on  Figs.~\ref{f:all-Fs} and \ref{f:Fmean}.

\subsection{Remark on resummation}
In Eq.~\eq{40}, we had written the scaling function ${\cal F}(x)$ in the exponential. Since our calculation is performed at first order in $\epsilon$, other forms are possible, 
\bea
\ca P_1'(x) &=& {\cal N} [x(1-x) ]^{\frac{1}{H}-2} 
\rme^{ \epsilon {\cal F}(x) } +\ca O(\epsilon^2)  \label{form75}\\
  &=& {\cal N} [x(1-x) ]^{\frac{1}{H}-2} 			\label{form76}
\big[ 1+ \epsilon {\cal F}(x) \big] +\ca O(\epsilon^2) \qquad \\ 
&=& {\cal N} [x(1-x) ]^{\frac{1}{H}-2} 
\frac1{ 1- \epsilon {\cal F}(x)  } +\ca O(\epsilon^2)  \\
&=& ... \nn
\eea
The question arises which one to choose. There are many good reasons to choose the form \eq{form75}: 
\begin{enumerate}
\item adding drift $\mu$ to Browian motion, the latter appears as an additive term in the exponential
\be\label{P+mu(x,y,t)}
P^\mu_+(x,y,t) =\rme^{\frac{\mu (y-x)}2 -\frac{\mu^2 t}{4}} P_+(x,y,t) \ .
\ee
\item 
the first-order correction \eq{A'} contains logarithmic terms, visible in \Eq{expansion69}. Having them in the exponential, they are resummed into power laws, according to 
\be
\rme^{\epsilon \ln (x)} = x^\epsilon .
\ee 
This is how in \Eq{40} the scaling function of the Brownian, $x(1-x)$, was changed into $[x(1-x)]^{\frac1H-1}$. At the same time, the scaling function $\ca F(x)$, defined in \Eq{Fofx}, becomes regular for $x\to 0$, as can be seen on \Eq{F-Taylor}. 
\item in field theory, perturbative corrections are in general, and most efficiently, calculated for the effective action, i.e.\ the log of the partition function. In a thermodynamic setting as the one here, the effective action can be interpreted as the free energy. 
\item
Finally, as the exponential function is always positive for real arguments, the form \eq{form75}   remains positive even when $\epsilon \ca F(x)$ becomes large. This is a necessary condition for a probability density. 
\end{enumerate}
For all these reasons, using the exponentiated version is the most natural choice, and the one chosen throughout this article. When corrections are large, which is especially important for universal amplitudes, we will compare   this choice with the linear extrapolation \eq{form76}. 

\subsection{Expectation of exit time}
The non-trivial 1-loop correction to  $\left< T_{\rm exit}(x)\right> $ given in Eq.~(\ref{Tabs0}) is $2\epsilon $ times 
\be\label{calB}
{\cal B}(x):=-\partial_s \left[ \tilde {\cal A}(x,s) + \tilde {\cal A}(1-x,s)\right]\Big|_{s=0}\ .
\ee
Note that this combination is much simpler than the unsymmetrized one, which will allow us to integrate it analytically. 
We find with the same variable transformations as above 
\be
{\cal B}(x) = \int\limits_{\rme^{-\sqrt{\Lambda}}}^1 \rmd r \left[ \frac{1}{ r \left(\frac{1}{r^{x-1}-1}+\frac{1}{1-r^x}\right) \ln
   ^2(r)}+\frac{(1-x) x}{r \ln (r)}\right] .
\ee
Both terms can be integrated, the first after taking two derivatives w.r.t.\ $x$. Integrating twice w.r.t.\ $x$, and fixing the lost terms of the form $a+b x$  by demanding, according to \Eq{32}, that ${\cal B}(0)={\cal B}(1)=0$ yields\footnote{Note that there are corrections in the boundary region of the form $\rme^{-\sqrt{\Lambda} x}/x$. These might be interpreted as the finite-discretization corrections seen in the simulations of section \ref{s:simul}. We did not try to make this statement quantitative.}\bea
{\cal B}(x) &=& \left[ \gamma_{\rm E}  (x-1)+1\right] x-x \ln (x) + \frac{1}{2} (x-1) x \ln (\Lambda ) \nn\\
&& +\psi ^{(-2)}(1-x)+\psi ^{(-2)}(x+1)-\ln
   (2 \pi )\ .
\eea
Taylor-expanding for small $x(1-x)$, we find
\bea
\frac{2 \epsilon {\cal B}(x)}{
\left< T_{\rm exit}(x)\right>_0} &=& - 4 \epsilon \left[\gamma_{\rm E}-1+\ln\!\Big(  {x(1-x)}{\sqrt \Lambda}\Big) \right] \nn\\
&&+ \ca O\big(x(1-x)\big)\ .
\eea
This is consistent with 
\bea
\left< T_{\rm exit}(x)\right> &=&\frac1{D_\epsilon}\left< T_{\rm exit}(x)\right>_0+ 2 \epsilon {\cal B}(x) +\ca O (\epsilon^2)\nn\\
&  \sim& [x(1-x)]^{\frac1H -1} + \ca O (\epsilon^2)\ .
\eea
Let us define 
\bea \label{FTdef}
&&\!\!{\cal F}_T(x) := \frac{2   {\cal B}(x)}{
\left< T_{\rm exit}(x)\right>_0} + 4 \left[ \log \!\Big(\sqrt{\Lambda } (1-x) x\Big)+\gamma_{\rm E} -1\right]  \nn\\
&&= \frac{4} {x(1-x)} \Big[ x^2+x \Big((1-x) \log (1-x)-x \log (x)\Big) \nn\\
&& ~~~~~~~~~~~~~~~+\psi
   ^{(-2)}(1-x)+\psi ^{(-2)}(x+1)-\log (2 \pi )\Big].~~~
\eea
This function is plotted on Fig.\ \ref{f:FT} (left), and numerically validated in Fig.~\ref{f:ampTboundary}, and more precisely in Fig.~\ref{f:FT+FT2}.

\subsection{Expectation of exit time squared, $\left< T_{\rm exit}(x)^2 \right> $}
The first-order correction to $\left< T_{\rm exit}(x)^2\right> $ given in Eq.~(\ref{Tabs0}) is $2\epsilon $ times 
\be
{\cal C}(x):=\partial_s^2 \left[ \tilde {\cal A}(x,s) + \tilde {\cal A}(1-x,s)\right]\Big|_{s=0}\ .
\ee
Again, this combination is much simpler than the unsymmetrized one. 
We find with the same variable transformations as above 
\bea\label{52}
{\cal C}(x) &=& \int_{\rme^{-\sqrt{\Lambda}}}^1 \rmd r\, \Bigg[ \frac{1}{r \left(\frac{r}{r-r^x}+\frac{1}{r^x-1}\right) \log ^4(r)}\nn\\
&& +\frac{x r^{1-x}+x
   r^{x-1}-x r^{-x}+r^{-x}-x r^x+r^x-2}{(r-1)^2 \log ^3(r)} \nn\\
   &&  +\frac{r^{1-x}+r^x+(r+1) (6
   (x-1) x-1)}{6 (r-1) r \log ^2(r)}\nn\\
   && +\frac{x^4-2 x^3+x}{3 r \log (r)}\Bigg] \ .
\eea
Anticipating that a good approximation is given by $\left< T^2(x)\right> \sim \left[\left< T_{\rm exit}^2(x)\right>_0 \right]^{\frac1H-1}$, we set with normalization ${\cal N}$
\be
\left< T_{\rm exit}^2(x)\right> = {\cal N} \left[\left< T_{\rm exit}^2(x)\right>_0 \right]^{\frac1H-1} \rme^{\epsilon {\cal F}_{T^2}(x)+\ca O(\epsilon^2)}\ .
\ee
This implies that up to a constant, which will notably depend on the UV-cutoff $\Lambda$,
\bea\label{FT2}
&& \!\!\!\!\!\!\!{ {\cal F}_{T^2}(x) = 2\, \ca C(X) + 4 \ln\big(\left< T_{\rm exit}^2(x)\right>_0 \big) + \mbox{const} }
\nn\\
& = & 2 \,\ca C(X)  + 4 \ln\big( x(1-x)(1+x-x^2) \big) + \mbox{const}\ .\qquad 
\eea
We did not succeed to integrate \Eq{52} analytically. A numerical integration can be done without   difficulty, and yields the points on Fig.~\ref{f:FT} (right). A fit with a symmetric polynomial of degree 8 (with a total systematic plus numerical deviation smaller than $10^{-3}$) reads 
\bea
{\cal F}_{T^2}(x)&\approx & \textstyle 1.26033 -3.73328 \left(x-\frac{1}{2}\right)^2 -4.16628
   \left(x-\frac{1}{2}\right)^4\nn\\ \textstyle
   && \textstyle +5.24129 \left(x-\frac{1}{2}\right)^6  -38.0198 \left(x-\frac{1}{2}\right)^8\ .
   \label{54}
\eea
Validation via a numerical simulation  is presented on Fig.~\ref{f:FT+FT2}.

\subsection{Estimation of time scales: The mean exit time}
Up to now, we considered universal functions, without explicit evaluation of the proper time scales. This is motivated by the observation that time scales  are often more sensitive to details of the implementation than amplitude ratios, as those encoded in the functions ${\cal F}(x)$, $\ca F_T(x)$, and $\ca F_{T^2}(x)$. Nevertheless, our formalism is able to compute universal amplitudes, a task we  turn to now.

We start by the simplest such observable, the mean exit time in the strip. By mean we understand an average over the starting position $x$, and the realization of the process. According to Eqs.~\eq{Tabs0}, \eq{calJ} and \eq{calB}
\bea\label{83}
\lefteqn{\int_0^1 \left< T_{\rm exit}(x)\right> \rmd x} \nn \\
&&=-\partial_s\Big|_{s=0} \int_0^1 \rmd x \,\left[\tilde J\Big(x,1,\frac s{D_{\epsilon}}\Big)-\tilde J\Big(x,0,\frac s{D_{\epsilon}}\Big) \right] \nn\\
&& \hphantom{=-\partial_s\Big|_{s=0} \int_0^1 \rmd x }+ 2 \epsilon\left[\tilde A(x,s) -\tilde A(1-x,s) \right]\nn\\
&=& \int_0^1  \frac{x(1-x)}{2D_\epsilon} + 2 \epsilon\, \ca B(x) \,\rmd x  \ .
\eea
Recalling the definition of $\ca F_T(x)$ in \Eq{FTdef}, this can be written as 
\bea
\lefteqn{\int_0^1 \left< T_{\rm exit}(x)\right> \rmd x} \nn \\
&=& \frac 1{12 D_\epsilon}+ \frac\epsilon2\, \int_0^1\rmd x \ca F_T(x) x(1-x)\nn\\
&& - 2 \epsilon \int_0^1   x(1-x) \left[ \ln (\sqrt \Lambda x(1-x) )+\gamma_{\rm E}-1\right] \rmd x\ .~~~~~~~
\eea
The terms in question are 
\bea
\label{85}
&&\frac12 \int_0^1 \ca F_T(x) x(1-x)\, \rmd x= -\frac{5}9 - 4 \zeta'(-1) = 0.106129... \nn\\\\
&& -2\int_0^1   x(1-x) \left[ \ln (\sqrt \Lambda x(1-x) )+\gamma_{\rm E}-1\right] \rmd x \nn\\
&&= -\frac{\log (\Lambda )}{6}-\frac{\gamma_{\rm E} }{3}+\frac{8}{9} = 0.696484... -\frac{\ln (\Lambda)}6\ .
\label{86bis}
\eea
This yields 
\bea\label{Texit}
\!\!\!\lefteqn{\int_0^1 \left< T_{\rm exit}(x)\right> \rmd x} \nn \\
&&~~~=\frac1{12}\Big[ 1+2 \epsilon  \Big(1 -  \gamma_{\rm E}-24 \zeta '(-1) \Big) + \ca O (\epsilon^2) \Big]\nn\\
&&~~~=\frac1{12}\Big[ 1+ 8.78578 \epsilon    + \ca O (\epsilon^2) \Big] \nn\\
&&~~~=\frac1{12}\exp( 8.78578 \epsilon )   + \ca O (\epsilon^2)  \ .
\eea
Note that all cutoff dependence has canceled, as expected. 
In the last two lines we   gave two alternative resummations: The linear, order-$\epsilon$ term, and an exponential resummation, which was beneficial in resumming  logarithms into a change in power-law. We will see later (Fig.~\ref{f:epxT+expT2}) that the numerically obtained result  lies between the two expressions. 

\subsection{Exit times in the limit of $x\to 0$}
Another interesting limit is  $x\to 0$,  also considered analytically in Ref.~\cite{GuerinLevernierBenichouVoituriez2016}.
Let us define the ratio
\be\label{R(x)}
R(x) := \frac{ \left< T_{\rm exit}(x)\right> }{[x(1-x)]^{\frac1H-1}}\ .
\ee
As in the preceding section, we derive for $x\to 0$
\be\label{92}
R(0):=\lim _{x\to 0} R(x)=  \frac 1 2 +\epsilon (1-\gamma_{\rm E}) +\ca O(\epsilon^2)\ .
\ee
A test is given   on Fig.~\ref{f:ampTboundary}. This limit agrees\footnote{Using in the supplementary material of \cite{GuerinLevernierBenichouVoituriez2016} $\psi(t) = 2D t^{2H}$ yields $\psi_0(t)=2D t$, and  $\psi_1(t) = 4D t \ln(t)$.  Eq.~(G48) (generalized  to  arbitrary $D$)    becomes  $T p_s(0)= \frac{x_0}{2D} + \epsilon \frac{x_0}{D}(1-\gamma_{\rm E}-2 \ln (x_0)+4 \ln(D))   + \ca O(\epsilon^2)$. 
Setting $D=1$, we get $ \lim _{x\to 0}T p_s(0)/x^{\frac1H-1} = \frac 12 + \epsilon (1-\gamma_{\rm E}) $, which agrees with $R(0)$ in Eq.~\ref{92}.
The order-$\epsilon$ correction extracted from the simulation data of  Ref.~\cite{GuerinLevernierBenichouVoituriez2016}    is consistent with this value.} 
with the equivalent object calculated in Ref.~\cite{GuerinLevernierBenichouVoituriez2016}. 
\begin{figure}
\Fig{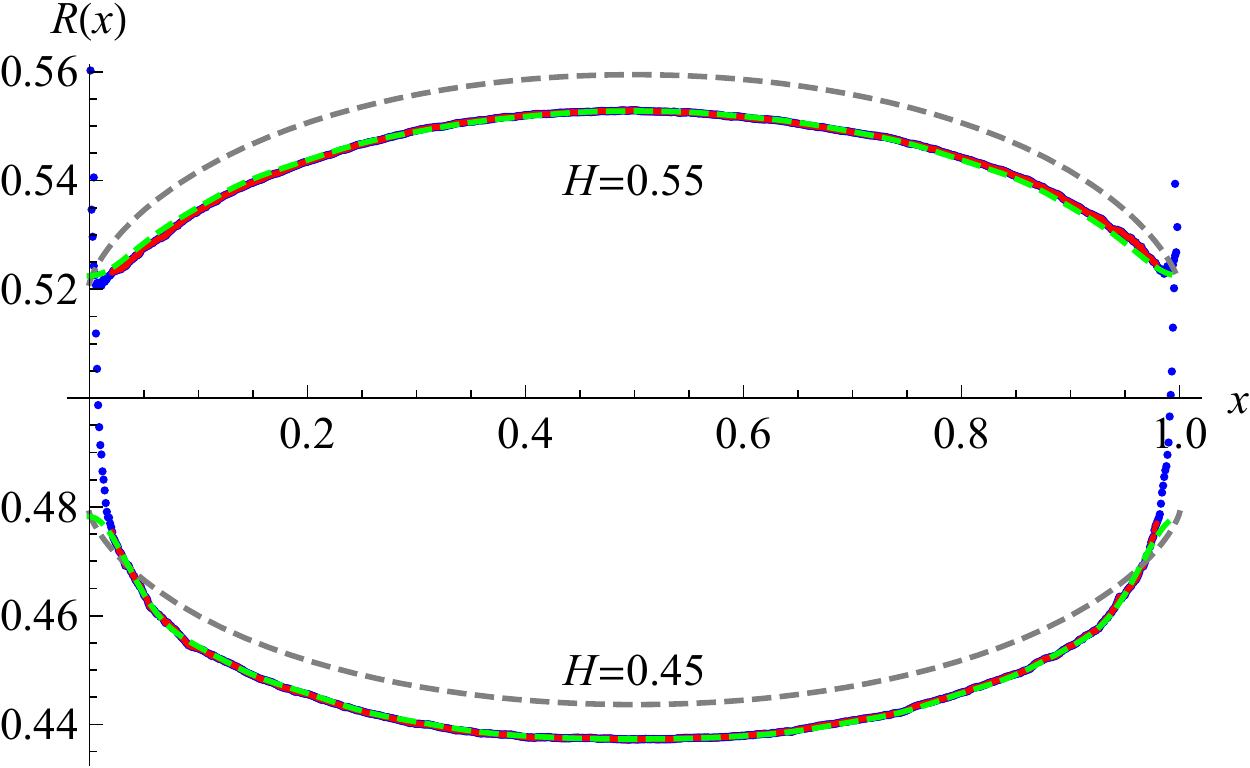}
\caption{The ratio ${R(x)}$ defined in \Eq{R(x)}. In grey are the analytical (parameter free) predictions, in red or blue the data points of numerical simulations, presented in section \ref{s:simul}. The red data points have been used to extract via a polynomial fit of order 40 (dashed green line) $R(0)$, with the result $R_{H=0.45}^{\rm num}(0)=0.478$, $R^{\rm num}_{H=0.55}(0)=0.522$. Analytically \Eq{92} yields  $R_{H=0.45}^{\rm ana}(0)=0.479$, $R^{\rm ana}_{H=0.55}(0)=0.521$. The deviation in the middle of the domain is of order $7 \times 10^{-3}$, consistent with an $\ca O(\epsilon^2)$ correction to $R(x)$ of amplitude 3. More robust tests of our formulas, focusing on the shape of $R(x)$, and the spatially averaged exit times,  are   presented in section \ref{s:Expectation of exit times and their squares}.}
\label{f:ampTboundary}
\end{figure}

\subsection{Time scales: The second moment of the exit time}
Analogously to the derivation of \Eq{83}, we have
\bea
\lefteqn{\int_0^1 \left< T_{\rm exit}(x)^2\right> \rmd x} \nn \\
&&=\partial_s^2\Big|_{s=0} \int_0^1 \rmd x \,\left[\tilde J\big(x,1,\frac s{D_{\epsilon}}\Big)-\tilde J\Big(x,0,\frac s{D_{\epsilon}}\Big) \right] \nn\\
&& \hphantom{=-\partial_s\Big|_{s=0} \int_0^1 \rmd x }+ 2 \epsilon\left[\tilde A(x,s) -\tilde A(1-x,s) \right]\nn\\
&&= \int_0^1  \frac{x(1-x)(1+x-x^2)}{12D_\epsilon^2} + 2 \epsilon\, \ca C(x) \,\rmd x \ .
\eea
The first integral is 
\be
\int_0^1  \frac{x(1-x)(1+x-x^2)}{12D_\epsilon^2} \rmd x = \frac1{60 D_\epsilon^2}\ .
\ee
The $x$-integral over $\ca C(x)$ defined in \Eq{52} can be done analytically. The remaining non-trivial integral reads
\bea
\int\limits_0^1 \ca C(x) \,\rmd x&&= \int\limits_{\rme^{-\sqrt{\Lambda}}}^1 \rmd r \bigg[ \frac{r+1}{3r (1-r) \log ^2(r)}  +\frac{r+1}{(r-1) r \log
   ^4(r)} \nn\\
&&\hphantom{\int_{\rme^{-\sqrt{\Lambda}}}^1 \rmd r }~~~+\frac{\frac{1}{3 r}-\frac{2}{(r-1)^2}}{\log ^3(r)}+\frac{1}{15 r \log (r)}\bigg].
\label{86}
\eea
This integral is hard to evaluate analytically. A precise numerical estimation can be obtained as follows: Taylor-expand the integral at small $r$, and integrate, to show that 
\be\label{87}
\int\limits_0^1 \ca C(x) \,\rmd x \simeq \mbox{const} + \frac{1}{3 \Lambda ^{3/2}}-\frac{1}{3 \sqrt{\Lambda }}+\frac{1}{6 \Lambda }-\frac{\log
   (\Lambda )}{30}\ .~~
\ee
The integral \eq{86} can then be integrated numerically. To this aim, one splits it into two pieces: The region close to $r=1$, for which one Taylor-expands the integrand around $r=1$ and then integrates symbolically. And the remaining region, with cutoff $\Lambda$. 
Subtracting the  terms in \Eq{87} from the numerically evaluated integral allows us to obtain the latter precisely already for relatively small $\Lambda$. The   result of this procedure is
\be
\int\limits_0^1 \ca C(x) \,\rmd x = 0.152119\, -\frac{\log (\Lambda )}{30} +\ca O \!\left(\frac1{\sqrt{\Lambda}}\right)\ .
\ee
This   yields for the second moment of the exit time
\bea\label{Texit2}
\int_0^1 \left< T_{\rm exit}(x)^2\right> \rmd x &=& \frac1{60}\Big[ 1+  16.5632 \epsilon + \ca O (\epsilon^2) \Big]\nn\\
&=& \frac1{60}\exp\Big( 16.5632 \epsilon + \ca O (\epsilon^2) \Big)\ .~~~~~~
\eea
Note that all cutoff dependence has canceled, as it should.
The result is confronted to numerical simulations on Fig.~\ref{f:epxT+expT2} (right).

\subsection{Corrections to $\left< T_1\right> $}
\label{s:T1-corr}
In \Eq{PT1-bare}, we had established that for  Brownian motion
\bea\label{PT1-bare-bis}
P_{T_1}(t) &=&  2(1+2 t\partial_t)  \int_0^1 \rmd x\, J\!\left( x,0,t\right)  \nn\\
   &=&  (1+2 t\partial_t)  \int_0^1 \rmd x\, \left[ J\!\left( x,0,t\right) - J\!\left( x,1,t\right)  \right]\ .~~~~~~~
\eea
The generalization to fBm at order $\epsilon$ is obtained as in the preceding sections as
\be\label{PT1-bare-cis}
{\cal P}_{T_1}(t)     =   \left(1+\frac tH\partial_t \right) \int_0^1 \rmd x\, \left[ \ca J\!\left( x,0,t \right) -\ca J\!\left( x,1,t\right)  \right]\ .
\ee
The factor of $1/H$ comes from the fact that   the derivative in \Eq{PT1-bare} was w.r.t.\ $m$, and the scaling variable now is $m/t^H$. 
We conjecture that this result remains valid to all orders in $\epsilon$, s.t.\be
{\cal P}_{T_1}(t)  =  \left(1+\frac tH\partial_t \right) \int_0^1 \rmd x\,{\cal P}_{\rm exit}(x,t)\ .
\ee
As a consequence, 
\be
\left< T_1^n\right>  = \left( 1+ \frac n H \right) \int_0^1\rmd x\, \left< T_{\rm exit}^n(x)\right>  \ .
\ee
For the first two moments, this yields
\bea
\left< T_1\right> &=&\frac{1}{4}+\epsilon  \left[\frac{1}{6} -12 \zeta '(-1)-\frac{\gamma_{\rm E}
   }{2}\right]+\ca O (\epsilon^2) \nn
\\
&=& \frac14 \left[ 1+ 7.45245 \epsilon +\ca O (\epsilon^2)\right] \nn\\
&=& \frac14 \exp\!\Big( 7.45245 \epsilon +\ca O (\epsilon^2)\Big)\ .
\label{T1-full}
 \\
\left< T_1^2\right> &=& \frac1{12} \left[ 1+ 14.9632 \epsilon +\ca O (\epsilon^2)\right] \nn\\
&=& \frac1{12} \exp\!\Big( 14.9632 \epsilon +\ca O (\epsilon^2)\Big)\ .
\label{T12-full}
\eea
As usual, we have given two possible resummations. This will be tested later, see Fig.~\ref{f:T1-full}.

\section{Numerical validation}
\label{s:simul}
\subsection{Algorithm}
\label{s:algo}
A numerical estimation of the calculated observables    is obtained using the discrete-time algorithm by Davies and Harte \cite{DaviesHarte1987}, as described in \cite{DiekerPhD,DiekerMandjes2003}. It generates  for fBm of a given  $H$  sample trajectories over a discretized time window $[0,1]$; the trajectories are 
 drawn from a Gaussian probability with covariance \eqref{eq:covariance}. Time and space are then rescaled in respect of \Eq{eq:covariance} s.t.\ not more than $10^{-4}$ of all samples fail to exit for a given $H$. (The time-scales in question are the upper times in the plots on Fig.~\ref{f:all-P-T1}.) While this induces a small systematic error, we can take advantage of the lin $\times $ log performance of the Davies-Harte algorithm \cite{DaviesHarte1987,DiekerPhD,DiekerMandjes2003}, whereas the execution time for a sequential generation of the sequence   grows quadratically in time. Given the necessary system size, this would be very inefficient. 
 
 \begin{figure}\Fig{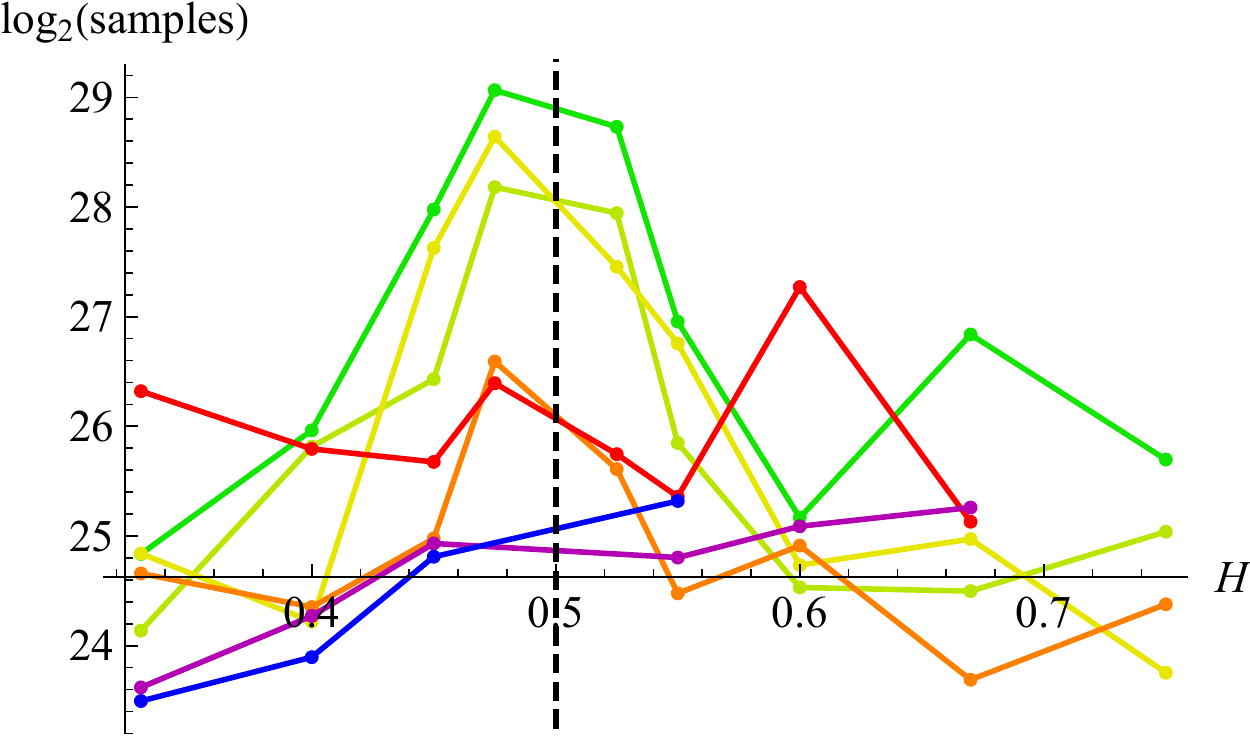}
\caption{Number of samples for each system size and value of $H$. The color code is as in Fig.~\ref{f:P'forH=0.33}. Since the measured signal is proportional to $\epsilon$, the error scales like the square root of the number of samples devided by $|\epsilon|$. Thus more samples are needed for $H$ close to $1/2$, and only small systems can be simulated for $H=0.475$ and $H=0.525$.}
\end{figure}
\subsection{Exit probability}
\begin{figure}[tb]
\fig{8.5cm}{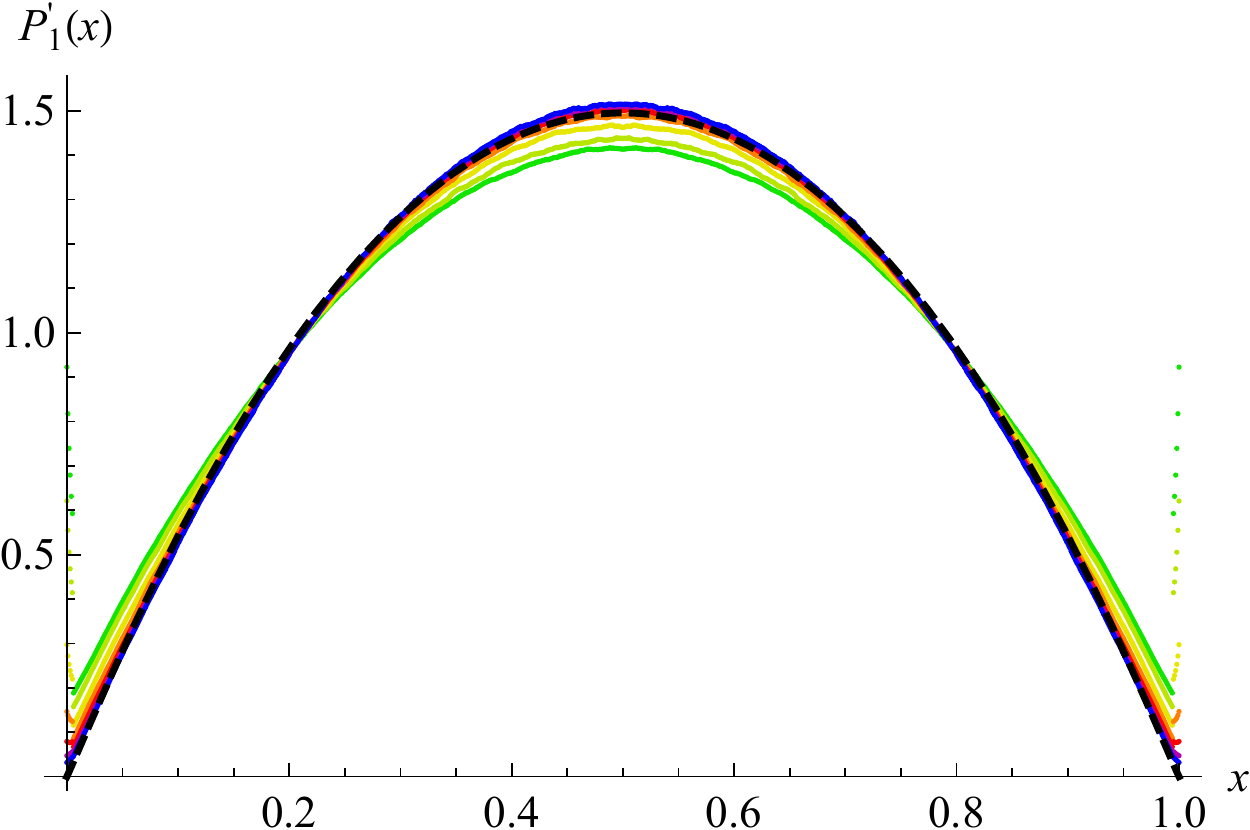}
\caption{The measured probability $P_1'(x)$ for $H=0.33$. The system sizes are (from bottom to top at $x=0.5$): $N=2^{13}$ (dark green), $2^{14}$ (green), $2^{16}$ (olive), $2^{18}$ (orange),  $2^{20}$ (red), $2^{22}$ (dark magenta), $2^{24}$ (blue). The dashed line is the scaling ansatz \eq{P'scaling} (i.e.\ almost a parabola).  Note the slow convergence for $x\to 0$ and $x\to 1$. Also note that the measured result for the largest system size   at $x=1/2$ is larger than the scaling ansatz \eq{P'scaling}.  This is equivalent to a positive curvature of the function $\ca F(x)$ defined in \Eqs{40} and \eq{Fofx}, and given in the first (top left) plot of Fig.~\ref{f:all-Fs}.}
\label{f:P'forH=0.33}
\end{figure}

\begin{figure*}
\!\!\!\raisebox{2.2cm}[0mm][0mm]{(a)}~~
\fig{8cm}{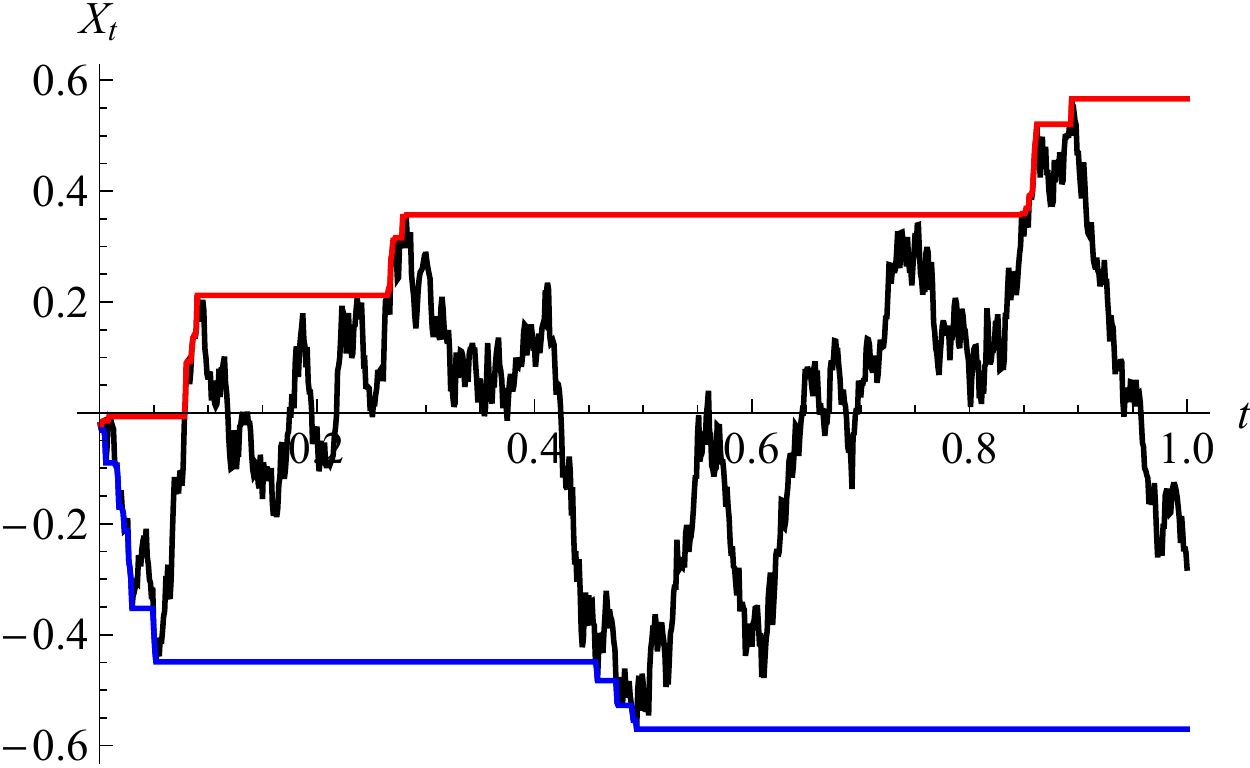}~~~~~~~~~~~~~\!\!\!\raisebox{2.2cm}[0mm][0mm]{(b)}\fig{8cm}{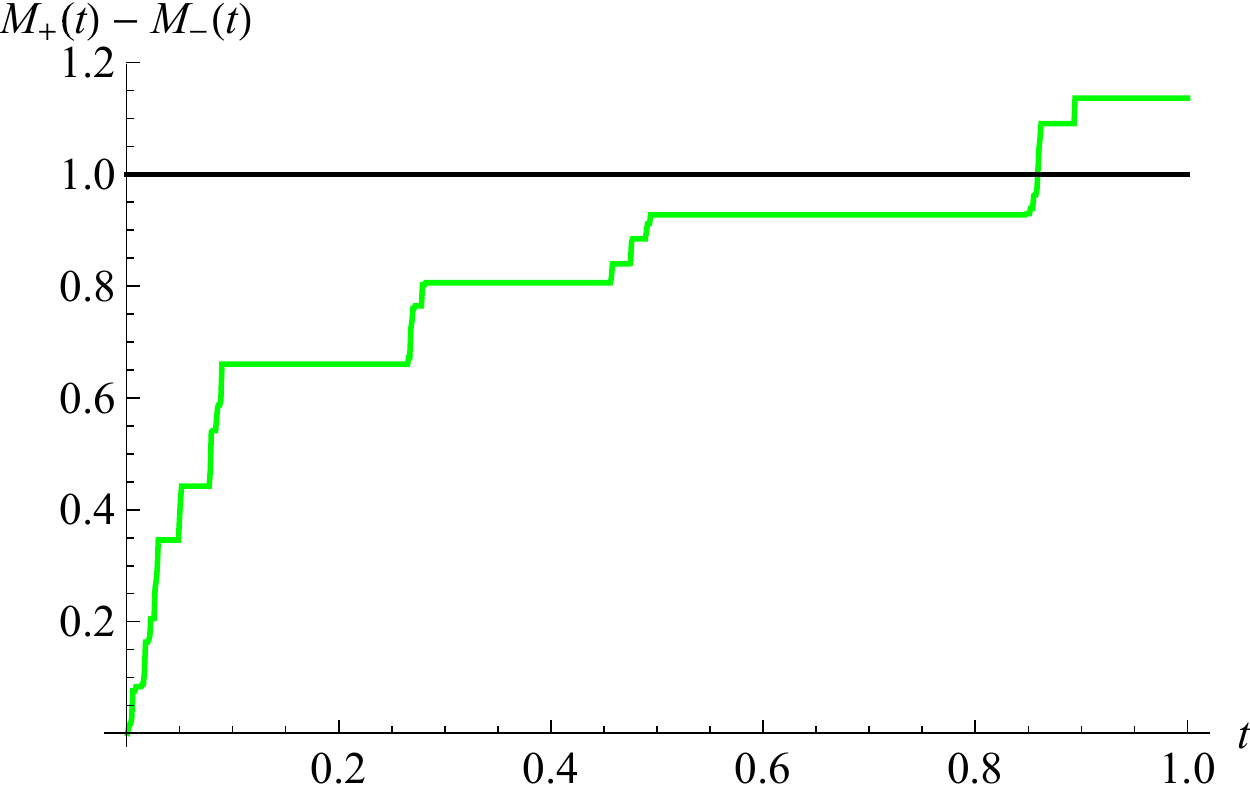}
\caption{(a) The random process $X_t$, with its running max (in red) and min (in blue), see main text. (b) The span, i.e.\ running max minus running min.}
\label{f:running-max-and-min}
\medskip

\noindent
\raisebox{5.0cm}[0mm][0mm]{(a)}~~\fig{8.3cm}{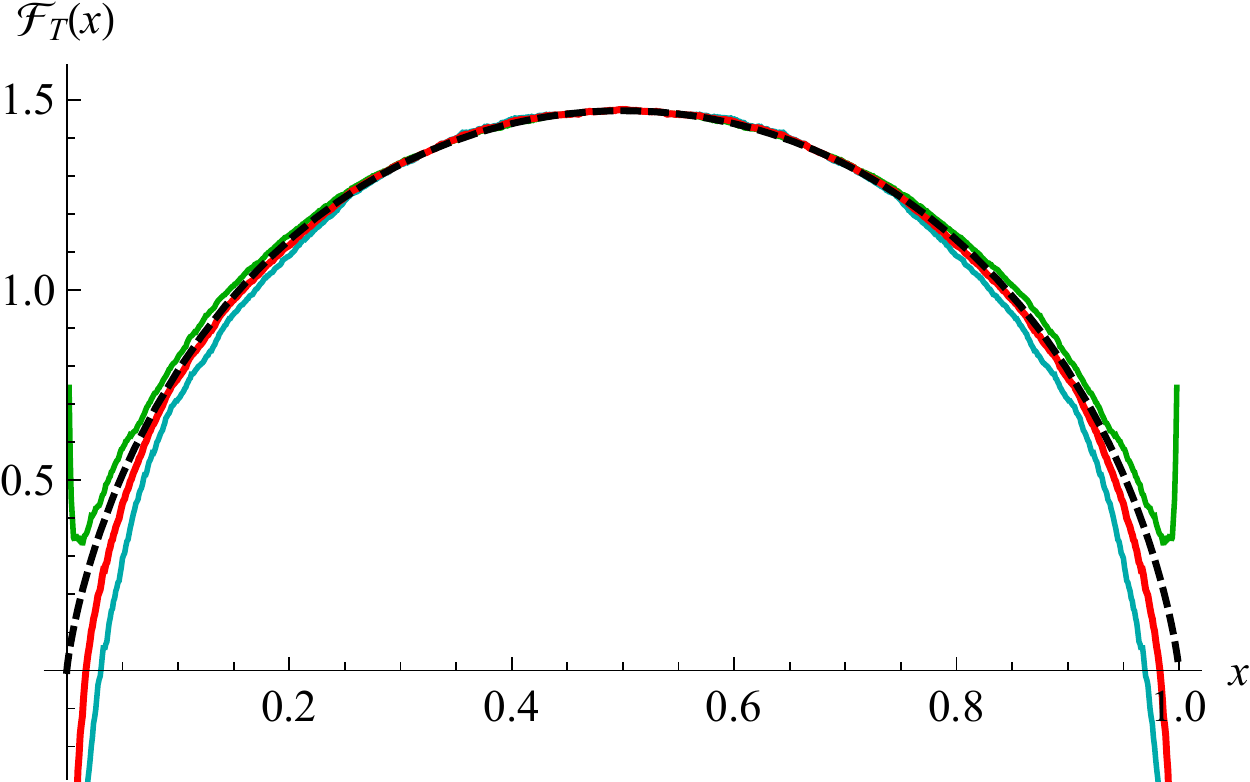}~~~\!\!\!\!\raisebox{5.0cm}[0mm][0mm]{(b)}~~\fig{8.5cm}{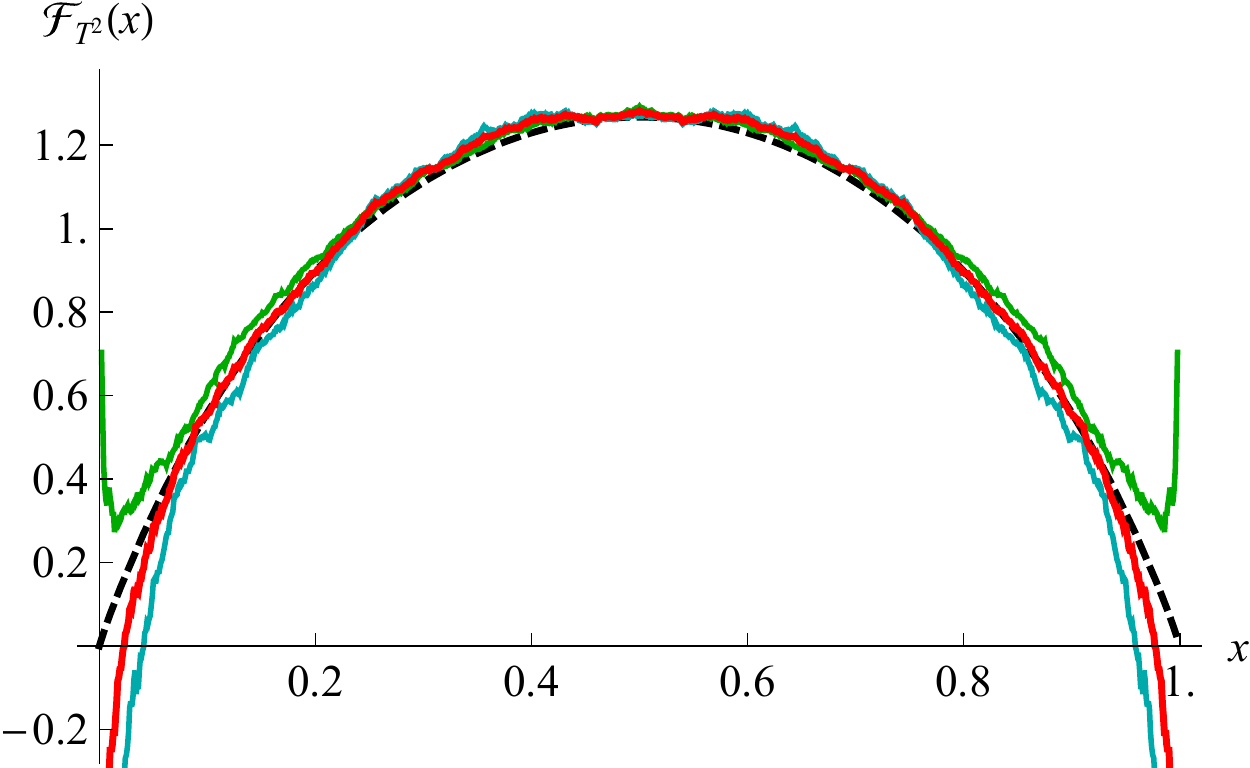}
\caption{(a) Mean (red) between the measured functions $ {\cal F}_T(x) $ for  $H=0.45$ (dark cyan, bottom line) and $H=0.55$ (dark green, top line). (b) ibid for ${\cal F}_{T^2}(x)$.
System size used was $N=2^{24}$, and the total number of samples is 
$1.64 \times 10^6$ for $H=0.45$, and $2.19\times 10^6$ for $H=0.55$.
} \label{f:FT+FT2}
\end{figure*}

\begin{figure*}[thbp!]
\newcommand{\figsize}{8.25cm}
\centerline{$H=0.33$ \hspace{8cm} $H=0.4$}\vspace*{-0.5cm}
\fig{\figsize}{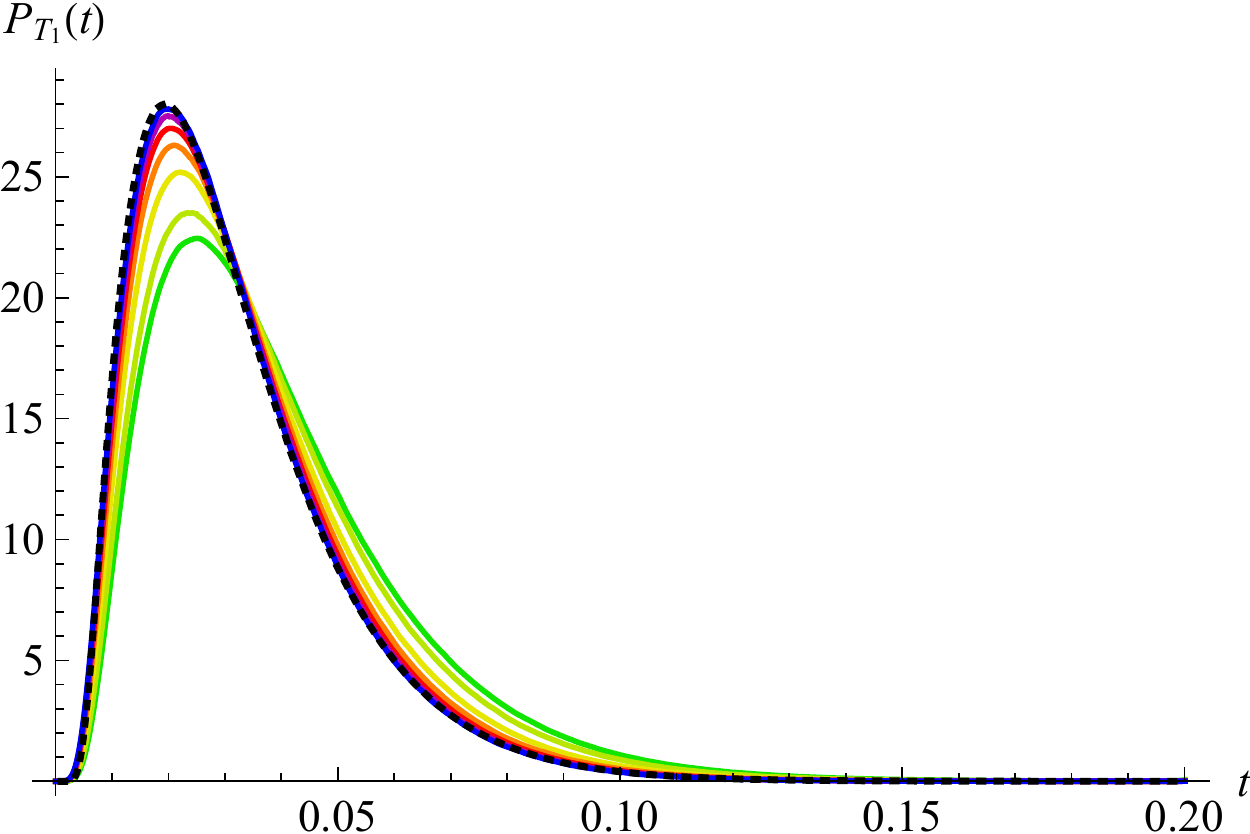}~~~~~~\fig{\figsize}{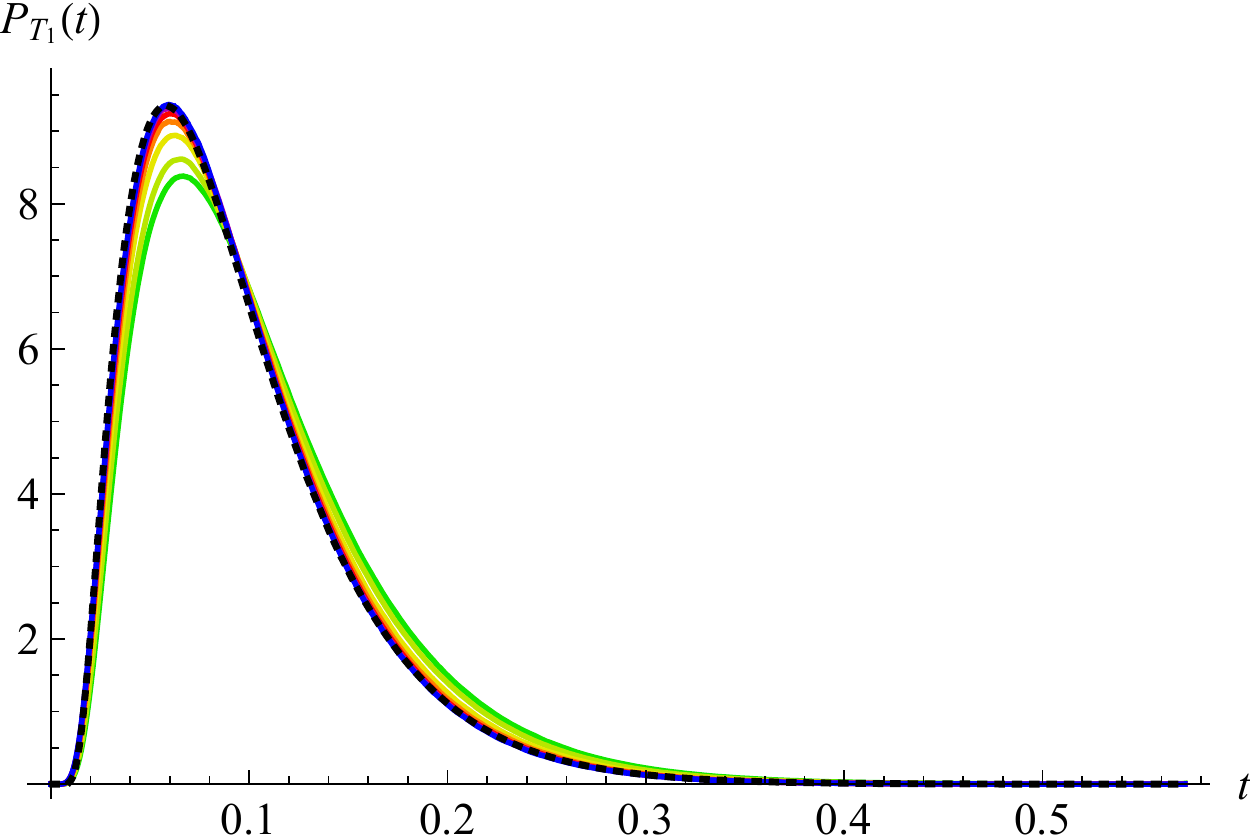}\\
\medskip

\centerline{$H=0.45$ \hspace{8cm} $H=0.475$}\vspace*{-0.5cm}
\fig{\figsize}{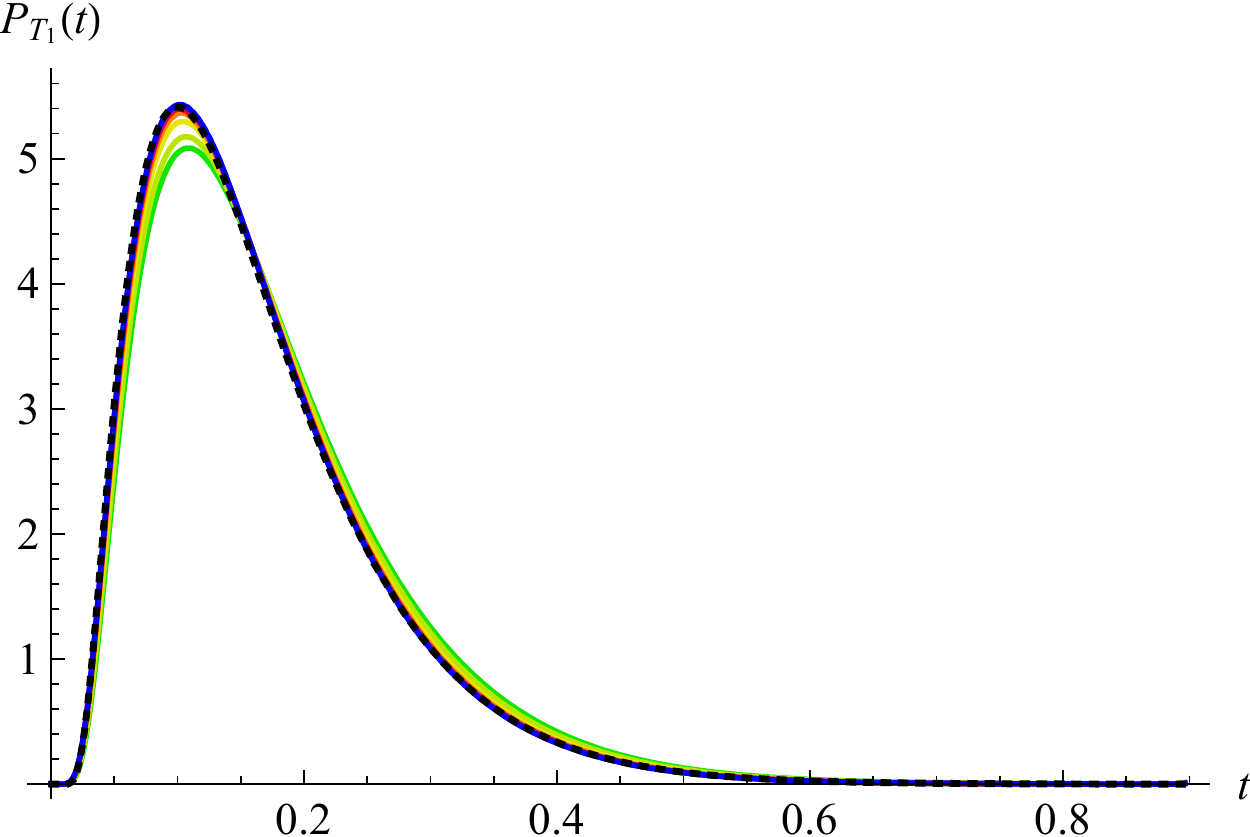}~~~~~~\fig{\figsize}{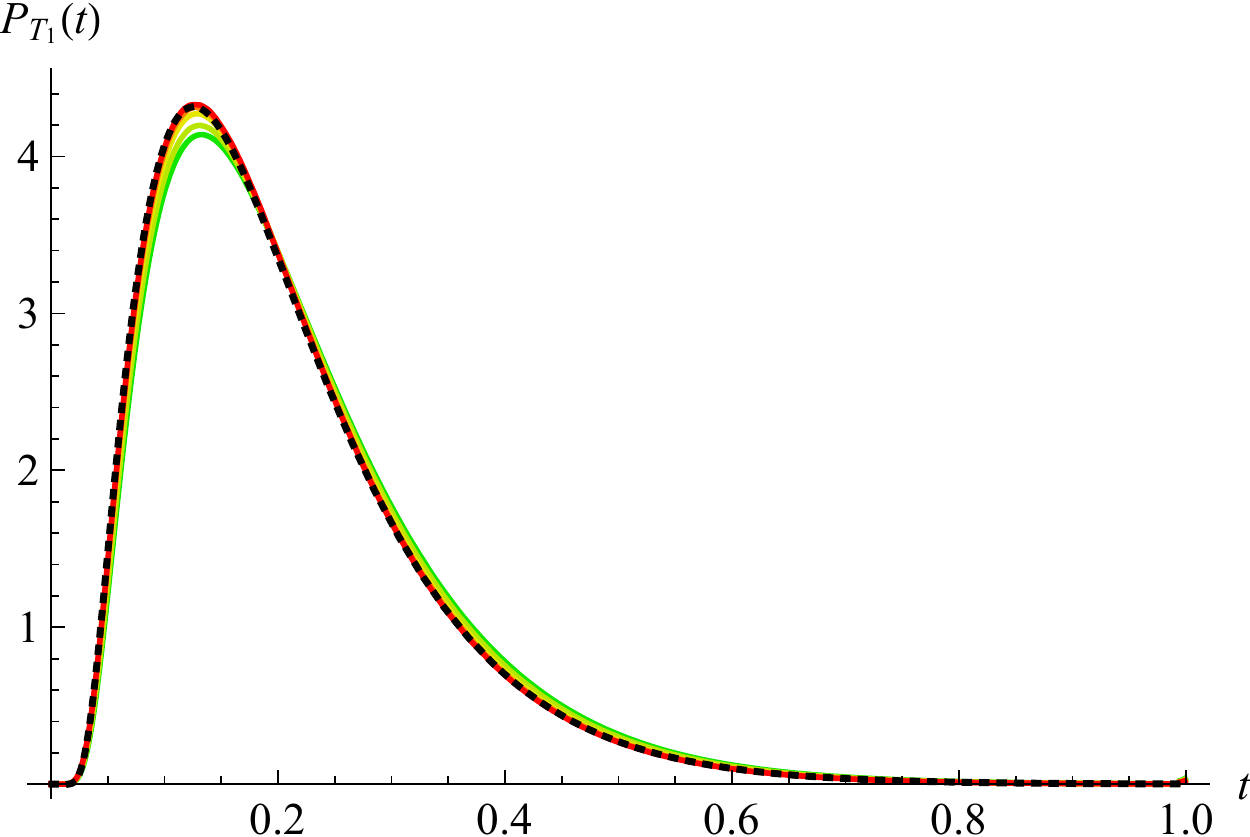}\\
\medskip

\centerline{$H=0.525$ \hspace{8cm} $H=0.55$}\vspace*{-0.5cm}
\fig{\figsize}{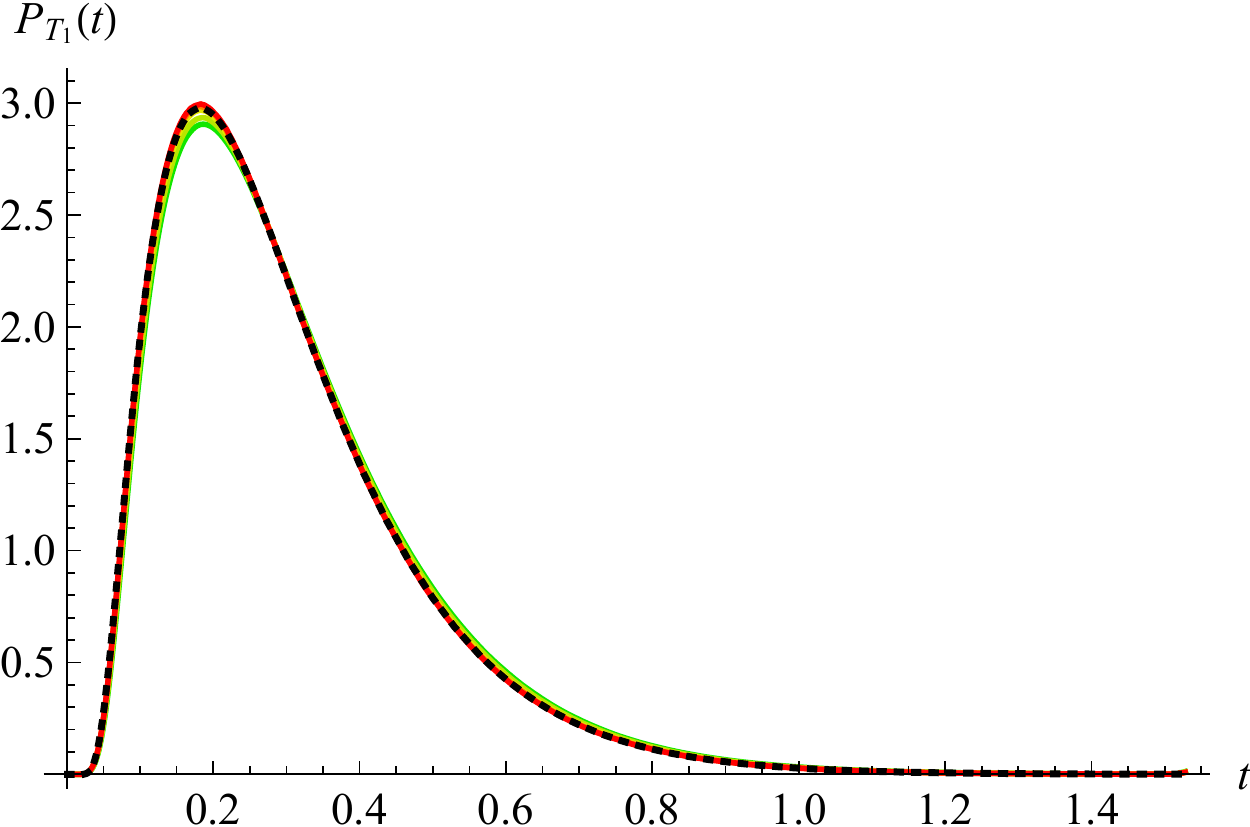}~~~~~~\fig{\figsize}{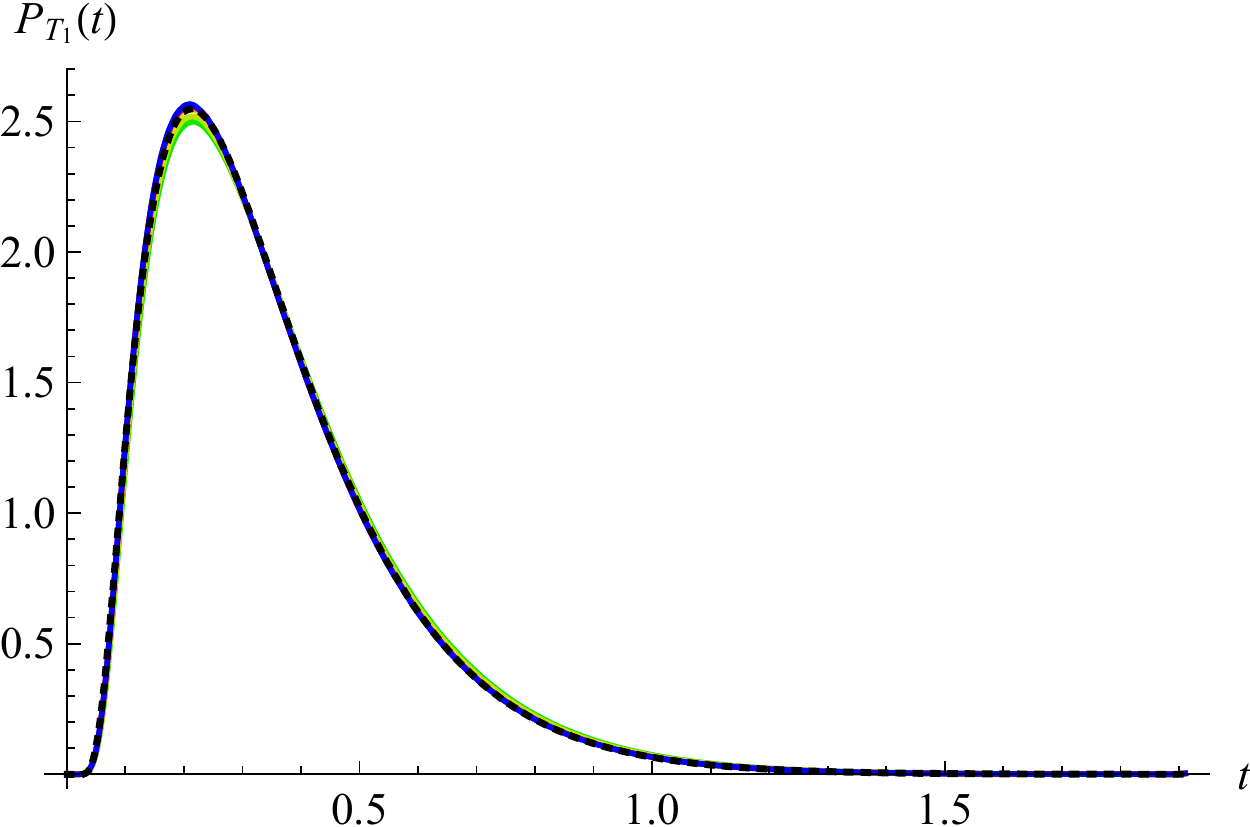}\\
\medskip

\centerline{$H=0.67$ \hspace{8cm} $H=0.75$}\vspace*{-0.5cm}
\fig{\figsize}{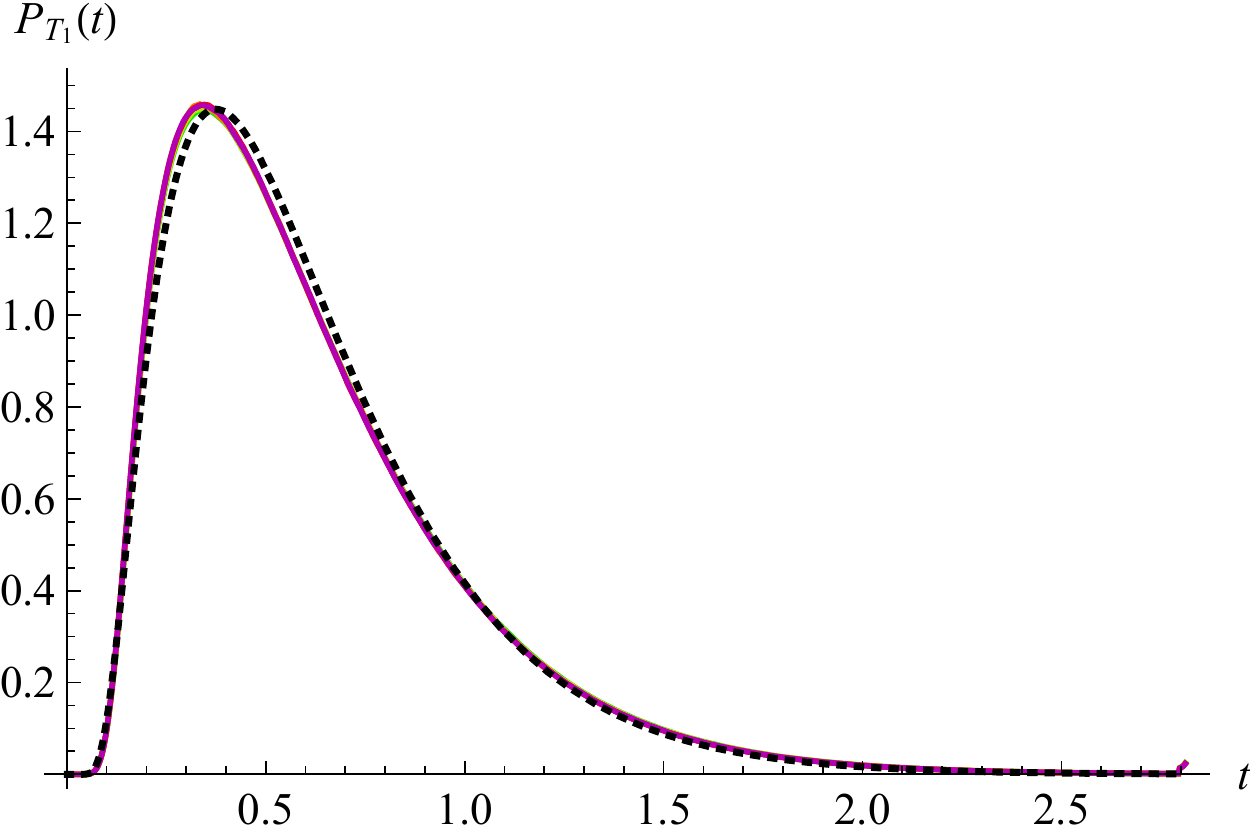}~~~~~~\fig{\figsize}{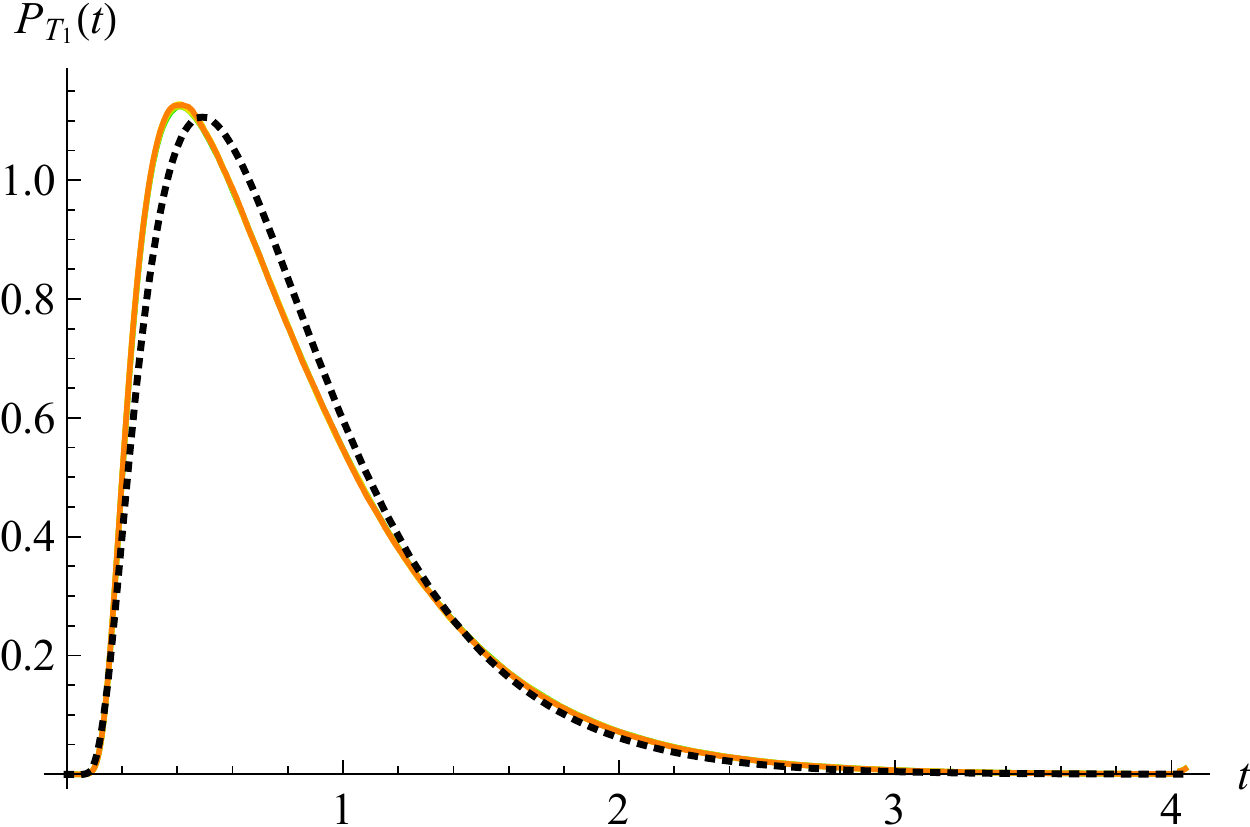}\\

\caption{Probability distribution of the numerically estimated time $T_{1}$ when the width of the process reaches 1, i.e.\ the process is absorbed irrespective of its starting position.  $N=2^{13}$ (dark green), $2^{14}$ (green), $2^{16}$ (olive), $2^{18}$ (orange),  $2^{20}$ (red), $2^{22}$ (dark magenta), $2^{24}$ (blue). The characteristic   time depends quite strongly on $H$. The result for $H=1/2$ is between the distributions for $H=0.475$, and $H=0.525$. The black dotted line is the analytic result for $H=1/2$, given in Eqs.~(\ref{PT1})--(\ref{PT1-Poisson}), rescaled so that the first moment $\left< T_1\right>$ is correctly reproduced. Small systematic deviations are visible for small and large values of $H$, especially $H=0.33$, and $H=0.75$.}
\label{f:all-P-T1}
\end{figure*}

In order to measure $P_1(x)$  one could start the process $X_t$ at $x$ and measure whether $X_t$ is first absorbed at $x=0$ or $x=1$. This is very inefficient, as for each $x$ one has to run a simulation, and repeat the latter until the statistics is good enough.  A slightly better strategy is to start with  $X_0=0$, generate $X_t$, shift it by $x$, and check for each $x$, whether it is first absorbed at $x=0$ or $x=1$. There is, however, a much more clever procedure, which we explain now, and which is illustrated on figure \ref{f:running-max-and-min}. 
Define for a  random process $X_t$   the running max and min, 
\bea
{ M}_+(t) &:=& \max_{0<t'<t} X_{t'}\ ,\\
{ M}_-(t) &:=& \min_{0<t'<t} X_{t'}\ .
\eea
The total width or span  $s(t):={ M}_+(t)-{ M}_-(t)$ grows monotonically. We are interested in the time $T_1$ when it attains $1$. Define
\bea
\label{T1}
T_1 &:=& \min_t \big({ M}_+(t)-{ M}_-(t)\ge 1\big)\ ,\\
x_0&:=& X_{T_1} \equiv \frac{X_{T_1}}{{ M}_+(T_1)-{ M}_-(T_1)}\ .
\label{x0}
\eea
If the process starts for $x>x_0$, it will first be absorbed by the upper boundary (at $x=1$), whereas if it starts for $x<x_0$, it will first be absorbed by the lower one (at $x=0$). Denote the probability distribution of $x_0$ by  $ P_{x_0}(x)$. It  satisfies
\be
P_{x_0}(x) = -P_0'(x) = P_1'(x)\ .
\ee
An example of the measurement of $P_1'(x)$ for $H=0.33$ is shown on figure \ref{f:P'forH=0.33}.
The most remarkable feature of this plot is the very slow convergence in system size towards the asymptotic curve, which via scaling is very close to a parabola, see Eqs.~\eq{P'scaling}--\eq{40}. 
This slow convergence can also be seen on the distribution $P_{T_1}(t)$ of the times $T_1$ defined in \Eq{T1}. This is plotted on Fig.~\ref{f:all-P-T1}.


\begin{figure*}[thbp!]
\newcommand{\figsize}{8.5cm}\fig{\figsize}{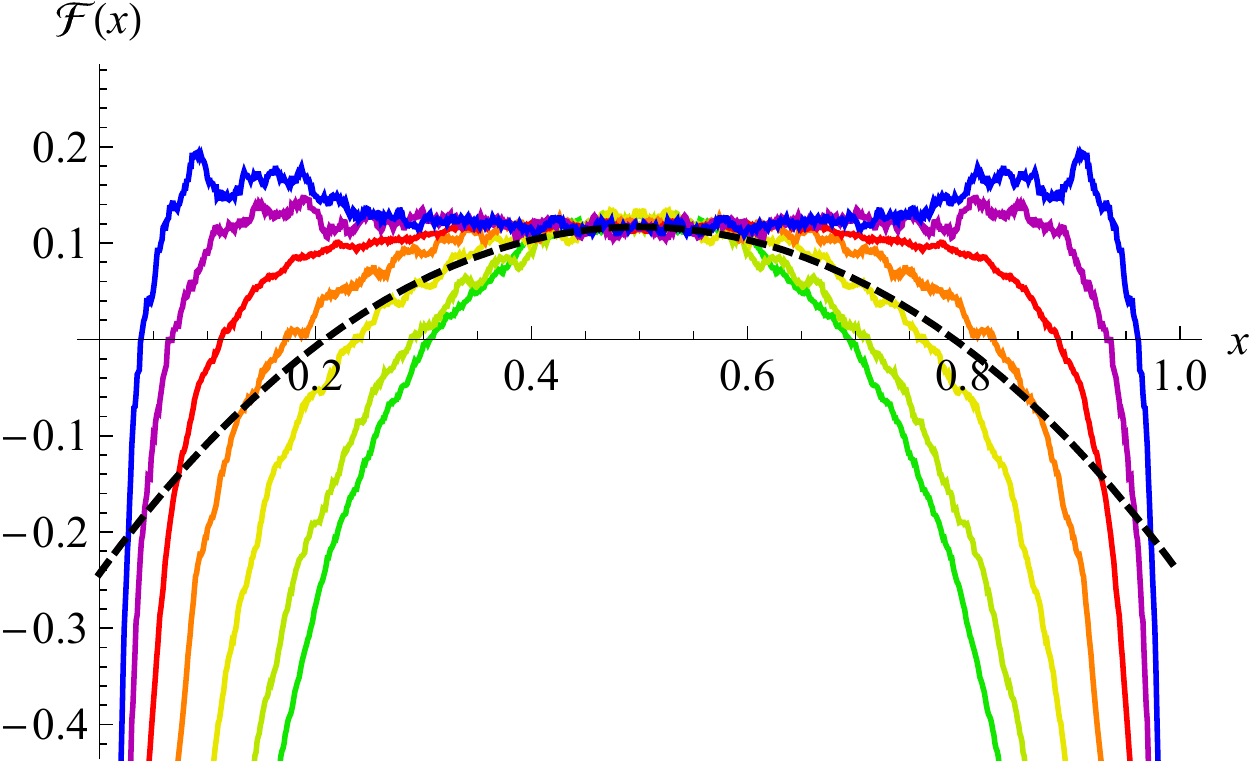}\hfill\fig{\figsize}{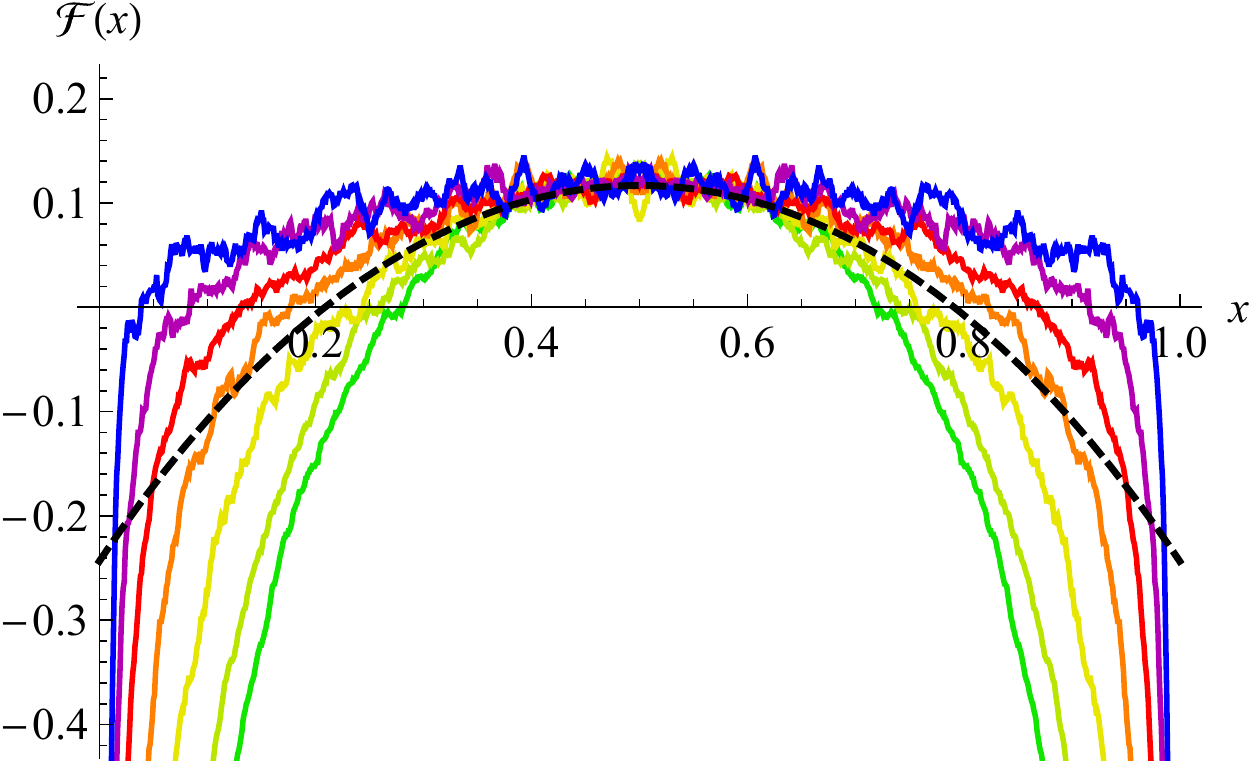}\\
\vspace*{-0.5cm}
\centerline{$H=0.33$ \hspace{8cm} $H=0.4$}\bigskip

\fig{\figsize}{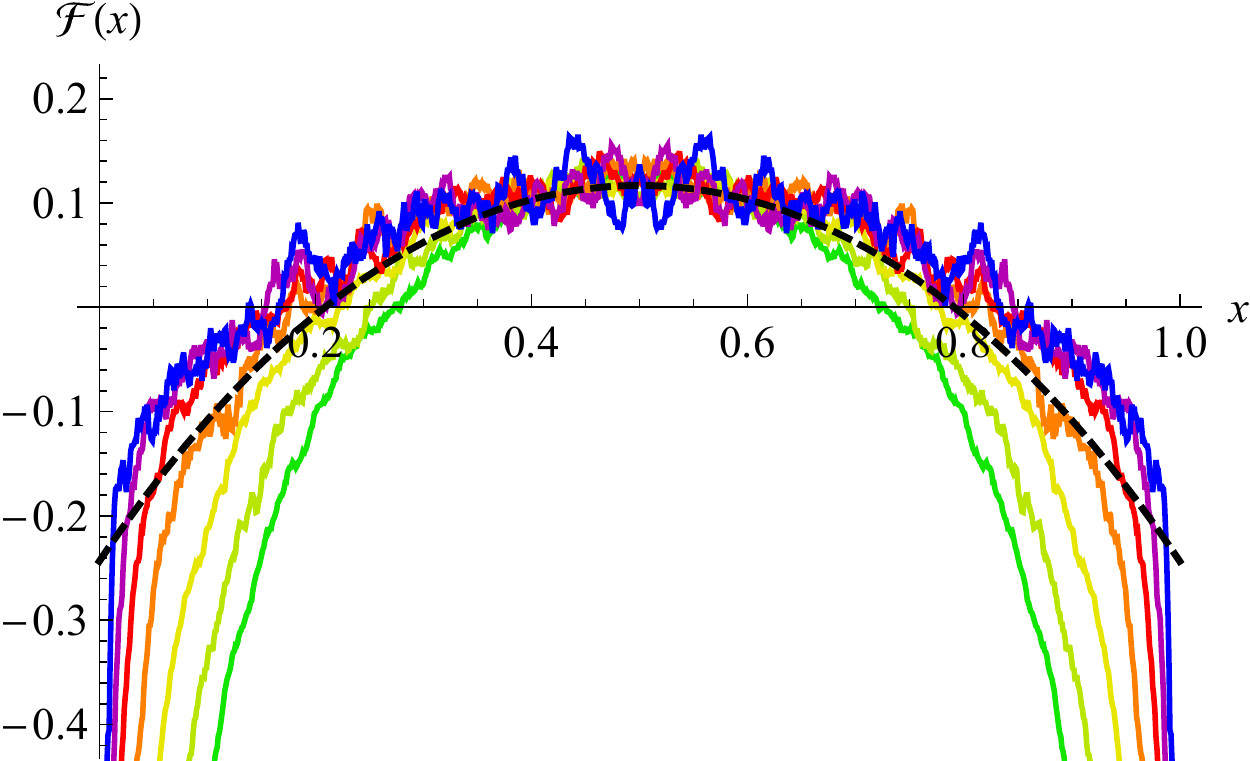}\hfill\fig{\figsize}{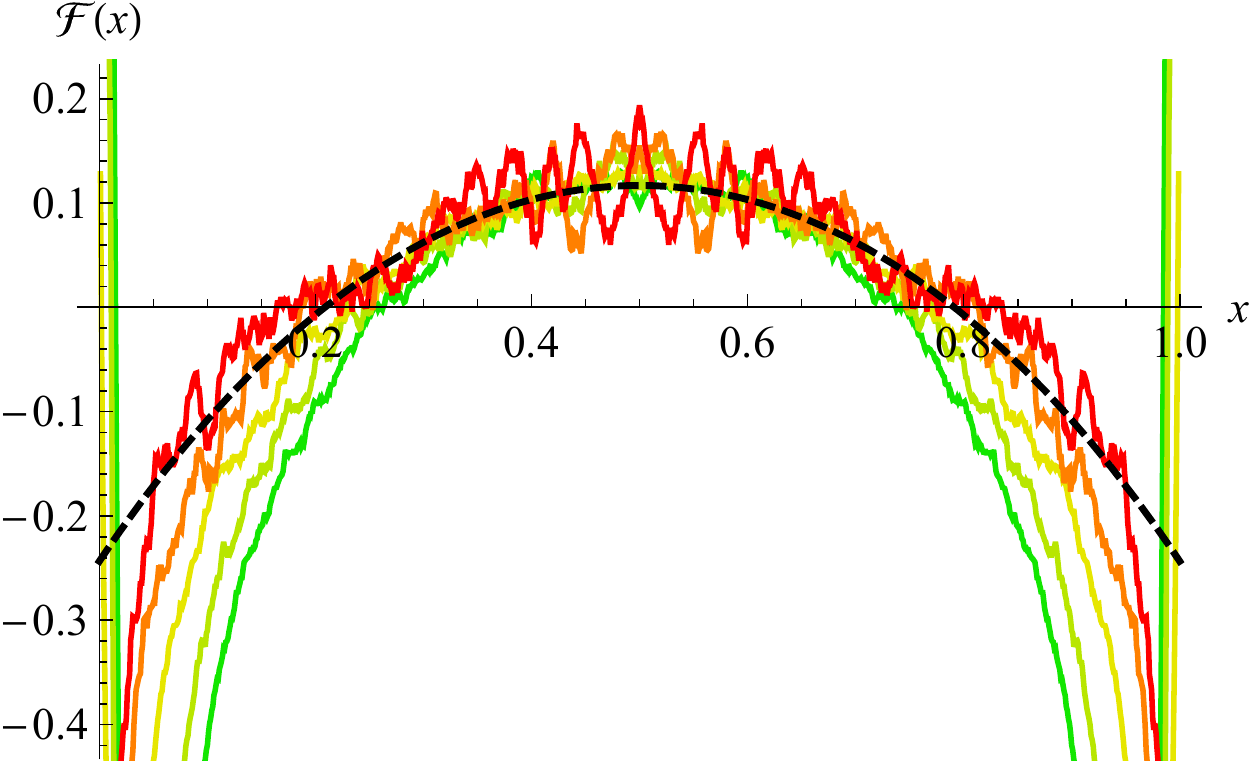}\\
\vspace*{-0.5cm}
\centerline{$H=0.45$ \hspace{8cm} $H=0.475$}\bigskip

\fig{\figsize}{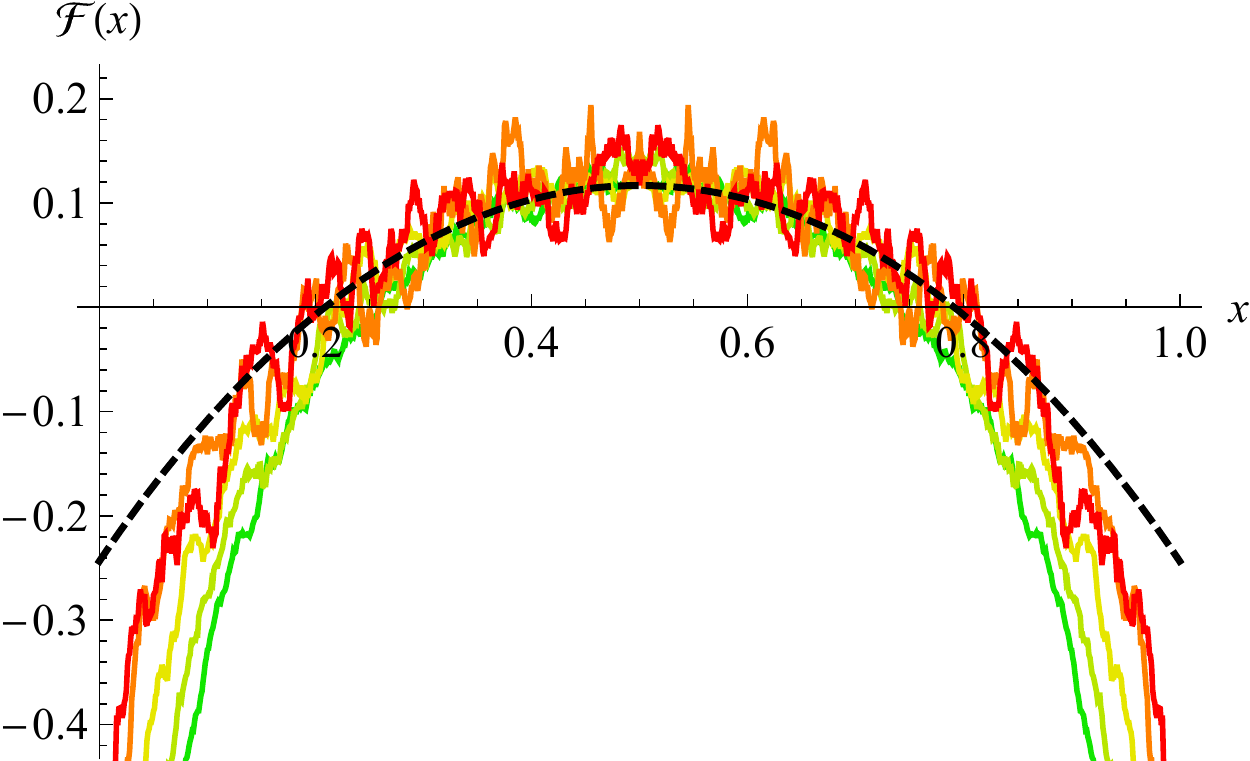}\hfill\fig{\figsize}{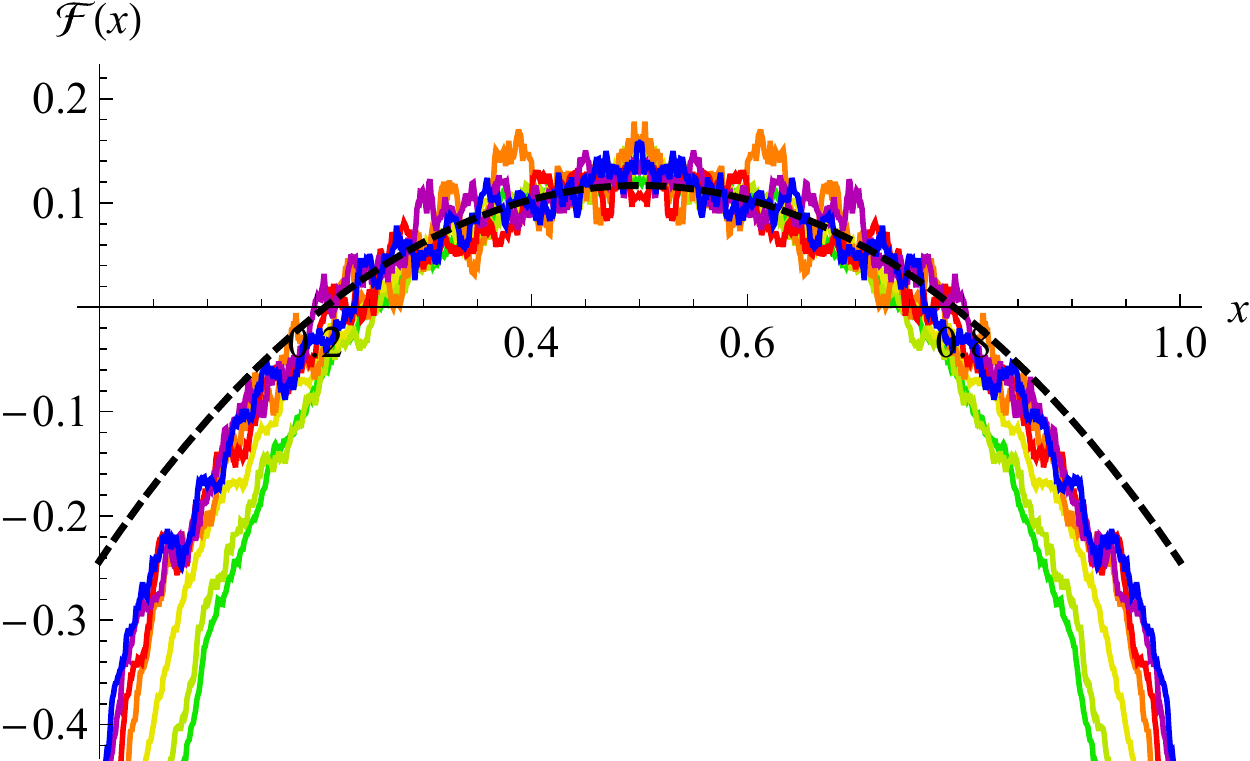}\\
\vspace*{-0.5cm}
\centerline{$H=0.525$ \hspace{8cm} $H=0.55$}\bigskip

\fig{\figsize}{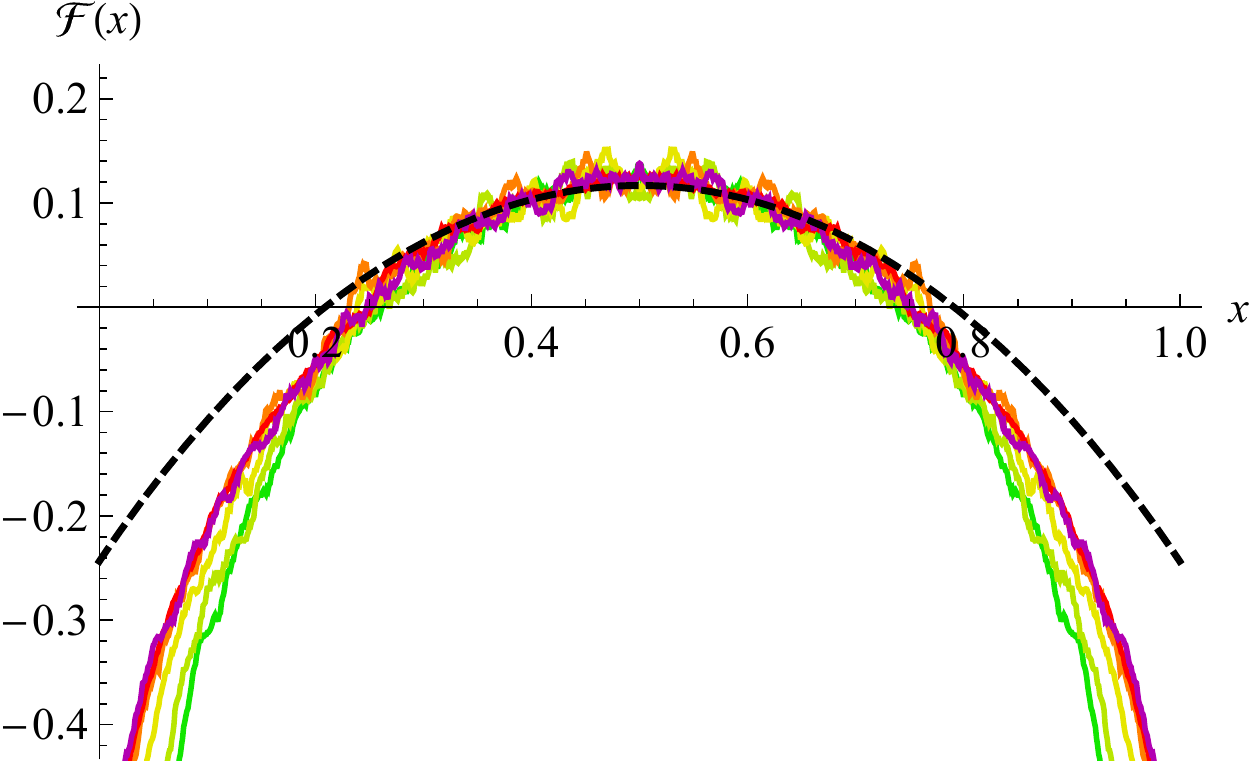}\hfill\fig{\figsize}{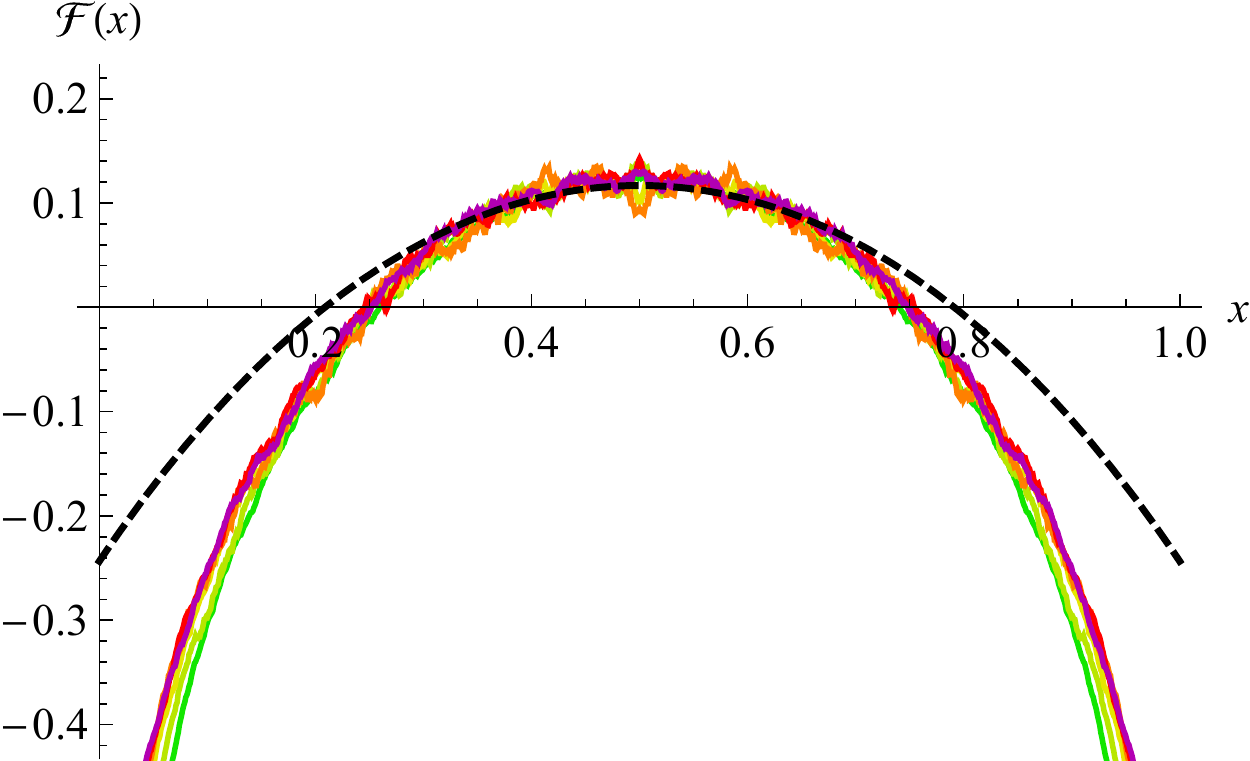}\\
\vspace*{-0.5cm}
\centerline{$H=0.6$ \hspace{8cm} $H=0.67$}\bigskip
\caption{The numerically estimated scaling function ${\cal F}(x)$ for (from top to bottom) $N=2^{13}$ (dark green), $2^{14}$ (green), $2^{16}$ (olive), $2^{18}$ (orange),  $2^{20}$ (red), $2^{22}$ (dark magenta), $2^{24}$ (blue). The black dashed line is the result of \Eq{Fofx}. For $\epsilon=\pm 0.025$, i.e.\ $H=0.475$ and $H=0.525$, due to the large statistics needed, we only simulated systems up to size $N=2^{20}$. For $H\ge 0.6$, convergence in system size is good, and we skipped the largest system $N=2^{24}$. Note that the scaling ansatzof \eq{P'scaling}, i.e.\ $\ca P_{\rm 1, scaling}'(x) \sim   [x(1-x) ]^{\frac{1}{H}-2}$ is equivalent to $\ca F(x)\equiv 0$.}
\label{f:all-Fs}
\end{figure*}

The slow convergence of $P_1'(x)$ can better be seen via the function $\ca F(x)$:
An estimate of the latter can   be extracted from the simulations, by inverting \Eq{40}, 
\be
{\cal F}^{\epsilon}_{\rm num}(x) := \frac1\epsilon \ln\left( P'(x)  [x(1-x)]^{2- \frac 1H}\right) +\mbox{const}\ .
\ee
The constant is  chosen s.t.\ ${\cal F}^\epsilon_{\rm num}(1/2) ={\cal F}(1/2) $.
According to our theory, 
\be
{\cal F}^\epsilon_{\rm num}(x) = {\cal F}(x)+ \ca O (\epsilon)\ .
\ee
Examples are given on Fig.~\ref{f:all-Fs} (we   suppressed all indices on $\ca F$). The case $H=0.33$ (upper left) corresponds to the plot on Fig.~\ref{f:P'forH=0.33}, with the same colors. One clearly sees that convergence in system size is   slow for all $H$, but especially for the smaller ones. It becomes better for larger values of $H$. 

\begin{figure*}
\fig{8.5cm}{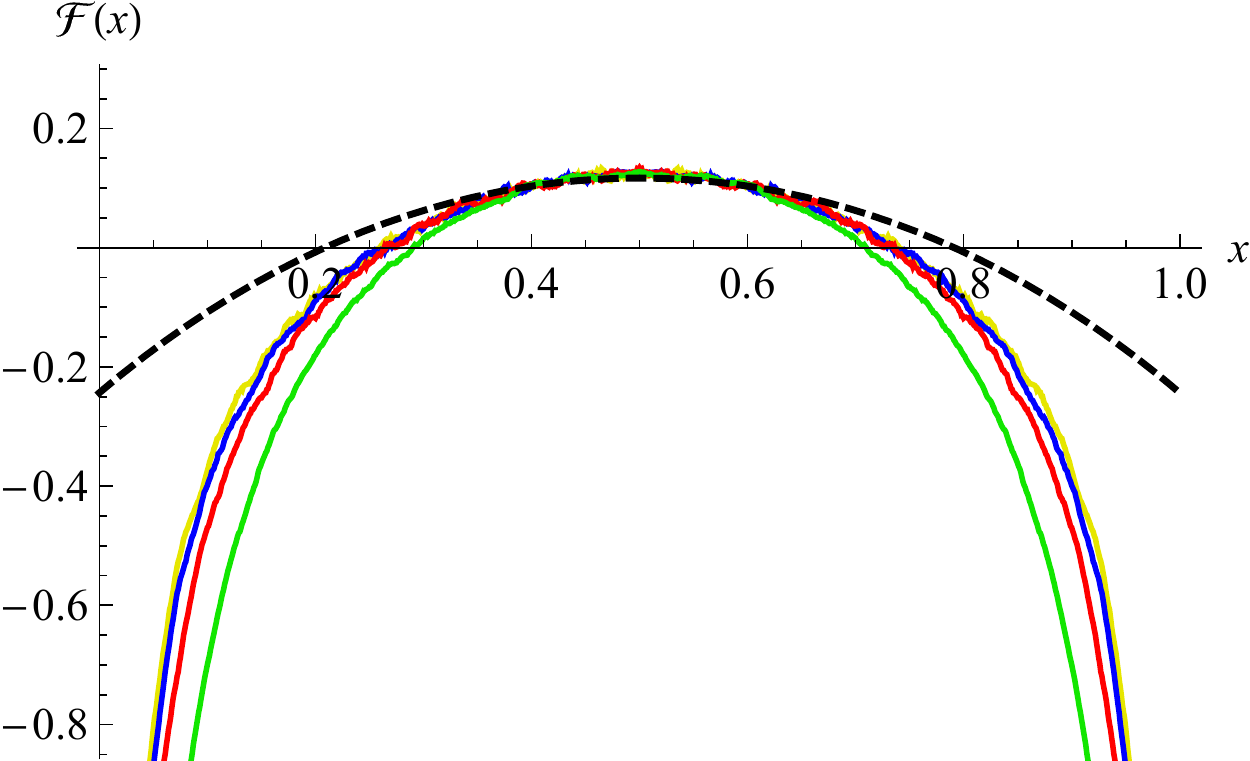}~~~~~~~\fig{8.5cm}{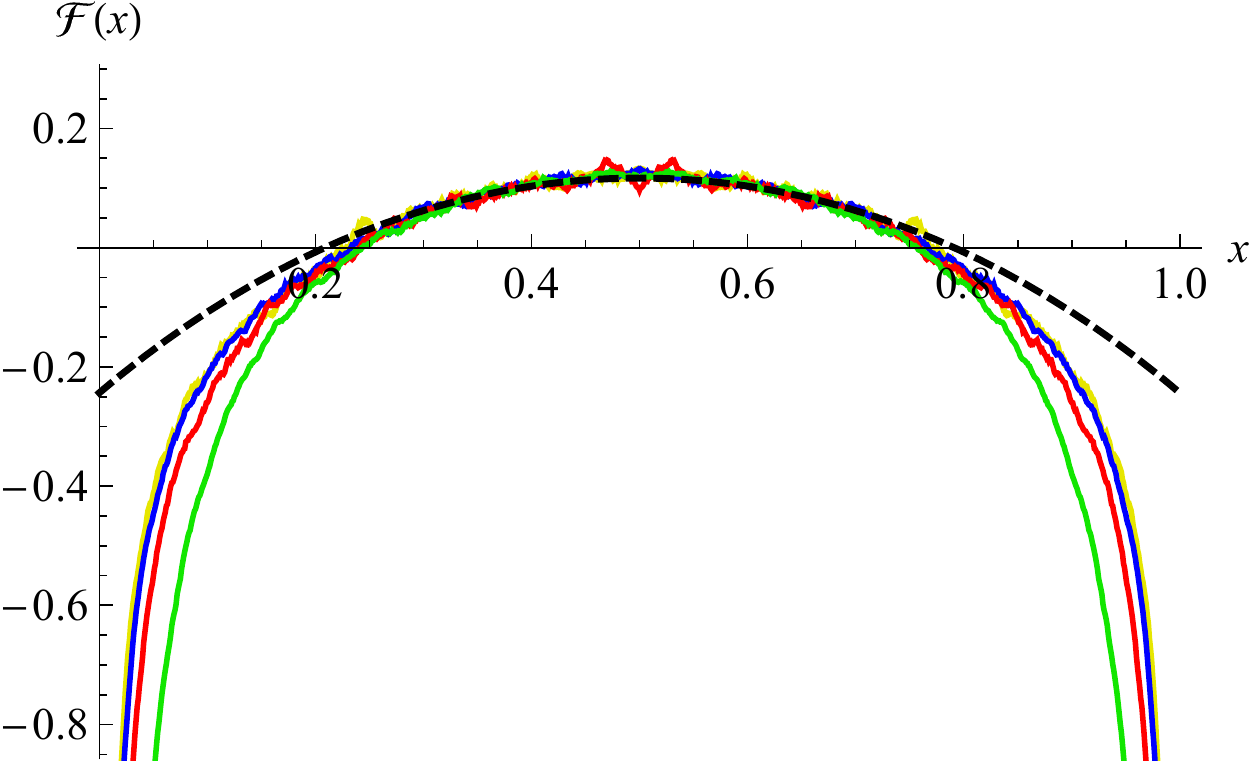}\\
\vspace*{-0.5cm}
\centerline{$N=2^{13}$ \hspace{7.9cm} $N=2^{16}$}\bigskip
\fig{8.5cm}{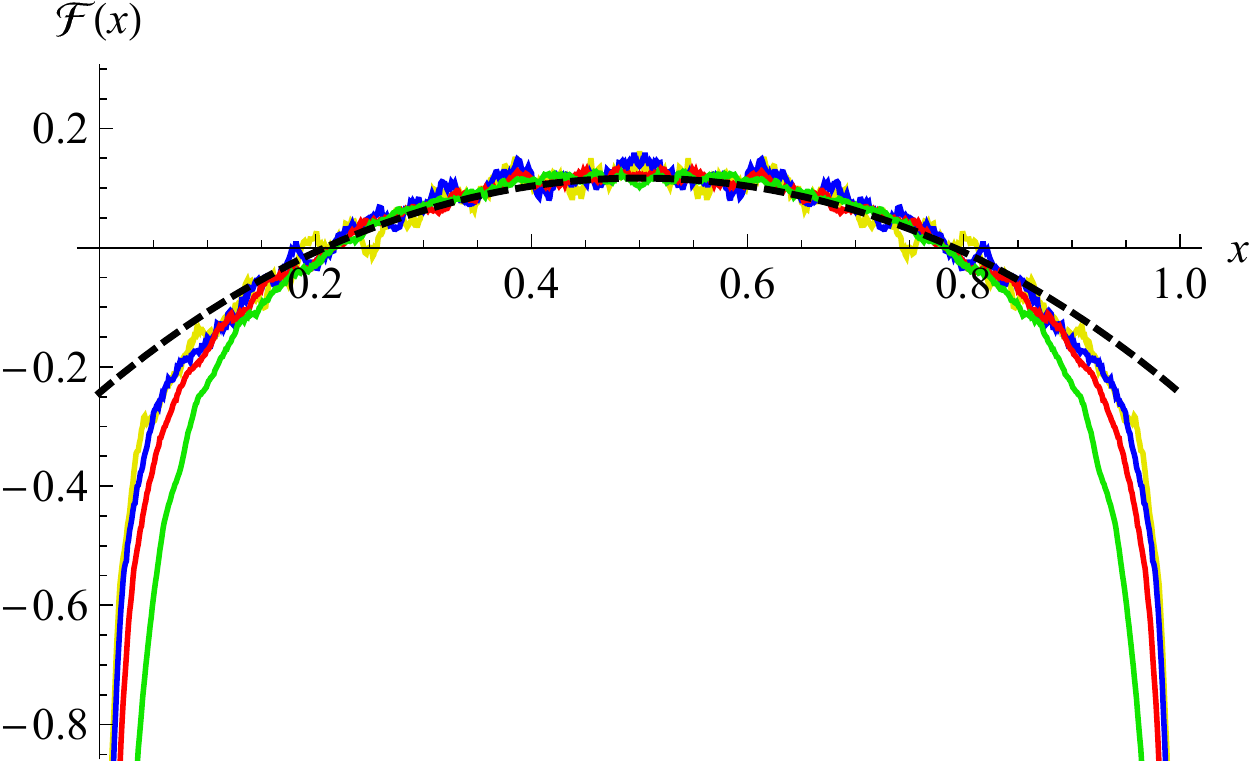}~~~~~~~\fig{8.5cm}{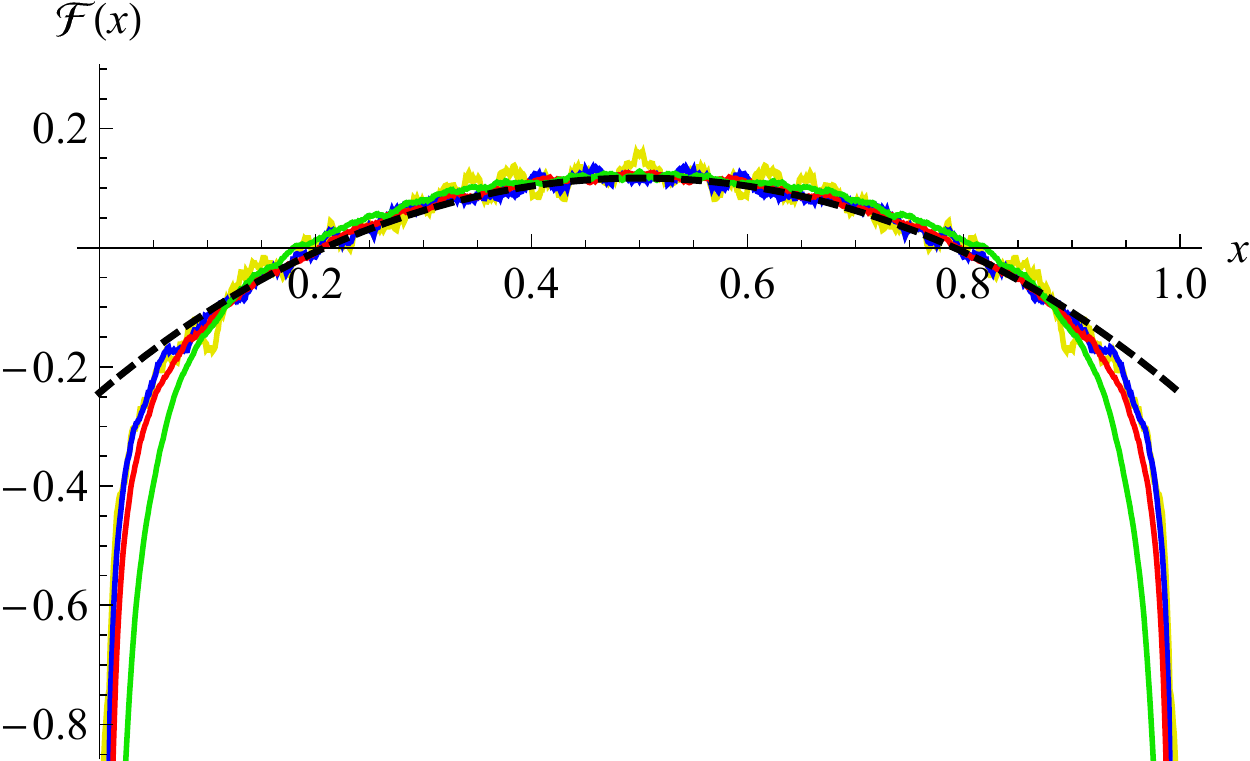}\\
\vspace*{-0.5cm}
\centerline{$N=2^{18}$ \hspace{7.9cm} $N=2^{20}$}\bigskip
\fig{8.5cm}{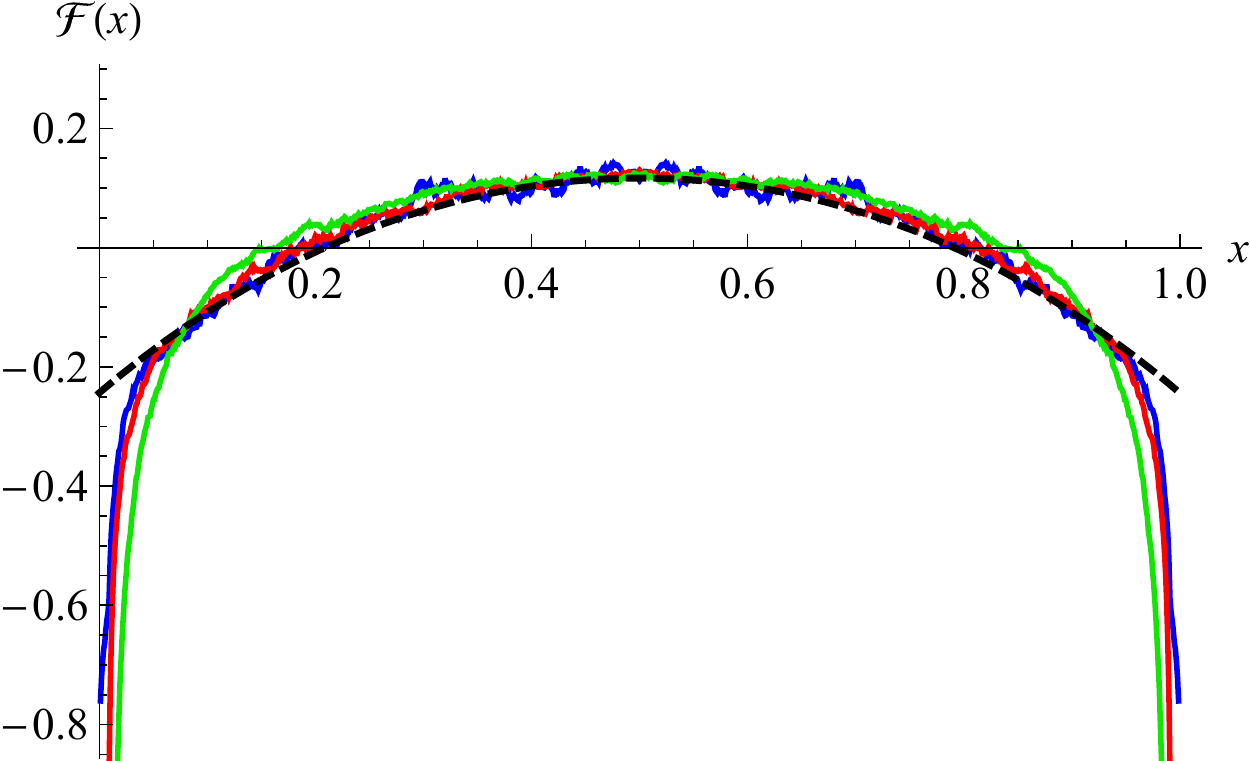}~~~~~~~\fig{8.5cm}{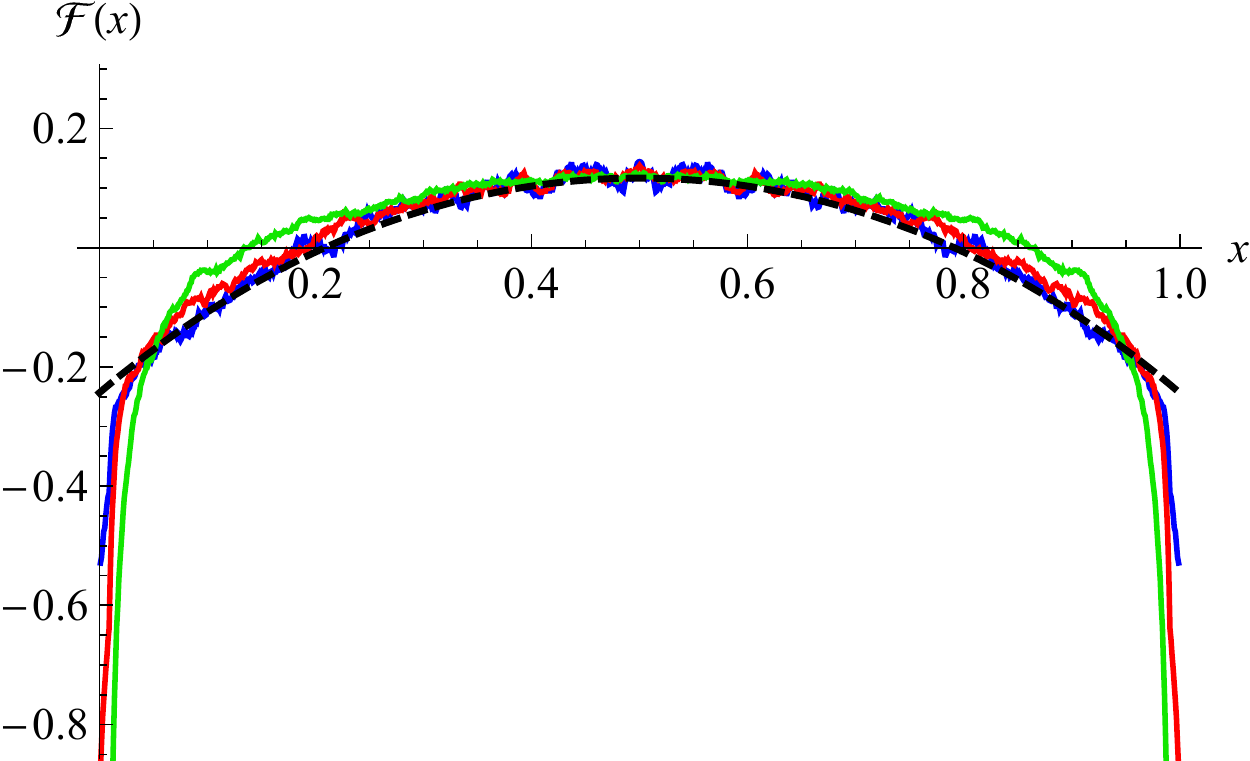}\\
\vspace*{-0.5cm}
\centerline{$N=2^{22}$ \hspace{7.9cm} $N=2^{24}$}\bigskip
\fig{8.5cm}{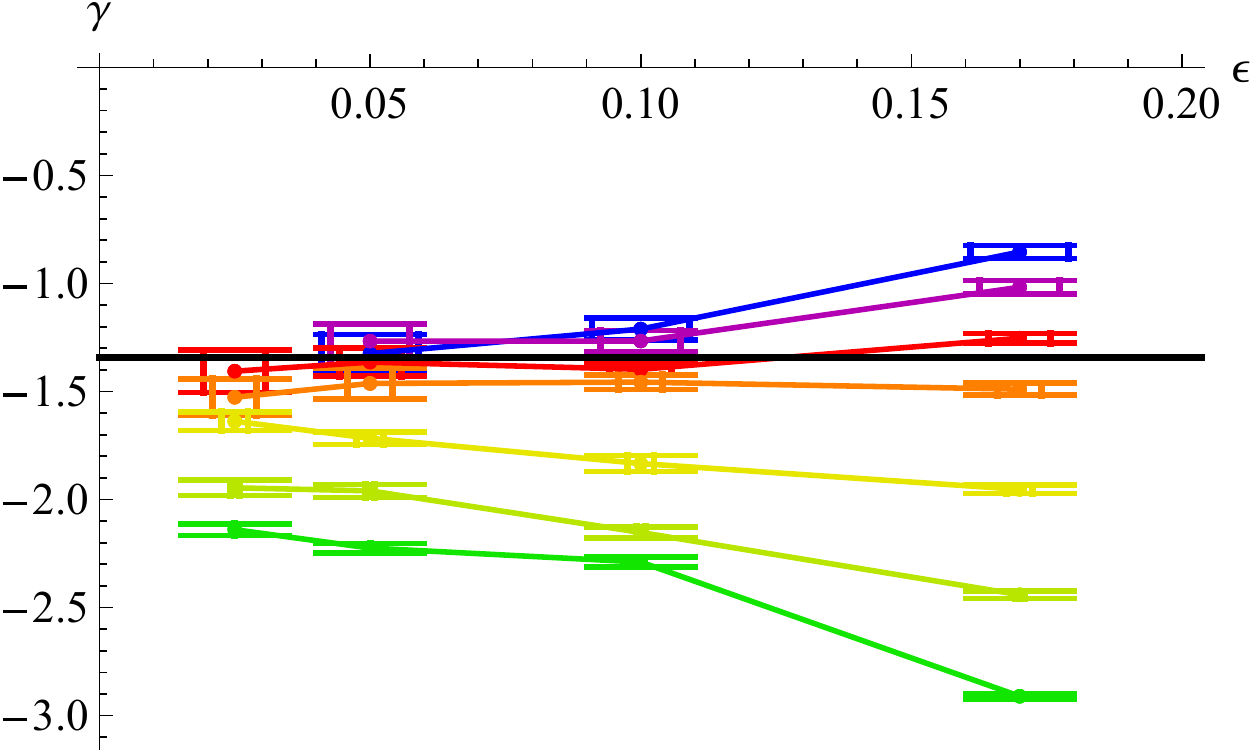}~~~~~~~\fig{8.5cm}{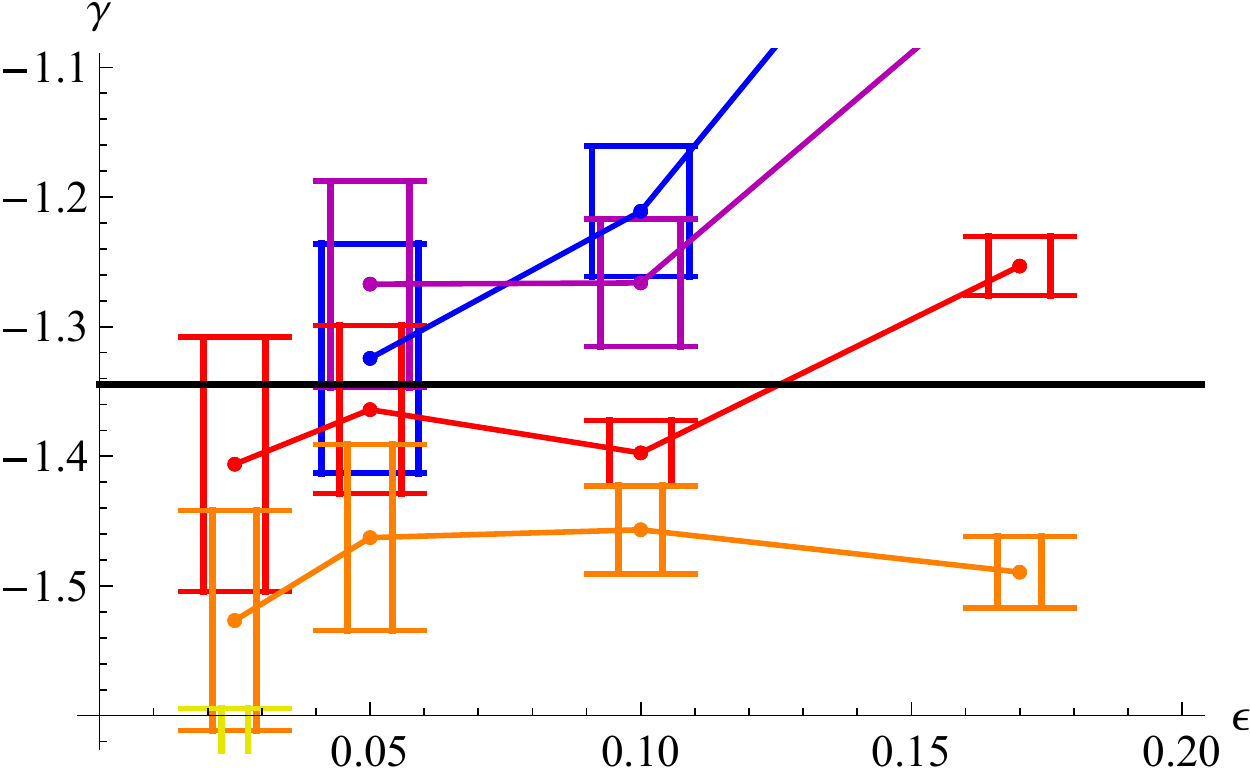}\\
\caption{Dependence of the scaling function ${\cal F}(x)$ on $N$, by taking the mean between $H=1/2\pm \epsilon$, for  $H=0.33/0.67$ (green), $H=0.4/0.6$ (red), $H=0.45/0.55$ (blue), $H=0.475/0.525$ (olive/yellow).
For $N=2^{24}$, we have replaced the estimate for $H\ge0.6$ by those of $N=2^{22}$, justified by the much better convergence in system size for these values of $H$, see figures  \ref{f:all-Fs}.
The last two plots show the extracted curvature $\gamma$ with system sizes increasing from bottom to top: $N=2^{13}$ (green), $N=2^{14}$ (bright green), $N=2^{16}$ (olive/yellow), $N=2^{18}$ (orange), $N=2^{20}$ (red), $N=2^{22}$ (violet), $N=2^{24}$ (blue, single dot with big error bars). For the error estimate see Fig.~\ref{f:gamma-estimate}. Note that the scaling ansatzof \eq{P'scaling}, i.e.\ $\ca P_{\rm 1, scaling}'(x) \sim   [x(1-x) ]^{\frac{1}{H}-2}$ is equivalent to $\ca F(x)\equiv 0$.}
\label{f:Fmean}
\end{figure*}

\begin{figure*}[thbp!]
\newcommand{\figsize}{8.5cm}
\centerline{$H=0.33$ \hspace{8cm} $H=0.4$}\vspace*{-0.5cm}
\fig{\figsize}{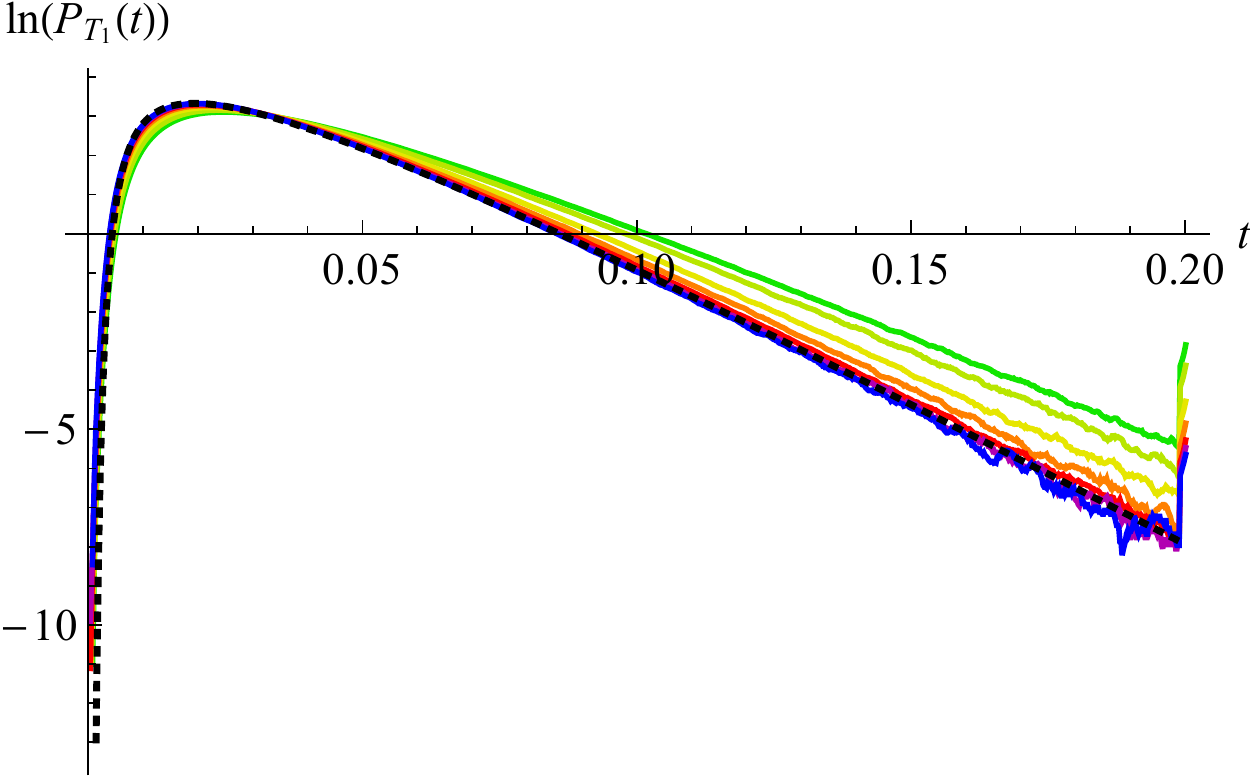}~~~~~~\fig{\figsize}{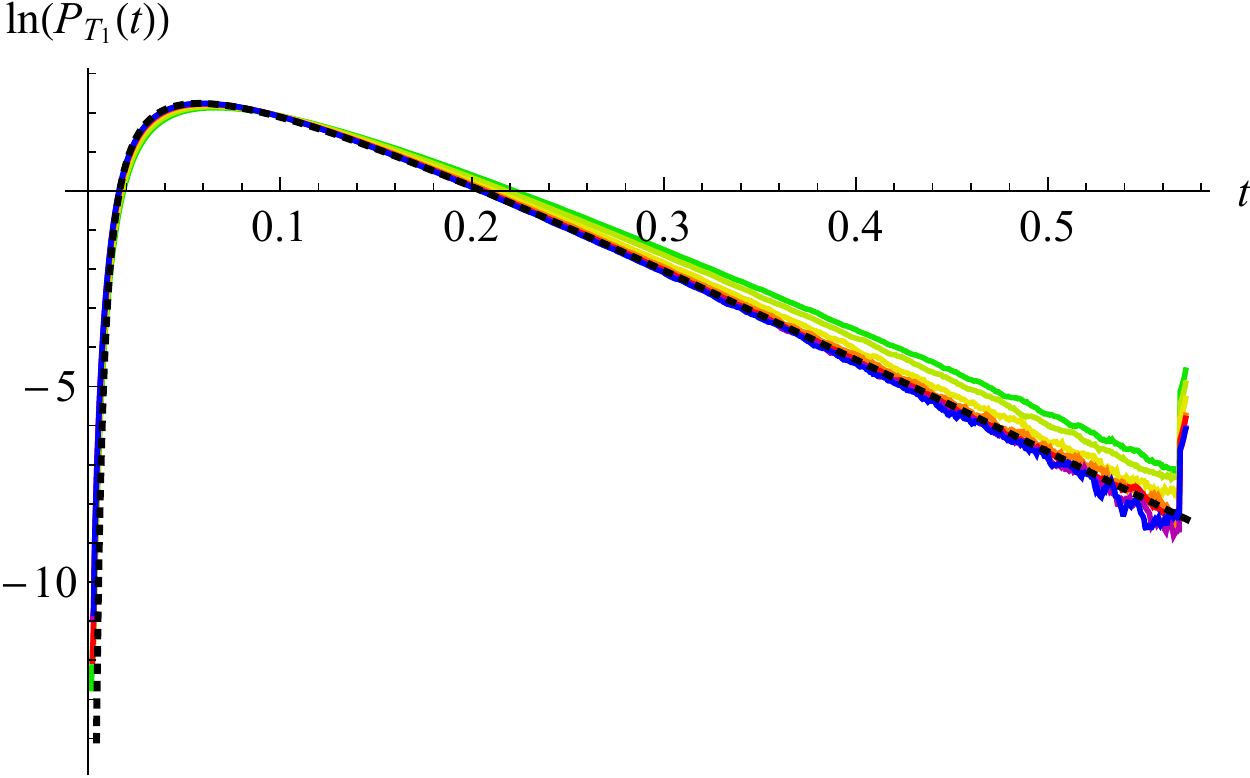}\\
\bigskip

\centerline{$H=0.45$ \hspace{8cm} $H=0.475$}\vspace*{-0.5cm}
\fig{\figsize}{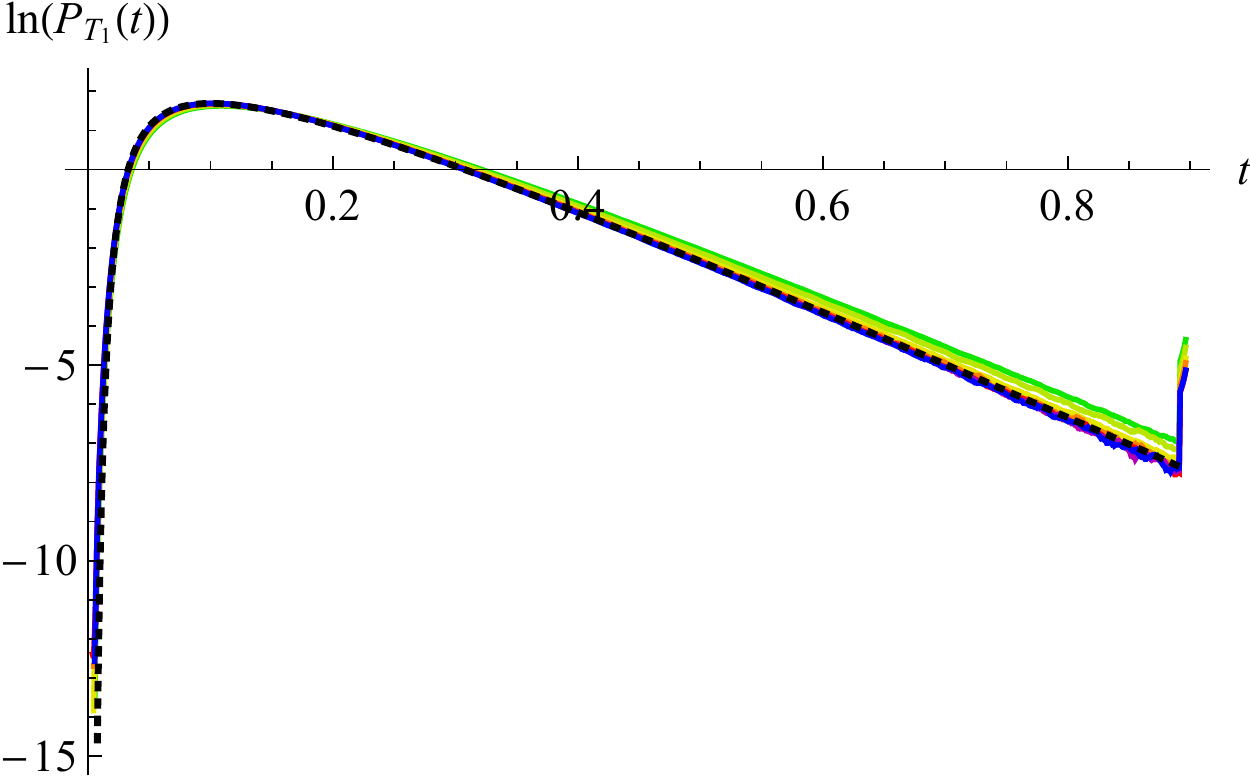}~~~~~~\fig{\figsize}{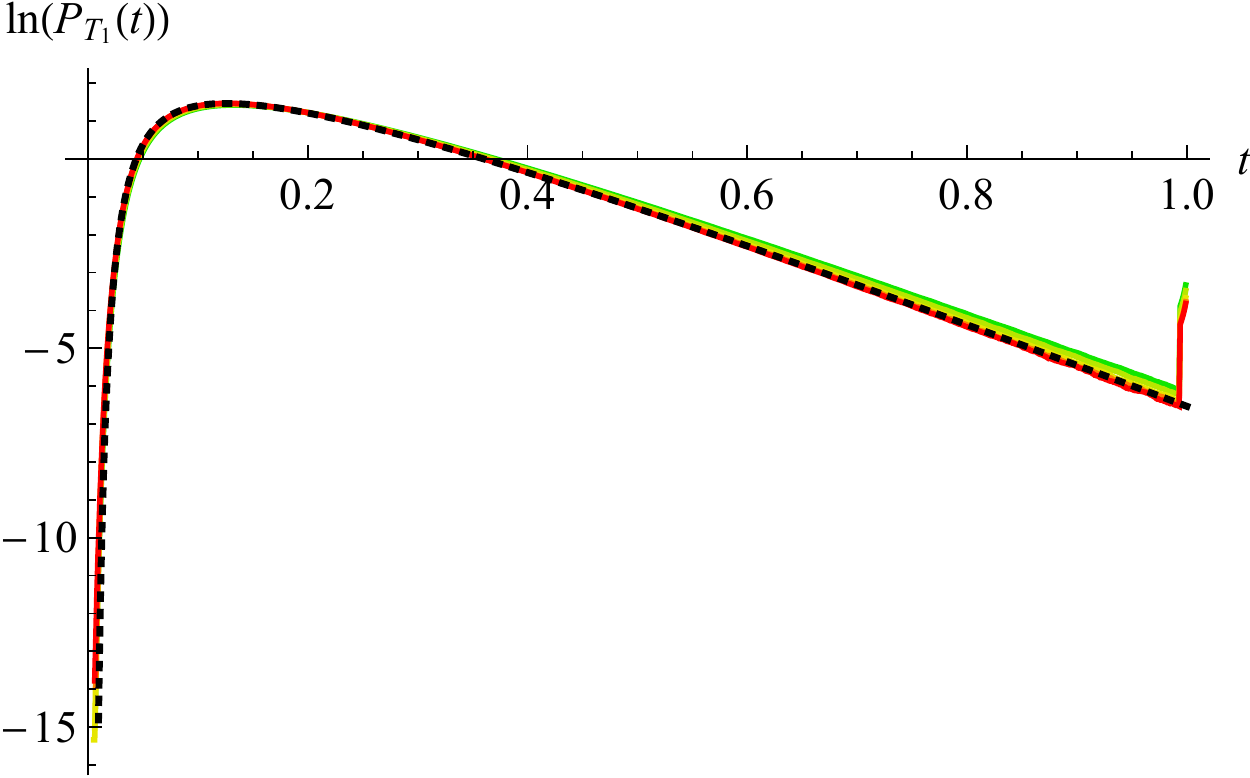}\\
\bigskip

\centerline{$H=0.525$ \hspace{8cm} $H=0.55$}\vspace*{-0.5cm}
\fig{\figsize}{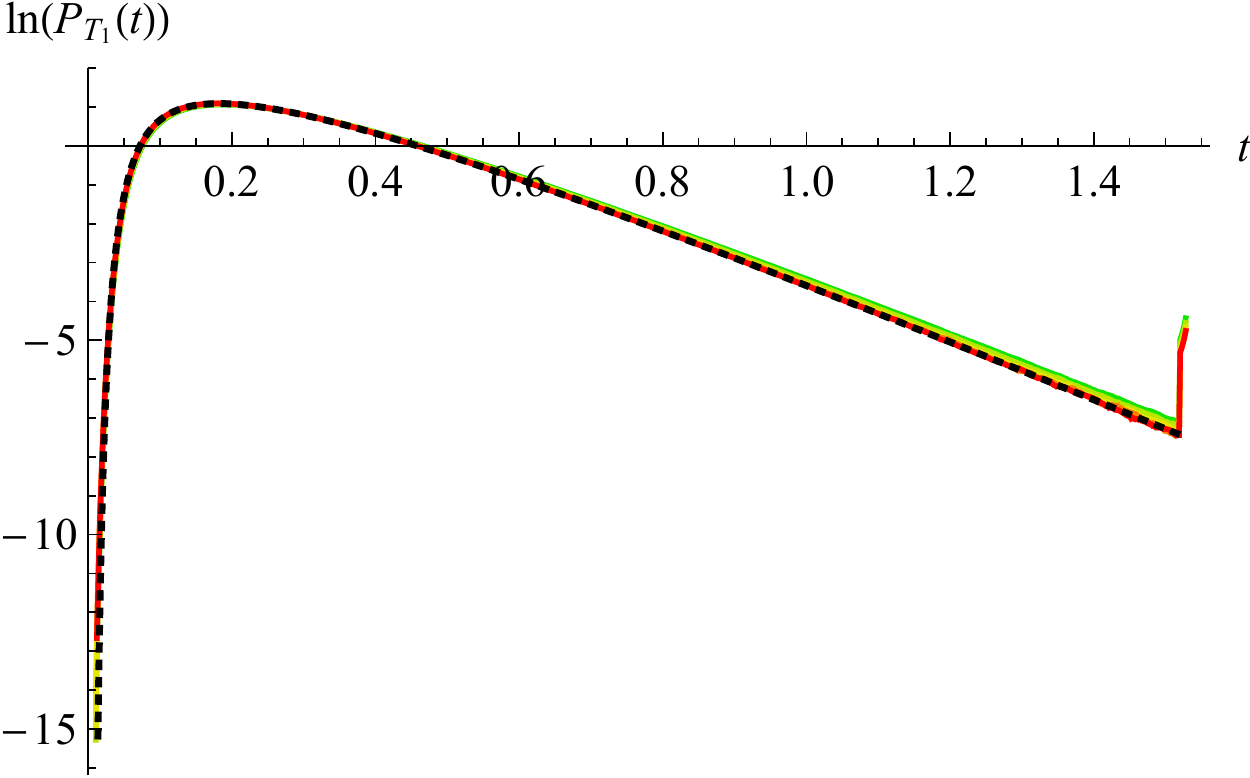}~~~~~~\fig{\figsize}{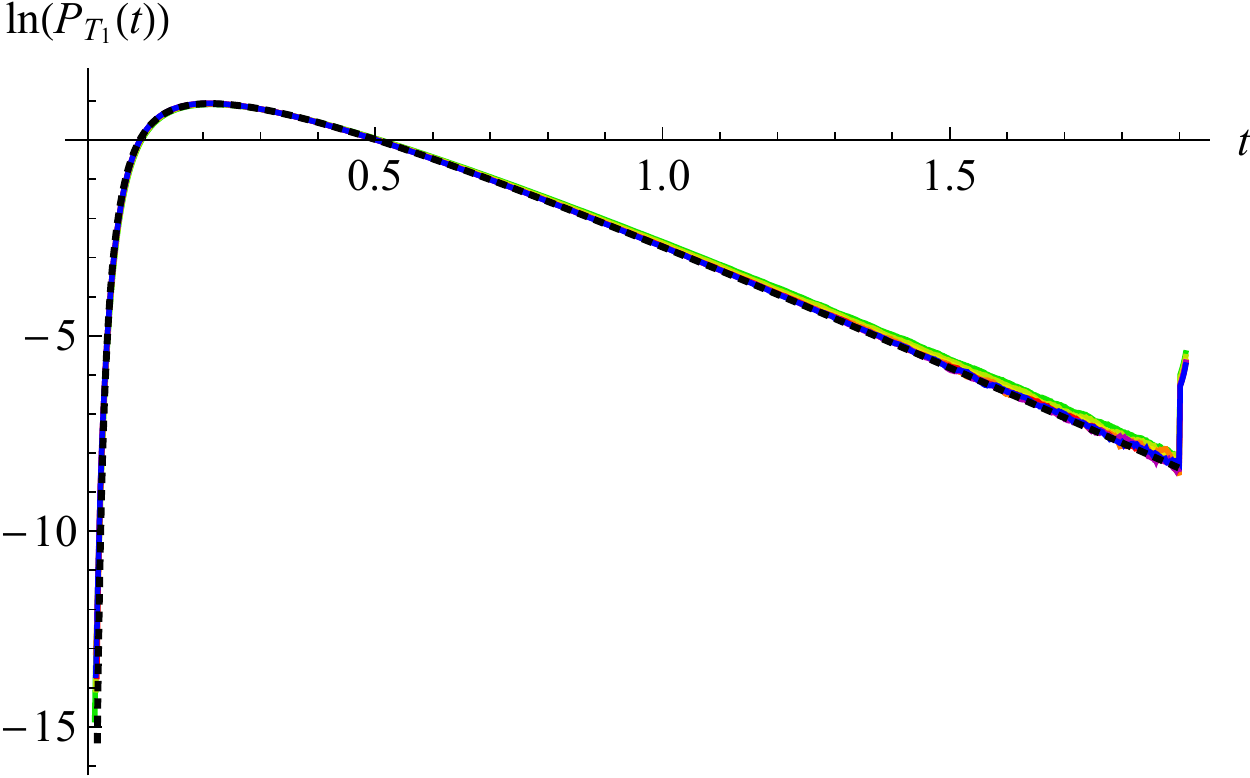}\\
\bigskip

\centerline{$H=0.67$ \hspace{8cm} $H=0.75$}\vspace*{-0.5cm}
\fig{\figsize}{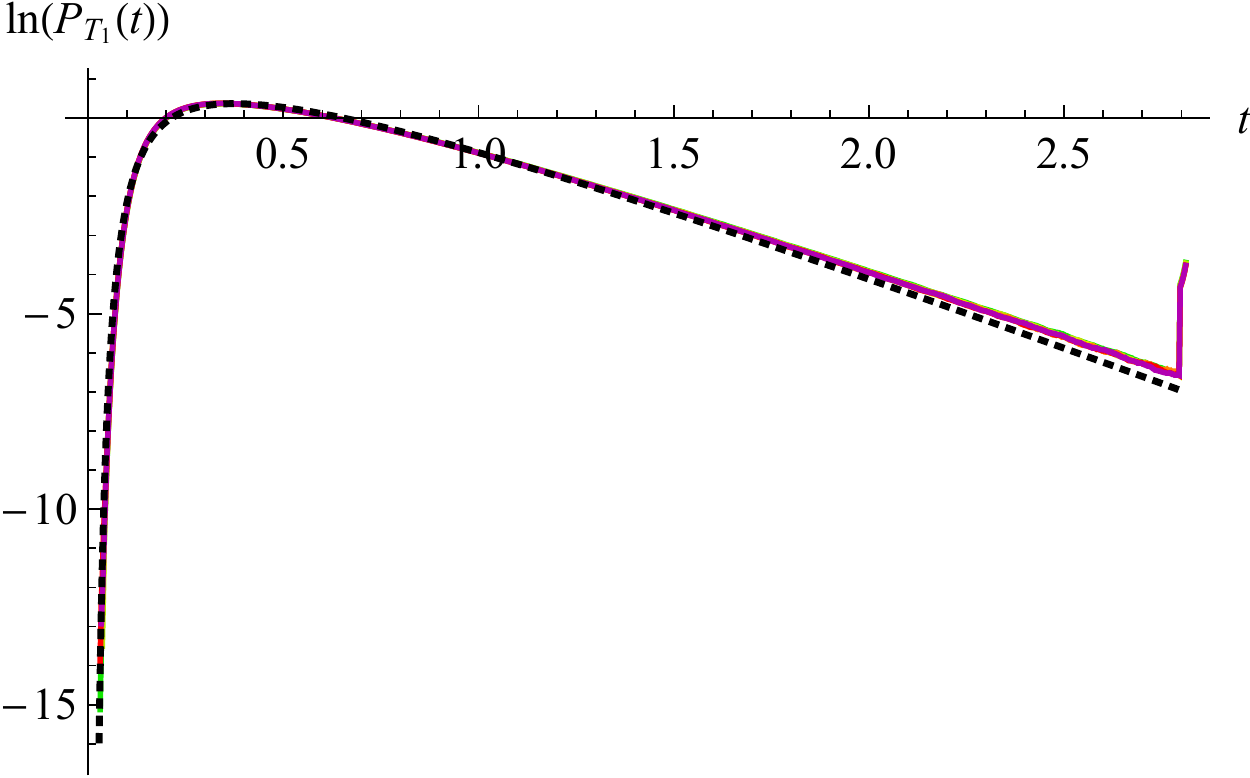}~~~~~~\fig{\figsize}{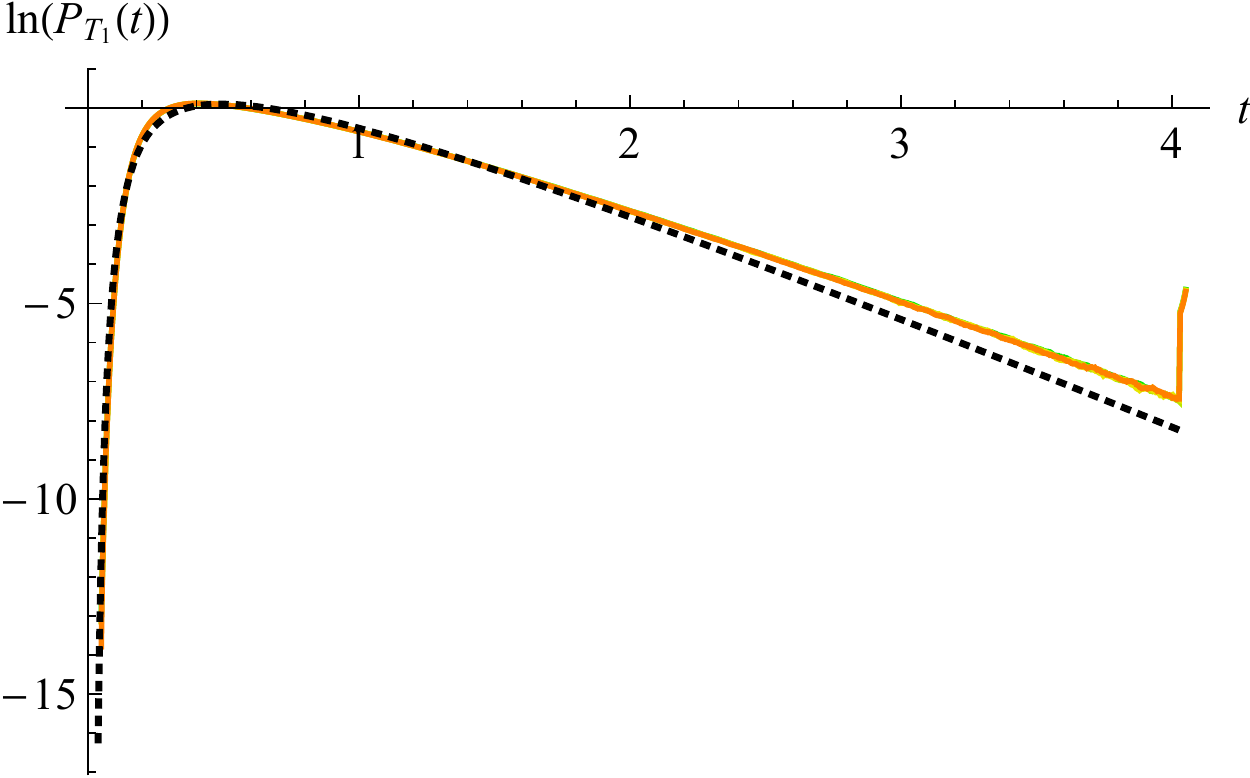}\\
\bigskip
\caption{The same as figure \ref{f:all-P-T1}, on a logarithmic scale. The little bump at large $T$ corresponds to all realizations which did not exit up to that time, i.e.\ it is the integrated tail.}
\label{f:all-P-T1-log}
\end{figure*}

We stopped our simulations after a total estimated 28 CPU years. It seems clear that measuring more than the curvature is illusory. 
We therefore defined 
\be\label{69}
\gamma:=\frac12 \partial_x^2 \ca F_{\rm num}^\epsilon(x)\Big|_{x=1/2}\ ,
\ee
and in practice measured it by fitting a polynomial of degree two in an $x$-range   from $x=0.25$ to $x=0.75$ for the smaller systems, to $x=0.15$ to $x=0.85$ for the largest systems.

Analytically, we obtained in \Eq{43} with the Catalan constant $C$, 
\be
\gamma = 16 (C-1) = 16 \sum_{n= 1}^{\infty} \frac{(-1)^n} {(2n+1)^2} \approx -1.34455\ .
\label{gamma-theory}
\ee 
Our direct numerical estimate for $\gamma$ is shown on Fig.~\ref{f:gamma-estimate}. One sees that extrapolation to $H=\frac 12$ is good only for  large systems. An alternative and more precise way to extract $\gamma$ is to define 
\be
\overline {\cal F}^\epsilon_{\rm num}(x):=\frac12 \left[ {\cal F}^\epsilon_{\rm num}(x)+{\cal F}^{-\epsilon}_{\rm num}(x)\right] \ .
\ee
This combination cancels the first subleading contribution in $\epsilon$; the result is plotted on Fig.~\ref{f:Fmean}. Again one sees that $\ca F(x)$ is well approximated for large system sizes. 
Our best estimate is 
\be
\gamma_{\rm num} = - 1.34 \pm 0.02 \ .
\ee

\subsection{Expectation of exit times and their squares}
\label{s:Expectation of exit times and their squares}
\begin{figure}[t]
\!\!\!\!\raisebox{5.1cm}[0mm][0mm]{(a)}\!\!\!\!\!\!\fig{4.2cm}{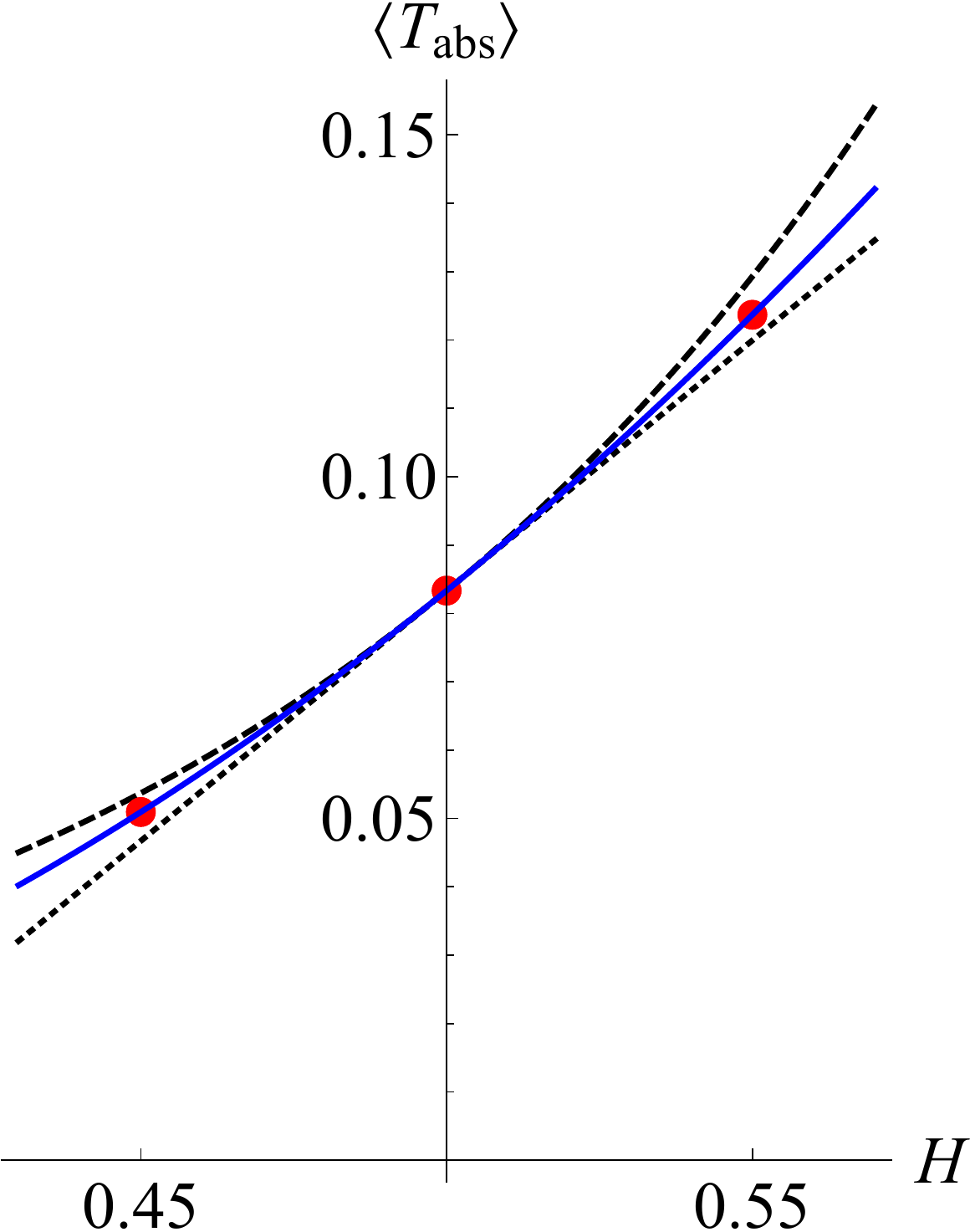}~~~\raisebox{5.1cm}[0mm][0mm]{(b)}\!\!\!\!\!\!\fig{4.2cm}{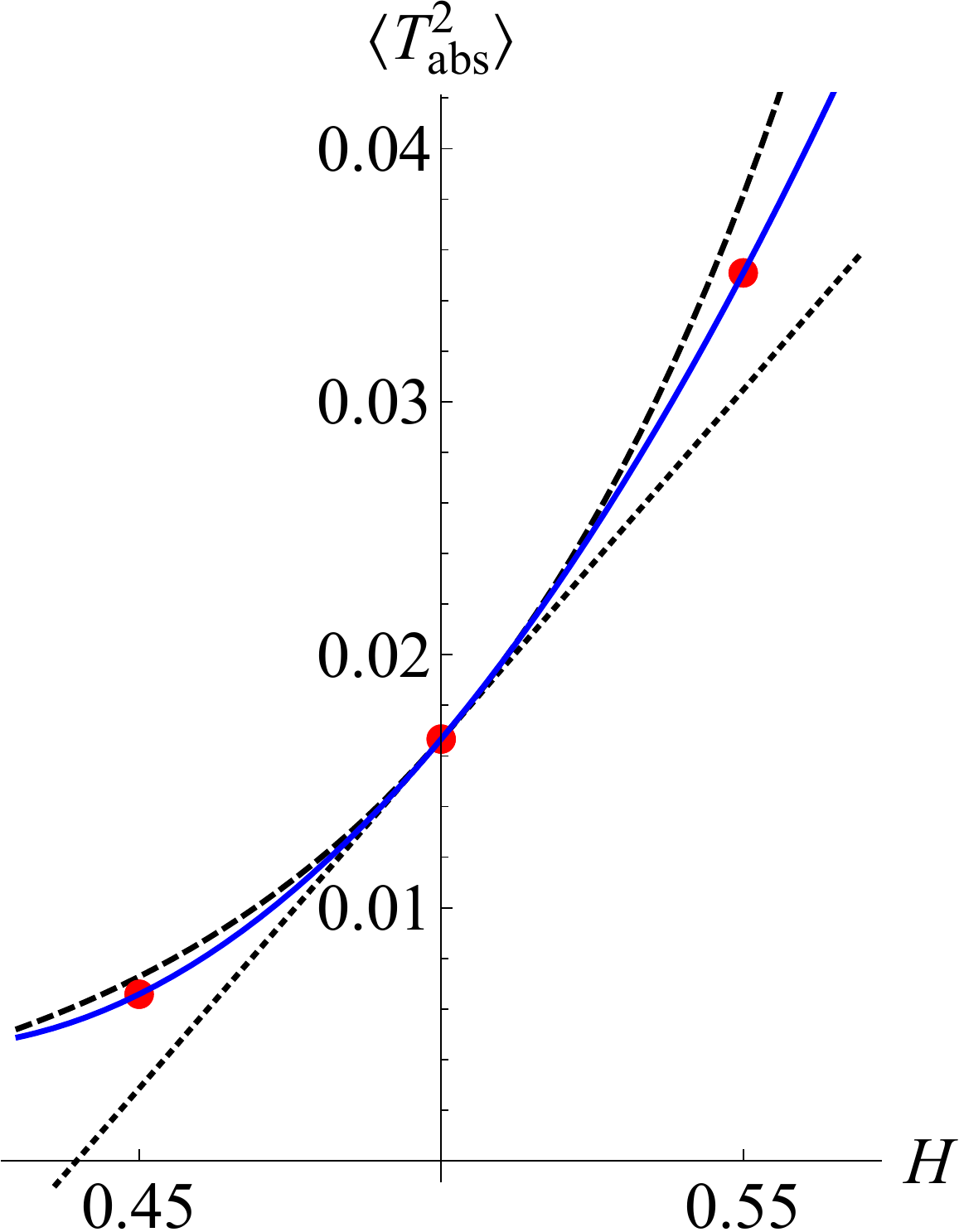}
\caption{(a) The mean, i.e.\ spatially averaged absorption time $\left< T_{\rm abs}\right>=\int_0^1 \rmd x\,\left< T_{\rm abs}(x)\right>$. (b) {\it ibid.} for the second moment. The black dotted line is the direct expansion in $\epsilon$, which  underestimates the true result. The dashed line is the exponentiation of this correction; it overestimates the result. The fit to a quadratic polynomial is given in Eqs.~\eq{Texitfit} and \eq{Texit2fit}.}
\label{f:epxT+expT2}
\end{figure}
Let us now turn to the exit times as a function of $x$. Measurements of the scaling functions $\ca F_T(x)$ and $\ca F_{T^2}(x)$ given in Eqs.~\eq{FTdef} and 
\eq{FT2}--\eq{54} are presented on Fig.~\ref{f:FT+FT2}, for $H=0.45$ and $H=0.55$. Their mean (in red) is a good approximation of the analytic curves (in black dashed). 
Simulations were performed for the largest system size at our disposal $N=2^{24}$, and $H=0.45$ as well as $H=0.55$. Having generated an fBm, we put its starting position at $x$, and then searched for the first instance when it was absorbed at either the upper or lower boundary. This procedure turned out to be rather time-consuming, and we only evaluated this function about $2\times 10^6$ times. 

To estimate the spatially averaged first two moments of the exit times, we fitted their numerically obtained values, supplemented with the analytically known values for $H=1/2$, with a polynomial of degree 2 in $H$. The result, shown on Fig.~\ref{f:epxT+expT2} is 
\begin{align}\label{Texitfit}
\!\!\!\int_0^1\rmd x \left< T_{\rm exit}(x) \right> =& \frac1{12}\left[ \textstyle 1+8.73  \epsilon +19.1
   \epsilon^2+\ca O  (\epsilon^3 )\right]  ,
   \\
   \!\!\!\int_0^1\rmd x \left< T_{\rm exit}^2(x) \right> =& \frac1{60}\left[ \textstyle 1+17.1  \epsilon +100
   \epsilon^2+\ca O  (\epsilon^3 )\right] .
   \label{Texit2fit}
\end{align}
Comparison to Eqs.~\eq{Texit} and \eq{Texit2} yields excellent agreement for the first moment of the exit time (coefficient $8.758$ to be compared to $8.73$), and still very good agreement for the second moment (coefficient $16.563$ as compared to $17.1$). The latter is difficult to estimate, as higher-order corrections are seemingly large.

Let us finally mention that for $H=1$ the probability that starting at $x$ the exit time is $t$, is given by
\be\label{PH=1}
P^{H=1}_{\rm exit}(t|x) = \frac1{\sqrt{2\pi }t^2}\left[ x \rme^{-\frac{x^2}{2t^2}} + (1-x)   \rme^{-\frac{(1-x)^2}{2t^2}}\right] \ .
\ee
Averaging over $x$ yields
\be\label{PH=1mean}
\int_0^1\rmd x \,P^{H=1}_{\rm exit}(t|x)   =\sqrt{\frac{2}{\pi }} \left[1-e^{-\frac{1}{2 t^2}}\right]\ .
\ee
All these distributions at $H=1$ are patologic. While they are normalizable, they have large tails which render already the first moment undefined. They are thus not a useful limit to test our formulas.

\subsection{The time the span reaches  1}
\begin{figure*}[b!t]
\!\!\!\!\!\!\!\!\raisebox{0cm}[0mm][0mm]{(a)}\!\!\!
\fig{8.5cm}{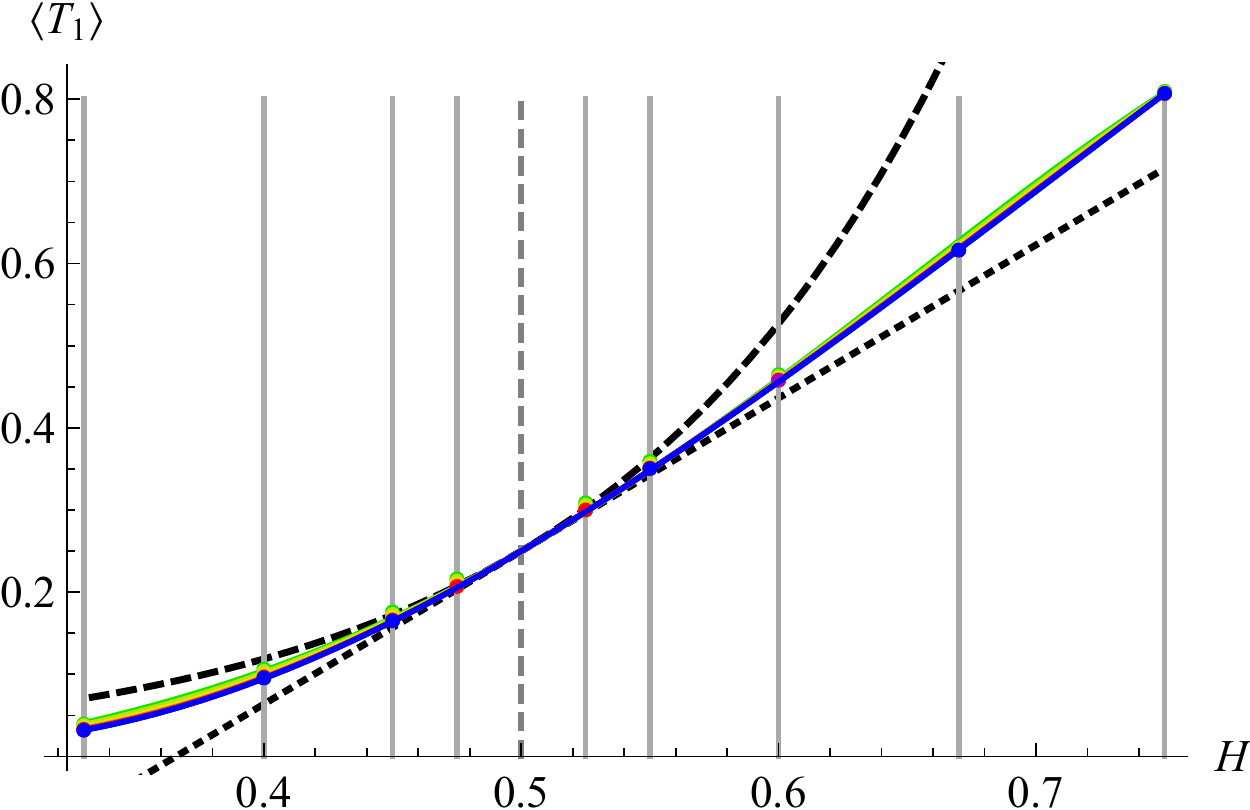}~~~~~~~~\!\!\!\!\raisebox{0cm}[0mm][0mm]{(b)}\!\!\!\!\!\!\fig{8.5cm}{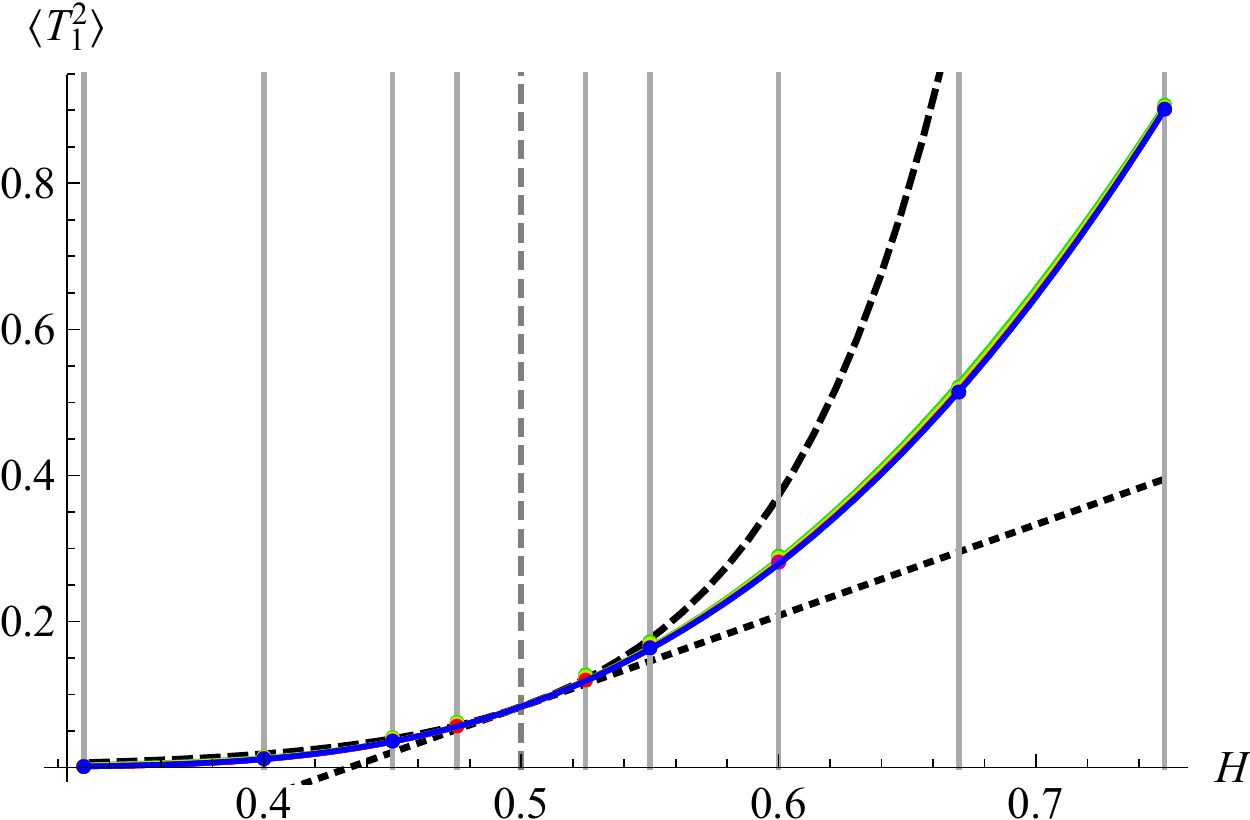}
\caption{Left: The expectation  $\left< T_1\right> $. The data-points with interpolation follow the color code of Fig.~\ref{f:all-P-T1}.  The black dotted line from \Eq{T1-full} is the direct expansion in $\epsilon$, which  underestimates the numerical result, while its exponentiation given by the dashed line   overestimates it. Right: {\em ibid.} for $\left<T_1^2\right> $, with the analytical result given in \Eq{T12-full}.}
\label{f:T1-full}
\end{figure*}

Our algorithm presented in section \ref{s:algo} to determine $P'_1(x)$ first determines the time the span (running max minus running min) reaches 1. For  Brownian motion, its probability distribution was given in Eqs.~(\ref{PT1}) and (\ref{PT1-Poisson}). As we have seen in section \ref{s:T1-corr}, it gets corrected for $H\neq 1/2$, in a way we are currently unable to obtain analytically. The question we   ask is how much does it differ from the result for Brownian motion? 
The most important effect is a change in time scale, which we estimated in Eq.~(\ref{T1-full}). Our numerical estimates, shown on Fig.~\ref{f:T1-full}
lead to 
\bea
\left< T_1 \right> &=& \frac14 \left[  
1 +7.4 \epsilon +10.1  \epsilon^2- 18 \epsilon^3 + ... \right] \ ,\\
\left< T_1^2 \right> &=& \frac1{12} \left[  
1 + 15 \epsilon +75  \epsilon^2+110 \epsilon^3 + ... \right]\ .\qquad 
\eea
This is in   good agreement with \Eqs{T1-full}-\eq{T12-full}, where the order-$\epsilon$ coefficients read $7.45$ and $14.96$, respectively.

In order to compare the full distributions, we     superimposed on the measured distribution the result for Brownian motion, rescaled s.t.\ the first moment $\left< T_1\right>$ is correctly reproduced. The result of this procedure is shown on figures \ref{f:all-P-T1} and \ref{f:all-P-T1-log}.
One can see deviations for large $|\epsilon|$, especially $H=0.75$, which on the whole rest  surprisingly  small.

Finally, it is easy to show analytically that for $H=1$, 
\be\label{8888}
P_{T_1}^{H=1}(t) = \sqrt {\frac2\pi} \frac{\rme^{-\frac1{2 t^2}}}{ t^2}\ .
\ee
As the distributions \eq{PH=1} and \eq{PH=1mean}, \Eq{8888} has a   large tail, leading  to  undefined moments.

\vspace*{-2mm}

\subsection{Finite-discretization effects}As we saw on Fig.~\ref{f:P'forH=0.33}, there are important finite-discretization corrections. This is even more visible on Fig.~\ref{f:all-Fs}, especially for $H=0.33$ (upper left corner). To better understand where this comes from, consider Fig.~\ref{f:running-max-and-min}. In the given example, the width $1$ is reached at $T_1\approx 0.85$, with $x_0\approx 0.58$. What is the error made, due to the finite discretization in time? 
Our argument will be made, as in the drawing, for a particle trajectory ``exiting at the upper boundary'', i.e.\ at $t_*$ the running max $M_+(t)$ is growing, whereas the running min $M_-(t)$ is constant. \begin{figure}[b!t]
\fig{8.5cm}{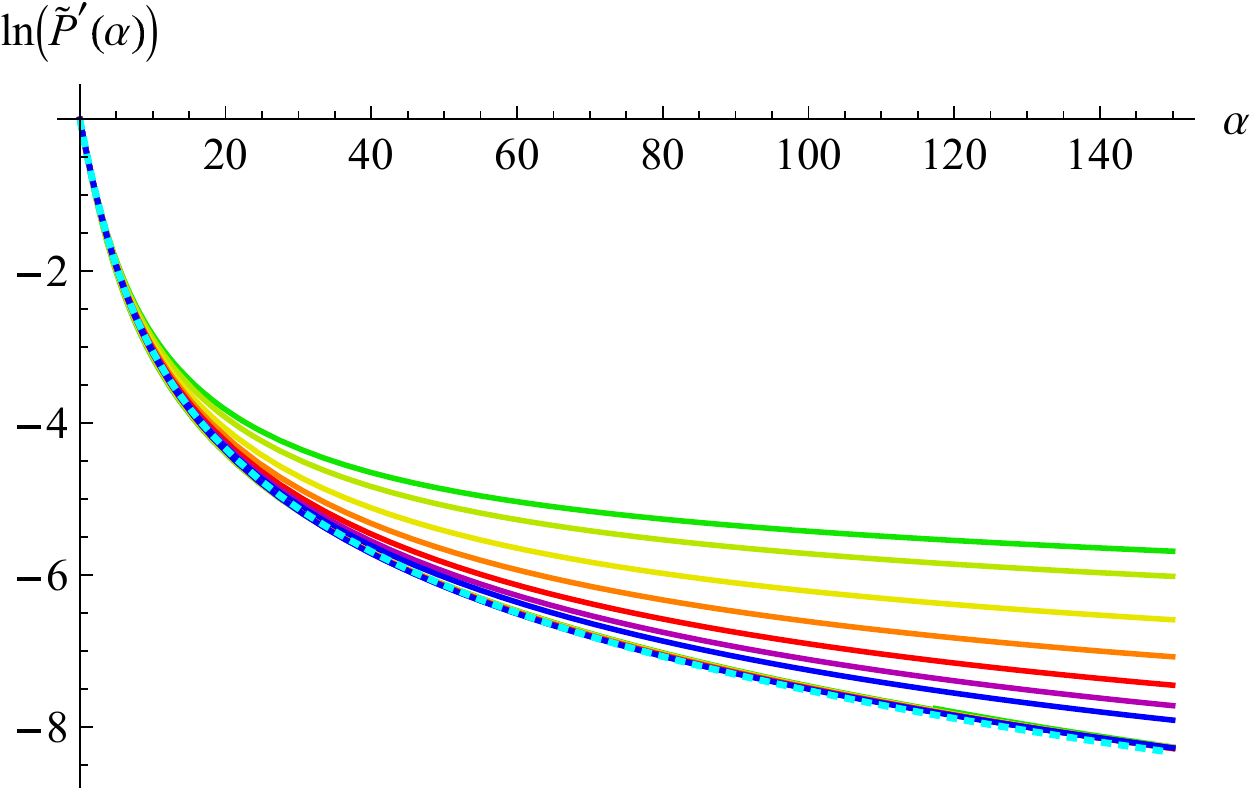}
\caption{Top 7 curves: The logarithm of  $\tilde P'(\alpha)$ (Laplace transform of $P'(x)$, as plotted on Fig.~\ref{f:P'forH=0.33}), for system sizes $N=2^{13}$ (top, green) to $N=2^{24}$ (blue). The lower solid curve are a scaling collapse of all 7 curves, using Eqs.~\eq{76} and \eq{77-bis}. The blue dashed line is the Laplace transform of the scaling ansatz \eq{P'scaling}, plotted on Fig.~\ref{f:P'forH=0.33}.}
\label{f:Palpha}

\end{figure}
The error in estimating the max is without consequences: while the true running max could be  underestimated, this would only result in a slight underestimation of $t_*$. The problem in estimating $M_-(t)$ is more severe: if we underestimate the true minimum by $\delta$, then   $x_0$ is 
\be
x_0 = M_-(t) +\delta\ .
\ee
Denote $P_N^{\rm miss}(\delta)$ the distribution of $\delta$. Close to the lower boundary, 
the   probability at system size $N$, $P'_{N}(x)$, is
\be\label{73}
P'_{\rm N}(x) = \int_{0}^\infty \rmd \delta \, P_N^{\rm miss}(\delta) P'_{N=\infty}(x-\delta)\ .
\ee
The function $P_N^{\rm miss}(\delta)$  should be a function of $\delta/\tau^{H}$, 
where $\tau=1/N$, thus
\be
 P_N^{\rm miss}(\delta) = P^{\rm miss}\left(  \delta N^H\right) \ .
\ee
Transforming to Laplace variables, \Eq{73}   reads (with the tilde indicating the Laplace transform)
\be
\tilde P'_{ N}(\alpha) = \tilde P'_{N=\infty}(\alpha)  \tilde P^{\rm miss}(\alpha/N^{H} )\ .
\ee
Taking a log and rearranging yields
\be\label{76}
\ln \left( \tilde P'_{N=\infty}(\alpha) \right)  = \ln\left( \tilde P'_{ N}(\alpha) \right) - \ln \left( \tilde P^{\rm miss}(\alpha/N^{H} ) \right) \ .
\ee%
We currently have no theory for $\tilde P_{\rm miss}(\alpha)$, but we find that a decent approximation for $H=0.33$ is given by 
\be\label{77-bis}
\ln \left (\tilde P_{\rm miss}(\alpha) \right)\approx 0.38 \ln \left(\tilde P_{\rm bridge}(1.7 \alpha ) \right) \ .
\ee
The function $P_{\rm bridge}(m)$ is the maximum of a fBm bridge for duration $T=1$, as given in Eq.~(90) of Ref.~\cite{DelormeWiese2016b}. 
Note that the numerical values of $0.38$ and $1.7$ are not significant. (Increasing one will decrease the other). 
As stated above, this formula is a guess, based on the following observations and hypothesis: Suppose that we   measured a maximum $m$ at time $t$, that the true maximum is between times $t$ and $t+1/N$, and that at time $t+1/N$ the process again achieves its maximum (it should be smaller than the maximum), then for Brownian motion the probability for the true maximum will be given by \eq{77-bis} with the numerical values put to 1. For a fBm, we again use the bridge process with the appropriate $H$. As for $H\neq 1/2$ the process is correlated, our ansatz induces a new error as it neglects correlations with the positions of the other points. The numbers induced above seem to  compensate for the approximations made.  

These arguments are illustrated on Fig.~\ref{f:Palpha}. The top seven curves show $\ln\!\big( \tilde P'_{ N}(\alpha) \big)$ for   system sizes $N=2^{13}$ (green, top), to $N=2^{24}$ (blue, second last curve from the bottom). The remaining lower curve is a scaling collapse estimating $\ln \big( \tilde P'_{N=\infty}(\alpha) \big) $, with $P_{\rm miss}(\alpha) $ given in \Eq{77-bis}. The dashed line (cyan) is the Laplace transform of the scaling ansatz \eq{P'scaling}.

Our analysis shows that {\em (i)} discretization corrections are important, {\em (ii)} they come from an underestimation of the extension of the process on the side at which it does not exit, {\em (iii)} there exists a correcting function one should be able to calculate analytically. The latter task is left for future research. 


\enlargethispage{2cm}
\section{Conclusion}
\label{s:conclusion}

In this article, we   considered the two-sided exit problem from a strip. We gave analytic results for the exit probabilities and times, in an expansion in $H-1/2$. While our numerical simulations confirm our findings, they also point to a fundamental problem: If the observation of the underlying process is not done  continuously but at a finite number of equally spaced times, the situation is exactly as in our numerical simulations. As a consequence, one may  not see the predicted analytic form, and not even the correct scaling laws. It will be important to   quantify these effects  to properly interpret experimental data once they appear. 

We also considered the probability distribution for the time the span (running max minus running mean), i.e. the area the process has visited, reaches 1. We gave analytic expressions of this probability for Brownian motion,  seemingly absent from the literature. For an fBM, evaluating corrections to its first moment analytically allows us to give a rather good approximation for this observable, without any adjustable parameter. 

Our results can   be extended to include drift, and to one or two reflecting boundaries. 

\bigskip
\medskip

\centerline{\bf ACKNOWLEDGMENTS}

\bigskip
It is a pleasure to thank Olivier Benichou, Mathieu Delorme, Alberto Rosso, Alejandro Kolton, Tridib Sadhu and Sidney Redner for stimulating discussions, and the two   referees for their constructive remarks.

\vfill

%
%

\ifx\doi\undefined
\providecommand{\doi}[2]{\href{http://dx.doi.org/#1}{#2}}
\else
\renewcommand{\doi}[2]{\href{http://dx.doi.org/#1}{#2}}
\fi
\providecommand{\link}[2]{\href{http://#1}{#2}}
\providecommand{\arxiv}[1]{\href{http://arxiv.org/abs/#1}{#1}}

\small

\tableofcontents

\end{document}